%% file: VS_PSC.tex
\documentclass[reprint,amsmath,amssymb,aps,twoside]{revtex4-2}

\usepackage{graphicx}
\usepackage{epstopdf}
\usepackage{dcolumn}
\usepackage{bm}
\usepackage{epsf}
\usepackage{verbatim}
\usepackage{amsmath}
\usepackage{amssymb}
\usepackage{color}
\usepackage{verbatim} 
\usepackage[normalem]{ulem} 
\usepackage{mathtools}
\usepackage{amsfonts}
\usepackage{relsize}
\usepackage{mwe,tikz}
\usepackage{xcolor}
\definecolor{darkgreen}{rgb}{0.0,0.5,0.0}
\usepackage{multirow}
\usepackage{threeparttable}
\usepackage{appendix}
\usepackage{aas_macros}
\usepackage{bibunits}
\usepackage[percent]{overpic}
\usepackage{fancyhdr} 
\usepackage{longtable}
\usepackage[normalem]{ulem}

\input{Definitions.tex}

\def\myfig#1{Figures/#1}

\def\DrawFig#1{#1}
\def\MYepsfbox#1{\includegraphics[width=18cm]{#1}} 
\def\MYSepsfbox#1{\includegraphics[width=8cm]{#1}} 

\newcommand{\Msun}{{M_{\odot}}}
\newcommand{\till}{{\mbox{--}}}
\newcommand{\tsh}{{\tau_{\rm sh}}}
\newcommand{\dgrdot}{{\overset{^\circ}{.}}}
\newcommand{\tauMax}{{\tau_{\rm max}}}
\newcommand{\thetaMin}{{\theta_{\rm min}}}
\newcommand{\thetaMax}{{\theta_{\rm max}}}
\newcommand{\tefh}{{\te_{500}}}
\newcommand{\Fermi}{{\emph{Fermi}}}

\newcommand{\Myc}{{\mathsf{c}}}
\newcommand{\Nclust}{{N}_{c}}
\newcommand{\NclustX}{{N}_{c,X}}
\newcommand{\NclustR}{{N}_{c,\nu}}
\newcommand{\sumclust}{\sum\limits_{\Myc=1}^{\Nclust}}

\newcommand{\NS}{\mathcal{N}}
\newcommand{\FS}{\mathcal{F}}
\newcommand{\MS}{\mathcal{M}}

\newcommand{\mach}{\Upsilon}

\newcommand{\CW}{CW }
\newcommand{\CWt}{\mbox{\tiny{CW}}}
\newcommand{\SW}{SW }
\newcommand{\SWt}{\mbox{\tiny{SW}}}

\newcommand{\Xflux}{{\rm {\,erg\,s^{-1}\, cm^{-2}}}}
\newcommand{\lumin}{{\rm {\,erg\,s^{-1}}}}
\newcommand{\myP}{{\mathbb{P}}}
\newcommand{\sigmaTS}{{\sigma_{\tiny{\mbox{TS}}}}}
\newcommand{\arcsec}{{''}}
\newcommand{\arcmin}{{'}}

\renewcommand\thesection{\arabic{section}}

\usepackage[colorlinks,citecolor=darkgreen]{hyperref} 

\numberwithin{equation}{section}

\newcommand{\MyTitle}{{Excess cataloged X-ray and radio sources at galaxy-cluster virial shocks}}
\newcommand{\MySTitle}{{Virial-shock sources}}
\newcommand{\MySAuthor}{{Ilani, Hou \& Keshet}}

\defaultbibliography{Virial}
\defaultbibliographystyle{apsrev4-2-author-truncate}

\begin{document}

\title{\MyTitle}

\pagestyle{fancy}
\fancyhead[RE]{\MySTitle\,\,\,\,\thepage}
\fancyhead[LE]{}
\fancyhead[LO]{\thepage\,\,\,\,\MySAuthor}
\fancyhead[RO]{}
\renewcommand{\headrulewidth}{0pt}

\author{Gideon Ilani}

\altaffiliation{Posthumously. Gideon Ilani was killed in action on December 10, 2023. This paper is based on his research.}
\affiliation{Physics Department, Ben-Gurion University of the Negev, POB 653, Be'er-Sheva 84105, Israel}

\author{Kuan-Chou Hou}
\altaffiliation{Institute of Astronomy and Astrophysics, Academia Sinica, PO Box 23-141, Taipei 10617, Taiwan}
\affiliation{Physics Department, Ben-Gurion University of the Negev, POB 653, Be'er-Sheva 84105, Israel}

\author{Uri Keshet}

\altaffiliation{keshet.uri@gmail.com}

\affiliation{Physics Department, Ben-Gurion University of the Negev, POB 653, Be'er-Sheva 84105, Israel}


\begin{abstract}
We detect a highly significant excess of X-ray (2RXS) and radio (NVSS, GMRT, VLSSr) catalog sources when stacked around MCXC galaxy clusters and groups, narrowly confined within $\lesssim100\mathrm{\,kpc}$ of the $\sim2.4 R_{500}$ virial shock radius (inferred from previous continuum stacking), with similar X-ray ($\sim4\sigma$ for $443$ clusters) and radio ($\sim4\sigma$ for $485$ clusters) characteristics ($>5\sigma$ joint).
The excess sources show $10$--$100$ kpc scales, $L_X(0.1\mbox{--}2.4\mbox{ keV})\simeq 10^{42\mbox{\scriptsize--}43}\mathrm{\,erg\,s^{-1}}$ or $\nu L_\nu(\nu=1.4\mathrm{\,GHz}) \simeq 10^{40{\mbox{\scriptsize--}}41}\mathrm{\,erg\,s^{-1}}$ luminosities, and a preferentially radial radio-polarization.
The narrow localization and properties of the excess identify these sources not as AGN, often invoked speculatively for excess X-ray sources at cluster outskirts, but rather as infalling gaseous clumps interacting with the virial shock, probably galactic halos and possibly outflow remnants.
The local excess of such discrete, radio-to-$\gamma$-ray sources around an object can probe its virial shock also at high redshifts and sub-cluster scales.
\end{abstract}

\maketitle

\begin{bibunit}[apsrev4-2-author-truncate]

\section{Introduction}
\label{sec:Intro}

Several studies have reported a radially-broad excess of X-ray sources in the outskirts of galaxy clusters.
These include
a $\gtrsim5\sigma$ excess at $1.5\lesssim r\lesssim3\Mpc$ radii in stacked \emph{Chandra} data around 24 relaxed, redshift $0.3<z<0.7$ MACS clusters  \citep{ruderman2005origin},
a $\sim4\sigma$ excess at $2\lesssim r\lesssim 3\Mpc$ in stacked \emph{XMM-Newton} data
around 22 luminous, $0.9<z\lesssim1.6$ clusters \citep{FassbenderEtAl12},
and a
statistically significant excess at normalized, $2<\tau\equiv r/R_{500}<4$ radii in stacked \emph{XMM}-LSS field data around 19 clusters  of $0.14<z<0.35$
\citep{koulouridis2014x}.
Here, the subscript $500$ refers (henceforth) to the radius around the center of a cluster enclosing a mass density $\rho$ that is $500$ times larger than the critical mass density of the Universe.
The excess sources were usually assumed to be associated with active galactic nuclei (AGN).
Their excess at the cluster periphery was tentatively attributed to an enhanced rate of major galaxy mergers \citep[\eg][and references therein]{ruderman2005origin, EhlerEtAl15}.
Part of the excess was speculated to arise from gravitational lensing of background quasi-stellar objects \citep{koulouridis2014x} or from foreground structures \citep{FassbenderEtAl12}.

In contrast, other analyses, including larger studies, have found no excess of peripheral sources.
In particular, no X-ray source excess was found beyond $r=1\Mpc$ using \emph{Chandra} data around 27 disturbed, $0.3<z<0.7$ MACS clusters \citep{ruderman2005origin} or 148 clusters of $0.1<z<0.9$ \citep{GilmourEtAl09},
and no excess of optical, IR or radio-selected cataloged AGN was found at the outskirts of 2300 infrared-selected clusters \citep[][]{mo2018massive}.
Searches for a source excess performed after normalizing distances to typical cluster scales such as $R_{500}$ did not show a consistent signal, either; for example, a local deficit of X-ray sources was found at $2<\tau<3$ around 14 clusters of $0.43<z<1.05$ \citep[][]{koulouridis2014x}.

More recently, a $\sim3\sigma$ excess of \emph{Chandra} X-ray sources was reported \citep{koulouridis2019high} in a fairly narrow, $2<\tau<2.5$ range of normalized radii around a small sample of five $z\sim 1$, mostly disturbed clusters.
This signal is dominated by two clusters, each with about ten excess $2.0<\tau<2.5$ sources with respect to neighbouring bins.
The excess sources show $10^{42.5}\lumin \lesssim L_{0.5\mbox{ \scriptsize keV}}^{8.0\mbox{ \scriptsize keV}} \lesssim 10^{44}\lumin$ luminosities, where energy or frequency subscripts (superscripts) denote low (high) band limits, henceforth.
These sources were again tentatively attributed to enhanced AGN activity associated with excessive galaxy mergers or gas stripping.

However, an excess of AGN over such a narrow range of radii at the outskirts of clusters would be unnatural, especially taking into account projection (and, in other
studies, also stacking) effects.
Indeed, the rate of galaxy mergers is thought to evolve quite gradually, rising well before the accreted galaxies reach the cluster outskirts, and remaining high after pericentric crossing \citep[\eg][]{VijayaraghavanRicker13}.
AGN activity can be affected by the intracluster medium (ICM) through several channels, notably, in this context, by ram pressure stripping of hot halo gas \citep[\eg][]{McCarthyEtAl08},
but these processes are thought to be important only once the galaxy approaches the central, $r\lesssim 1\Mpc$ region of the cluster \citep[][and references therein]{VijayaraghavanRicker13}.

Therefore, a sharp spike in the radial profile of the density of sources would indicate a localized, sudden change in the environment.
Interestingly, the structure formation, or virial, shock of a galaxy cluster, long anticipated to lie in the $2<\tau<3$ range according to cosmological simulations \citep[\eg][]{KeshetEtAl03, SchaalSpringel15}, provides a natural candidate for such a sharp environmental transition.
Indeed, as an infalling object crosses the virial shock (VS), it should see the ambient pressure suddenly jump by orders of magnitude as the infalling surrounding plasma is violently slowed down and heated, accompanied by the generation of magnetic fields and shock-accelerated relativistic particles (cosmic rays, CRs) carrying a substantial fraction of the elevated thermal energy.
While crossing a VS is unlikely to trigger an AGN,
and the processes by which it could lead to the emergence of bright excess X-ray or radio sources may be a-priori unclear,
the VS could explain at least the position and sharpness of the signal.

Moreover, recent studies have identified stacked VS signals in precisely the same $2<\tau<2.5$ region harboring the excess \citep{koulouridis2019high} X-ray sources.
Such stacking analyses include a $>5\sigma$ signal at $2.2\lesssim \tau \lesssim 2.5$ in \emph{Fermi}-LAT \gama-ray data around 112 extended clusters, interpreted as CR electrons (CREs) inverse-Compton (IC) scattering cosmic microwave background (CMB) photons \citep{ReissEtAl17, reiss2018detection};
a $\gtrsim5\sigma$ synchrotron counterpart at $2.4\lesssim\tau\lesssim2.6$ in OVRO-LWA radio data around 44 massive clusters \citep{HouEtAl23};
a significant drop around $2.0\lesssim\tau\lesssim2.6$ in \emph{Planck} $y$-parameter around 10 galaxy groups \citep{PrattEtAl21}; and
a projected $y$-parameter drop starting at $\tau\simeq2$ in South Pole Telescope data around 500 clusters \citep{Anbajagane2022}.
Such stacked VS signals corroborate the IC, synchrotron, and Sunyaev-Zel'dovich (SZ) signals detected separately or simultaneously in individual nearby clusters \citep{KeshetEtAl17, keshet2018evidence, HurierEtAl19, keshet20coincident}, and roughly coincide with the splashback feature inferred from a localized drop in the logarithmic slopes of galaxy density profiles \citep{MoreEtAl16, ShinEtAl19}.
Interestingly, stacked data suggest a secondary, weaker and broader IC \citep{reiss2018detection}, synchrotron \citep{HouEtAl23}, and possibly also SZ \citep{Anbajagane2022} signal around $\tau\simeq 6$; if verified, this would suggest an elongated averaged VS morphology of $\sim2.5$ ellipticity ratio, as indicated in Coma \citep{KeshetEtAl17, keshet2018evidence}.

Following these VS detections and reports of excess peripheral X-ray sources, we carry out an all-sky analysis of cataloged X-ray and radio sources in the peripheries of galaxy clusters and groups (henceforth clusters, for brevity).
To this end, X-ray sources from the Second \emph{ROSAT} All-Sky Survey (RASS) Source Catalog \citep[2RXS;][]{2RXS_paper} and
radio sources from the National Radio Astronomy Observatory (NRAO) Very Large Array (VLA) Sky Survey \citep[NVSS;][]{NVSS_paper} are stacked around extended clusters selected from the Meta-Catalog of X-ray detected Clusters of galaxies \citep[MCXC;][]{PiffarettiEtAl11}.
We find an excess of sources in both catalogs (with respect to the field, dominated by foreground or background sources seen in projection), with striking similarities between X-ray and radio signals, localized surprisingly narrowly near the predetermined VS radius.
The radio signal is verified also in the Giant Metrewave Radio Telescope (GMRT) All-Sky $150\MHz$
\citep{GMRT_paper},
and the Very Large Array (VLA) Low-Frequency ($74\MHz$) Sky Survey Redux (VLSSr)
\citep{VLSSR_2014}
source catalogs, used to estimate the radio-source spectra.
We then statistically characterize the properties of the excess sources,
and constrain the physical processes by which the VS could lead to such a local excess of X-ray and radio sources.

The paper is organized as follows.
We describe the cluster and source catalogs, along with our selection criteria, in \S\ref{sec:data_prep}, and outline our stacking and modelling methods in \S\ref{sec:analysis}.
The results are presented and phenomenologically modelled in \S\ref{sec:results}.
The properties of the excess sources are inferred statistically in \S\ref{sec:characteristics}, and their physical origin is explored in \S\ref{sec:origin}.
Finally, our results are summarized and discussed in \S\ref{sec:discussion}.

The Supplementary material provides supporting and auxiliary information regarding the spectra of the excess X-ray sources
(\S\ref{app:X_ray_emission_models}), limitations on source size estimates (\S\ref{app:cat_properties}), the evaluation of excess confidence levels (CLs) under Poisson statistics (\S\ref{app:Poisson}), various convergence and sensitivity tests (\S\ref{app:sensitivity_and_methods}), additional properties of the excess sources (\S\ref{App:characteristics}), and the absence of a significant VS signal from 2RXS--NVSS pairs (\S\ref{subsec:joint_xray_radio_sources}) or identified AGN (\S\ref{app:AGN_identification}).
All (field+excess) sources found near the VS are listed in \S\ref{app:SourceTables}.

We adopt a $\Lambda$CDM cosmological model with a Hubble constant $H_0=68\km\se^{-1}\Mpc^{-1}$, a mass fraction $\Omega_m=0.3$, and a $76\%$ hydrogen mass fraction, yielding a mean particle mass $\bar{m}\simeq 0.59m_p$.
The center of a galaxy cluster is defined as its X-ray peak, and an adiabatic index $\Gamma=5/3$ is assumed for the ICM.
Confidence intervals are $68\%$ containment projected for one parameter.

\section{Data preparation}
\label{sec:data_prep}

Here, we describe the catalogs and our data preparation procedure.
The MCXC cluster catalog is discussed in \S\ref{subsec:MCXC}, the 2RXS catalog of X-ray sources (subscript $X$)
in \S\ref{subsec:2RXS}, and the NVSS radio catalog (subscript $\nu$) in \S\ref{subsec:NVSS}.
Table \ref{tab:psc} summarizes the main properties of the catalogs.

\subsection{Cluster catalog and selection criteria}
\label{subsec:MCXC}

The MCXC catalog of galaxy clusters \citep{PiffarettiEtAl11}, based on a compilation of the \emph{ROSAT} all-sky survey and \emph{ROSAT} serendipitous catalogs, is particularly suitable for our purposes as it facilitates a uniform normalization to a fixed overdensity.
The catalog, with 1743 nearby clusters, was used in previous stacking analyses \citep{reiss2018detection, HouEtAl23}, and provides the redshift and characteristic length scale $R_{500}$ of each cluster, so the corresponding angular scale $\theta_{500}$ can be estimated.

We apply the following
nominal cuts to the MCXC,
in order to select the clusters most suitable for source stacking and to avoid contaminations.
\begin{enumerate}
    \setlength\itemsep{0.3em}
    \item
    Sufficiently extended clusters to resolve the virial ring.
    The source catalog resolutions are $45\arcsec$ (FWHM in radio) and $\sim 2\arcmin$ (detection cell in X-rays), whereas the thickness of the virial ring in continuum emission was estimated as $\sim\theta_{500}/3$ \citep{reiss2018detection, HouEtAl23}, so we
    conservatively require $\theta_{500}>3\cdot 2\arcmin=6\arcmin$.
    This cut is independently justified as it coincides with the MCXC turnaround angle, below which the number of detected clusters diminishes.
    \item
    Clusters that are less extended than typical foreground structures.
    The power spectrum of sources in the 2RXS (NVSS) catalog shows substantial structure on $\gtrsim 3\dgr$ ($\gtrsim 1\dgr$) scales,
    suggesting extended foregrounds.
    To avoid contaminating the VS region with such putative structures, we impose an upper, $\theta_{500}\le 0\dgrdot3$ limit.
    \item
    Sufficiently massive galaxy groups.
    In order to focus on galaxy clusters and rich galaxy groups, but avoid individual galaxies and poor groups which may not show the same type of peripheral sources,
     we impose a lower $M_{500}\ge 10^{13}\Msun$ mass limit.
    \item
    Avoiding confusion in regions of a high sky-projected density of clusters.
    The MCXC shows such high densities, with nearly overlapping projected virial radii, in two regions, around Galactic coordinates $\{l,b\} \sim \{315^\circ,32^\circ\}$ and $\sim \{12^\circ,50^\circ\}$ \citep{reiss2018detection}. We avoid these regions, by excluding all clusters within $5^{\circ}$ of these two sky positions.
    This cut is in practice redundant, as the other cuts already
    exclude all clusters in these two busy regions.
    \item
    Avoiding large Galactic structures.
    The Loop-I and Fermi bubble regions are excluded from the nominal analysis, to avoid any possible effects they may have on the catalogs.
    For this purpose, we adopt simplified models of these two structures.
    Loop-I is modeled as a disk centered on $\left\{l,b\right\} =  \left\{329,17.5\right\}$, with a radius of $55^{\circ}$ \citep[e.g.][]{Haslam1970, Dickinson2018}.
    The Fermi-bubble is crudely modeled as the region defined by $\left|b\right| \le 55^{\circ}$ and either $l<30^{\circ}$ or $l>330^{\circ}$.
    \item
    Additional cuts are imposed on the Galactic latitude $|b|$, to avoid the bright Galactic plane, and on the declination, $\mbox{Dec}$.
    These cuts are specific to each source catalog, so are detailed below.
\end{enumerate}

The above cuts leave a reduced sample of $\NclustX = 443$ clusters for stacking X-ray sources, and a sample of $\NclustR =485$ clusters for radio stacking.
These clusters have redshifts in the range $ 0.017<z<0.24$, and masses in the range $10^{13}\Msun<M_{500}< 9 \times10^{14}\Msun$; these and additional sample properties are provided in Table \ref{tab:psc}.
The MCXC
includes $\sim 50$ ($\sim 90$) clusters which harbour another MCXC cluster inside (twice) their virial radius, about $80\%$--$90\%$ of which survive our cluster cuts.
These clusters with nearby neighbours are not excluded from our nominal analysis, but we verify
in \S\ref{app:sensitivity_and_methods} that excluding close pairs does not alter our results significantly.

\subsection{X-ray source catalog and cuts}
\label{subsec:2RXS}

\begin{center}
\begin{figure}[b!]
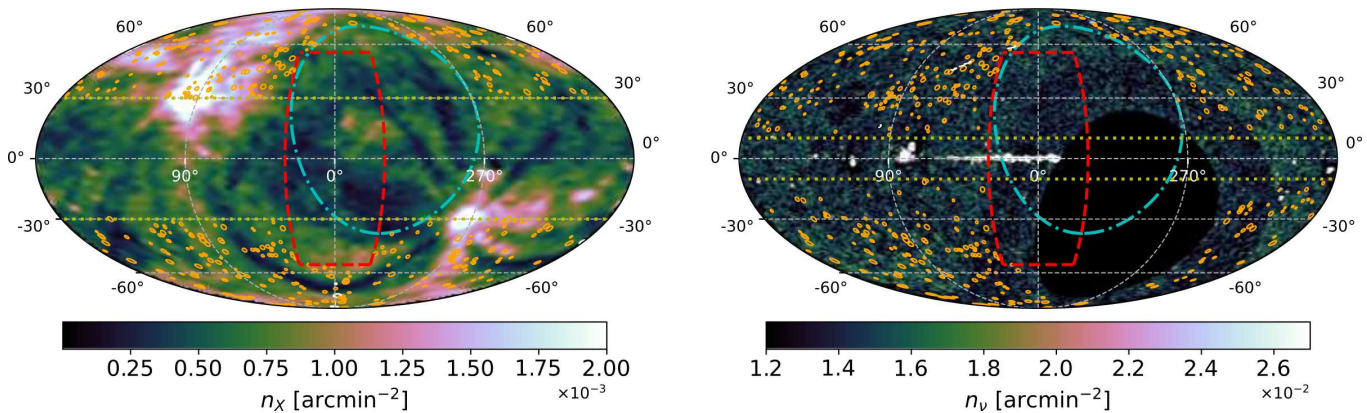

    \DrawFig{
    \centerline{
    \MYepsfbox{\myfig{src_clust_cut_LowRes.eps}}}
    }
	\caption{\label{fig:X_ray_srcs}\label{fig:radio_ray_srcs}
    Hammer-Aitoff all-sky projection of source density (cubehelix \citep{Green11_Cubehelix} color map) and cluster sample (orange circles of radius $5\tefh$), for X-ray (left panel; 2RXS) and radio (right panel, NVSS; note diminished exposure at low declinations, with no $\mbox{Dec}<-40\dgr$ sources) analyses.
    Also shown are the cuts of the Galactic plane (dotted yellow lines), the Fermi bubbles (dashed red curves), and Loop-I (dot-dashed cyan).
    }
\end{figure}
\end{center}

The 2RXS catalog is most suitable for stacking X-ray sources, as it is presently the deepest and cleanest X-ray all-sky survey available \citep{2RXS_paper}.
The catalog is based on RASS observations using the position-sensitive proportional counter (PSPC),
providing $\sim 1.3\times 10^5$
sources in the $[0.1,2.4]\keV$ energy range.
The projected density $n_X$ of X-ray sources on the sky is shown in Fig.~\ref{fig:X_ray_srcs} (left panel).
Table \ref{tab:moments_table} summarizes some relevant properties of the catalog and its sources (in column 3).

Softer X-rays undergo more photo-absorption by neutral hydrogen atoms in the Galaxy.
The resulting optical depth exceeds
unity for $\epsilon\simeq 0.1 \keV$  photons at $\left|b\right|\lesssim 30^{\circ}$ Galactic latitudes.
Therefore, for the nominal X-ray analysis, we place an extended cut on the Galactic latitude of the clusters, retaining only those with $\left|b\right|\ge 30^{\circ}$.

The number of sources in a catalog diminishes as one increases the threshold on the minimal likelihood $\mathcal{L}$ of source detection, defined such that the false-positive source probability is $e^{-\mathcal{L}}$.
The decline in the fraction of spurious sources as $\mathcal{L}$ increases has been estimated using
simulations \citep{2RXS_paper}.
The 2RXS catalog compromises on $\mathcal{L}\ge 6.5$, corresponding to $\gtrsim3\sigma$ source detection with a $\sim30\%$ fraction of spurious sources in the catalog.
Aiming for a smaller, $\sim10\%$ fraction of spurious sources, we adopt $\mathcal{L} \ge 8$ in our nominal analysis, corresponding to $\gtrsim3.4\sigma$ source detection.

\onecolumngrid
\twocolumngrid

\par\null\newpage
\clearpage

\onecolumngrid
\begin{center}
\begin{table}[t!]
    \caption{
    Catalog and stacking properties.
    \label{tab:moments_table}\label{tab:psc}
    }
    \vspace{0.3cm}
    \begin{center}
    \begin{tabular}{l@{\hskip 0.4cm}llll}
        & Property & Symbol\hspace{0.5cm} & 2RXS & NVSS\\
        \hline
        \multirow{7}{*}{\rotatebox{90}{Catalog}} & Photon range & $\epsilon,\nu$ & $0.1\till2.4\,\rm keV$ & $1.44 \pm 0.05\,\rm GHz$ \\
        & Sky coverage & & All-sky & $\mbox{Dec}>-40\dgr$ ($82\%$ sky) \\
        & Minimal flux$^a$  & $F_{\rm min}$ & $10^{-13}\,\Xflux$ &  $3.4\mJy$ ($99\%$ compl.) \\
        & Detection threshold & & $\mathcal{L}>6.5$ & $I_\nu>2 \mJy\rm{\,beam^{-1}}$ \\
        & Base map resolution & & $\sim 45\arcsec$ (pixel size)
        & $45\arcsec$ (FWHM)\\
        & Source size (see \S\ref{app:cat_properties}) & $r,\{a,b\}$ & $r\geq r_{psf}\simeq 40''$  & $14''<2b<2a<286''$ \\
        & Total source number & & $1.3 \times 10^5$ & $1.8\times10^6$\\
        \hline
         \multirow{4}{*}{\rotatebox{90}{Cuts }} &
         Latitude cut
         & $b$ & $|b|>30\dgr$ & $|b|>10\dgr$ \\
        & Cluster declination cut & Dec & --- & $\mbox{Dec}> -35\dgr$ \\
        & Source flux$^a$ cut & $F^*$ & $F_X<F_X^* \equiv 7\times10^{-12}\Xflux$ & $F_{\nu}>F_{\nu}^*\equiv100\mJy$ \\
        & Source likelihood cut & $\mathcal{L}$ & $\mathcal{L}>8$ & --- \\
        \hline
         \multirow{4}{*}{\rotatebox{90}{Clusters$^b$ }} & Number of clusters & $\Nclust$  & $\NclustX=443$ & $\NclustR=485$ \\
        & Redshift median (range) & $z$ & $0.077$ ($0.017\till0.228$) & $0.081$ ($0.017\till0.233$) \\
        & Angular radius median (range) & $\theta_{500}$ & $0\dgrdot15$ ($0\dgrdot10\till0\dgrdot30$) & $0\dgrdot14$ ($0\dgrdot10\till0\dgrdot30$) \\
        & Mass median (range) & $M_{500}$ & $1.5$ ($1.0\till8.7$) $\times10^{14}M_\odot$ & $1.6$ ($1.0\till9.0$) $\times10^{14}M_\odot$  \\
        \hline
        \multirow{4}{*}{\rotatebox{90}{Results$^b$}} & Mean field source density & $f$ & $5360\sr^{-1}$ & $5760\sr^{-1}$ \\
        & ICM-region sources (expected field) & $\NS$ ($\FS$) & 107
        ($\sim$82) & 102
        ($\sim$92) \\
        & VS-region sources (expected field) & $\NS$ ($\FS$) & 119
        ($\sim$77) & 128
        ($\sim$87) \\
        & VS SW-overdensity (significance)  & $\delta_{\rm SW}$ & $0.54$ ($4.4\sigma$ Poisson-corrected) & $0.47$ ($4.0\sigma$) \\
        \hline
    \end{tabular}
    \begin{tablenotes}
    $^a$ --- Energy flux (specific energy flux) in X-rays (radio). The cuts are explained in the text. \\
    $^b$ --- After cuts. Source numbers $\NS$ are summed over all clusters. VS sources are listed in \S\ref{app:SourceTables}. \\
    \end{tablenotes}
    \end{center}
\end{table}
\end{center}
\twocolumngrid

We define the X-ray energy flux $F_X \equiv F_{0.1\mbox{ \scriptsize keV}}^{2.4\mbox{ \scriptsize keV}}$
as the specific flux $F_{\epsilon}$ integrated over the \emph{ROSAT} band.
The catalog provides flux estimates based on three different emission models: a power-law spectrum, the optically-thin thermal plasma emission model Mekal \citep{MeweEtAl86}, and optically thick, black-body emission.
In most or all relevant cases, the model fits are not sufficiently accurate to identify which emission model is most appropriate.
Therefore, we henceforth adopt the power-law estimate of $F_X$, which is a simpler fit available for the largest number of sources; our results are not sensitive to this choice (see \S\ref{app:sensitivity_and_methods}).
We nevertheless consider below the fitted source temperatures, as well as
the fitted photon power-law index $\Gamma_X$ defined by $F_\epsilon\propto \epsilon^{1+\Gamma_X}$.

\begin{center}
\begin{figure}[h!]
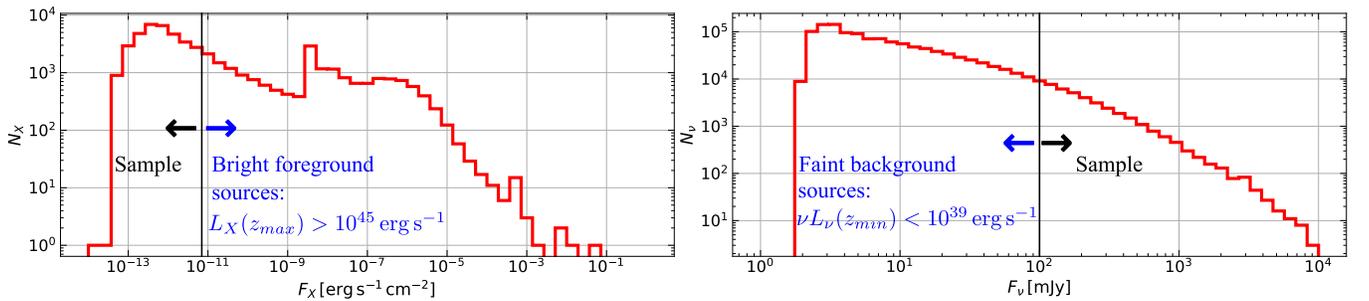

    \centerline{
    \DrawFig{\MYepsfbox{\myfig{all_typ_flux.eps}}}}
	\caption{\label{fig:all_typ_flux}
    The logarithmic energy flux $F_X$ (specific flux $F_\nu$) histogram of 2RXS (NVSS) sources in the analyzed sky is shown in the left (right) panel,
    along with our nominal flux thresholds (black vertical lines with explanatory labels).
    Our results are not sensitive to the precise choice of thresholds; see text.
  	}
\end{figure}
\end{center}

Figure \ref{fig:all_typ_flux} (left panel) shows the energy flux histogram of X-ray sources in the analyzed sky region.
The 2RXS catalog is dominated by foreground sources brighter than the stacked clusters themselves.
To reduce foreground confusion, we impose an upper flux limit that removes sources whose luminosities would exceed the typical, $L_X \simeq10^{45} \lumin$ luminosity of the entire ICM even around the most distant ($z_{\rm max}\simeq0.23$) clusters in our sample,
thus retaining only
$F_X<F_X^*\equiv 7\times10^{-12}\Xflux$ sources.
Note that this $F_X^*$ cutoff is already quite close to the 2RXS $\sim 10^{-13} \Xflux$ sensitivity threshold \citep{2RXS_paper}.
In \S\ref{app:sensitivity_and_methods}, we show that our results are not sensitive to the precise choice of flux cut;
for example,
a high $F_X^*=10^{-10} \Xflux$
changes the significance of excess virial sources by only $\sim 2\%$.
In fact, a significant, $\sim4\sigma$ signal persists even with no flux cut at all, despite the strong foreground.

\onecolumngrid
\twocolumngrid


\subsection{Radio source catalog and cuts}
\label{subsec:NVSS}

We use the extensive NVSS catalog for radio sources, unless otherwise stated.
This catalog covers only $\sim 82\%$ of the sky, but radio waves are not substantially absorbed in the galaxy, so the catalog can be used at lower latitudes than in X-rays.
Consequently, NVSS allows for the stacking of a similar and even slightly larger number of clusters than with 2RXS.
The projected density $n_\nu$ of catalog radio sources on the sky is shown in Fig.~\ref{fig:radio_ray_srcs} (right panel).
Table \ref{tab:moments_table} summarizes some relevant properties of the catalog and of its sources (in column 4).

The NVSS catalog covers the northern sky at declinations $\mbox{Dec} > -40^{\circ}$ (J2000).
Figure \ref{fig:radio_ray_srcs} shows an apparent decline in $n_\nu$ with decreasing $\mbox{Dec}$ even somewhat north of this declination.
Therefore, we impose a corresponding cut on the declination of clusters used for radio stacking,
nominally keeping only those with $\mbox{Dec}>-35\dgr$.

While Galactic absorption does not impose a substantial difficulty at the $1.4\GHz$ frequency of the NVSS, there is a marked excess of Galactic foreground sources at $|b|\lesssim 3\dgr$. To avoid these sources, we place a
conservative cutoff on the latitudes of clusters used in the radio analysis, retaining only those with $|b| \ge 10\dgr$.

The NVSS catalog does not provide information concerning the likelihood of radio source detection, but it does imposes a minimal $I_\nu>2\mJy\text{ beam}^{-1}$ intensity threshold, leading to a specific flux (\ie spectral flux density) distribution with $F_\nu\gtrsim 2.5\mJy$; for sources above $3.4 \mJy$, the NVSS completeness is $99\%$  \citep{NVSS_paper}.

A histogram of the specific flux of NVSS sources is shown in the right panel of figure \ref{fig:all_typ_flux}.
This catalog is strongly dominated by faint background sources (in contrast to the bright foreground source of 2RXS),
due in part to star-forming galaxies, whose luminosity function peaks near $\nu L_{\nu} \sim 10^{39}\lumin$ \citep[\eg][]{radio_Lum}; for the most part, these sources are not expected to be associated with the low-redshift clusters in our sample.
We therefore impose a lower specific-flux limit that removes sources with luminosities lower than this peak luminosity even around the nearest ($z_{\rm min}\simeq 0.017$) clusters in our sample, thus retaining
only sources of $F_{\nu}\gtrsim F_{\nu}^* \equiv 100 \mJy$.
In \S\ref{app:sensitivity_and_methods}, we show that changing the radio flux cutoff by $50\%$
has only a $\sim 15\%$ effect on the significance of our results, and our choice is conservative.
Interestingly, the signal persists even if one avoids a flux cut entirely, thus retaining the large number of faint background sources, but expectedly diminishes to the $\sim2\sigma$ level.

\section{Stacking analysis method}
\label{sec:analysis}

We normalize sky distances around each cluster by its $\theta_{500}$, radially bin the data onto circular annuli about the center of the cluster, and stack the resulting normalized, binned, radial distribution over all clusters in the sample.
Stacking is performed for X-ray and radio sources separately, although we later consider also the joint distribution and the significance of the joint excess.
The following procedure thus pertains separately to the stacking of 2RXS and NVSS sources, as well as GMRT and VLSSr sources considered later.

For each cluster, we define the normalized radius $\tau \equiv \te/\tefh$, where $\te$ is the angular distance from the center of the cluster.
Alternatively, in order to generate stacked 2D maps (shown in \S\ref{sec:results} and in \S\ref{app:sensitivity_and_methods}), \ie without radial binning,
we replace $\tau$ by its two-dimensional analog ${\bm \tau} \equiv \{\tau_x, \tau_y \}$.
For the purpose of stacking, we define the region of interest (ROI) as
$\tau<\tauMax=10$, choosing $\tauMax$ to be about
four times larger than the VS radius inferred from stacking {\Fermi}-LAT or LWA data.
Our results are not sensitive to the choice of $\tauMax$ and all other stacking parameters, as shown in \S\ref{app:sensitivity_and_methods}.

In order to test for a putative local excess of sources and determine its significance, one must estimate the sky-projected density of background or foreground sources, henceforth referred to colloquially as field sources, in the vicinity of each cluster.
In our nominal approach, we consider a fixed projected density of field sources around each cluster, measured outside its ROI, in the so-called field region defined by $\thetaMin<\theta<\thetaMax$.
We choose $\thetaMin=3\dgr$ slightly larger than the ROI radius, \ie $\thetaMin>\tauMax \theta_{500}$, of any cluster in our sample.
We take $\thetaMax=5\dgr$ sufficiently larger than $\thetaMin$, for good statistics and to overcome foreground fluctuations over $\gtrsim 1\dgr$ scales.
Alternative choices of $[\thetaMin,\thetaMax]$, as well as more elaborate, non-constant field models, are examined in \S\ref{app:sensitivity_and_methods}.

When testing for virial-shock signals, it can be useful to examine, in addition to the field region outside the VS, also the region just inward of the shock, beyond its range of influence.
For this purpose, we designate the range $1.0<\tau<1.5$ as a so-called ICM region.
The upper limit on this region is chosen sufficiently far from the virial radius to reduce the projected effects of the shock.
The lower limit is chosen to balance sufficient statistics with limited evolutionary effects and central-source contamination.

For a typical $\Delta\theta\sim \theta_{500}/3$ resolution for X-ray sources (and $\sim$twice better resolution for radio sources)
around our least extended clusters, and for consistency with \citep{reiss2018detection} and \citep{HouEtAl23}, we designate $\Delta \tau=0.25$ as our nominal radial bin size, but explore also smaller and larger bins.
We therefore designate the anticipated VS region as the $2.25<\tau<2.5$ bin, where the strongest VS signal is expected based on continuum \gama-ray \citep{reiss2018detection} and radio \citep{HouEtAl23} detections in stacked MCXC clusters.

For each catalog, $\tau$ bin, and cluster $\Myc$,
we compute the
solid angle $\Delta \Omega (\tau,\Myc)$ and
number $\NS(\tau,\Myc)$ of sources
falling within the bin.
The dimensionless
projected
density of catalog sources can now be estimated as
\begin{equation}
n(\tau,\Myc) = \frac{\NS(\tau,\Myc)}{\Delta\omega(\tau,\Myc)} \coma
\end{equation}
where
\begin{equation}
\Delta\omega(\tau,\Myc)\equiv \frac{\Delta\Omega(\tau,\Myc)}{\theta_{500,\Myc}^2}
\end{equation}
is the solid angle normalized to a square $\theta_{500}^2$ pixel.
The number $\FS(\tau,\Myc)$ of field sources anticipated in the bin is
extrapolated from the field region, giving the analogous dimensionless
projected
field density \begin{equation}
f(\tau,\Myc) = \frac{\FS(\tau,\Myc)}{\Delta \omega(\tau,\Myc)}\coma
\end{equation}
and the implied fractional overdensity
\begin{equation}
\delta(\tau,\Myc) \equiv \frac{\NS(\tau,\Myc)-\FS(\tau,\Myc)}{\FS(\tau,\Myc)} = \frac{n(\tau,\Myc)-f(\tau,\Myc)}{f(\tau,\Myc)} \fin
\end{equation}
In our nominal,
uniform field model, with a constant dimensionless
projected
density $f(\Myc)$ around each cluster $\Myc$, the field
distribution becomes $\FS(\tau,\Myc) = f(\Myc) \Delta\omega(\tau,\Myc)$.

In the $\FS\gg 1$ limit, one may approximate its distribution as normal,
as we verify using control cluster samples in \S\ref{subsec:ControlSample1}.
We may then infer the
Z-test significance
of any $\NS>\FS$ excess with respect to the field-only null hypothesis,
\begin{equation} \label{eq:SignificanceGeneral}
S \simeq\frac{\NS-\FS}{\sqrt{\FS\,}} \fin
\end{equation}
In practice, individual cluster bins at small radii typically show $\FS<1$, so the significance $S$ should be estimated from Poisson statistics (see \S\ref{app:Poisson}) or from control clusters (see
\S\ref{subsec:ControlSample1}).
One may co-add the
$N_{\mbox{\tiny{cat}}}=2$ catalogs,
similarly estimating the joint Z-test significance
\begin{equation}
 \label{eq:coadSsw}
     S_{\mbox{\tiny{joint}}} \simeq \frac{1}{\sqrt{N_{\mbox{\tiny{cat}}}}}\sum_{i}^{N_{\mbox{\tiny{cat}}}} S_{i},
 \end{equation}
 where $S_{i}$ is the significance profile in each catalog.

There is some freedom in the method in which these quantities may be stacked over clusters.
Following previous MCXC stacking analyses \citep{reiss2018detection, HouEtAl23}, we consider two extremes.
In the so-called source-weighted (SW) co-addition method, discussed in \S\ref{subsec:SW},
direct quantities (source numbers, solid angles) are stacked separately over all clusters and then combined to infer derived quantities (source densities, overdensities, excess significance levels).
In the cluster-weighted (CW) co-addition method, discussed in \S\ref{subsec:CW},
derived quantities are first computed on a cluster-by-cluster basis, and only subsequently stacked.
The results in \S\ref{sec:results} show good agreement between the two methods in the VS region, indicating that many clusters contribute fairly evenly to the stacked signal.

The nominal significance levels $S$ obtained above and below, based on approximating the distribution as normal and using standard error propagation,
are generalized for Poisson statistics (see \S\ref{app:Poisson}), and tested using samples of control clusters, as described in \S\ref{subsec:ControlSample1}.
We present simplified models for the radial dependence of the excess in \S\ref{subsec:ExcessModels}, and discuss in \S\ref{subsec:TS_sig} the procedure of fitting these
models to the data and assessing them using the TS-test.

\subsection{Source-weighted co-addition}
\label{subsec:SW}

The SW stacking method assigns an equal weight to each source, rather than to each cluster.
In each normalized angular $\tau$ bin, we thus separately co-add, over all clusters, both the number of sources,
\begin{equation} \label{eq:NStau}
\NS(\tau) \equiv \sum\limits_{\Myc=1}^{\Nclust} \NS(\tau,\Myc) \coma
\end{equation}
and the solid angle
\begin{equation}
\Omega(\tau) \equiv \sum\limits_{\Myc=1}^{\Nclust} \Delta \Omega(\tau,\Myc) \coma
\end{equation}
where $\Nclust$ is the number of clusters in the sample.

The dimensionless projected
number density of sources, co-added over clusters in the bin $\tau$, can now be computed as
\begin{equation}
    n_{\rm \SWt}(\tau) =\frac{\mathcal{N}(\tau)}{\bar{\omega}(\tau)}\coma
\label{eq:den2}
\end{equation}
where the dimensionless $\bar{\omega}(\tau) \equiv \Omega(\tau)/\bar{\te}_{500}^2$ is obtained by normalizing to the mean $\bar{\te}_{500}^2\equiv \Nclust^{-1}\sum_{\Myc}\te_{500}$;
this normalization has
no effect on our subsequent results.
The statistical uncertainty in $n_{\rm \SWt}$ may be estimated as
\begin{equation}\label{eq:sigma_n_SW}
    \sigma[n_{\rm \SWt}(\tau)] \simeq
    \frac{\sqrt{\NS(\tau)}}{\bar{\omega}(\tau)}\fin
\end{equation}
Source-weighted estimates such as $n_{\rm \SWt}(\tau)$ are somewhat skewed toward extended clusters or clusters in strong-field regions, which may harbour a larger number of sources and so may dominate the numerator of Eq.~\eqref{eq:den2}.
This bias is offset in part by the comparable extents of all clusters in our sample, and by avoiding regions with known contaminations.

The above approach is applied analogously to field sources, modelled with some distribution
$\FS(\tau,\Myc)$ based on the field region.
In analogy with Eqs.~\eqref{eq:NStau}--\eqref{eq:den2}, the co-added number of field sources is
\begin{equation} \label{eq:DefFsw}
\FS(\tau) \equiv \sum\limits_{\Myc=1}^{\Nclust} \FS(\tau,\Myc) \coma
\end{equation}
with co-added dimensionless
projected density
\begin{equation}
    f_{\rm \SWt}(\tau) =\frac{\FS(\tau)}{\bar{\omega}(\tau)} \fin
\end{equation}
The significance of a local excess of sources with respect to the
field-only null hypothesis can now be estimated as in
Eq.~\eqref{eq:SignificanceGeneral},
\begin{equation}
	\label{eq:sig_sw}
    S_{\rm \SWt}(\tau) = \frac{\NS(\tau)-\FS(\tau)}{\sqrt{\FS
    (\tau)}}\coma
\end{equation}
in the $\FS(\tau)\gg1$ limit of a
normal distribution.

Finally, the overdensity $\delta_{\rm \SWt}$ of sources with respect to the field can be estimated as
\begin{equation}
\label{eq:excess_source}
     \delta_{\rm \SWt}(\tau) = \frac{n_{\rm \SWt}(\tau) }{f_{\rm \SWt}(\tau) } - 1 \coma
\end{equation}
with an associated uncertainty
\begin{eqnarray}\label{eq:sigma_delta_SW}
 \sigma[\delta_{\rm \SWt}(\tau)] & \simeq & \frac{n_{\rm \SWt}(\tau)}{f_{\rm \SWt}(\tau)}\sqrt{\frac{\NS(\tau)+\FS(\tau)}{\NS(\tau)\FS(\tau)}}  \\
 & \simeq & \left[1+\delta_{\rm \SWt}(\tau)\right] \sqrt{\frac{2 }{f_{\rm \SWt}(\tau)\bar{\omega}(\tau)}}  \fin \nonumber
\end{eqnarray}
The second approximation holds
when the excess is small compared to the field;
it is shown here for completeness, but is not used in practice below.

\subsection{Cluster-weighted co-addition}
\label{subsec:CW}

The CW stacking method assigns the same weight to each cluster, rather than to each source.
In a given cluster $\Myc$, the significance of a local excess of sources with respect to the field-only null hypothesis is estimated as
\begin{equation}\label{eq:S_of_c_CW}
    S(\tau,\Myc) \simeq \frac{\NS(\tau,\Myc)-\FS(\tau,\Myc)}{\sqrt{\FS(\tau,\Myc)} } \fin
\end{equation}
Under the null hypothesis, we may approximate $S(\tau,\Myc)$ as a random variable of zero mean and unit standard deviation.
The CW
co-added significance of the excess over the field may now be estimated as
\begin{equation}
	\label{eq:sig_cw}
    S_{\rm \CWt}(\tau) = \frac{1}{\sqrt{\Nclust}}\sumclust S(\tau,\Myc)
    \fin
\end{equation}
Cluster-weighted estimates are somewhat skewed by clusters with low field values, where $|S(\tau,\Myc)|$ can have artificially large values.

In the same method,
we co-add densities over the sample of clusters to obtain the
average stacked dimensionless
projected density
\begin{equation}
    n_{\rm \CWt}(\tau)
    =\frac{1}{\Nclust}\sum_{\Myc=1}^{\Nclust}n(\tau,\Myc) \coma
\label{eq:den1}
\end{equation}
with the associated uncertainty
\begin{equation}\label{eq:sigma_n_CW}
    \sigma[ n_{\rm \CWt}(\tau)]
    = \frac{1}{\Nclust}\sqrt{ \sum\limits_{\Myc=1}^{\Nclust}  \frac{n(\tau,\Myc) }{\Delta\omega(\tau,\Myc)} }\fin
\end{equation}
The co-added dimensionless
projected density of field sources then become
\begin{equation}
f_{\rm \CWt}(\tau)=\frac{1}{\Nclust} \sumclust f(\tau,\Myc)\fin
\end{equation}

The local overdensity of sources above the field may
be defined in each cluster as
\begin{equation}
\label{eq:excess_cluster_j}
    \delta(\tau,\Myc) = \frac{n(\tau,\Myc)}{f(\tau,\Myc)}-1 \coma
\end{equation}
so the stacked overdensity is
\begin{equation}
\label{eq:excess_cluster}
    \delta_{\rm \CWt}(\tau) = \frac{1}{\Nclust}\sum\limits_{\Myc=1}^{\Nclust}\delta\left(\tau,\Myc\right) \coma
\end{equation}
with an associated uncertainty
\begin{equation} \label{eq:signaCW}
     \sigma\left[ \delta_{\rm \CWt}(\tau)\right] =
          \frac{1}{\Nclust}\sqrt{\sumclust \frac{n^2(\tau,\Myc)}{f^2(\tau,\Myc)\NS(\tau,\Myc)}}
     \coma
\end{equation}
where we neglected fluctuations in $\FS$.

\subsection{Poisson and control sample simulations}
\label{subsec:ControlSample1}

In the absence of any signals and source correlations, the statistical distribution of the small number of source counts in each radial bin within each cluster would follow a Poisson distribution, with a source density fixed by the foreground.
However, the statistical significance and uncertainty estimates in
Eqs.~\eqref{eq:SignificanceGeneral}--\eqref{eq:coadSsw},
\eqref{eq:sigma_n_SW},
\eqref{eq:sigma_delta_SW}--
\eqref{eq:sig_cw},
\eqref{eq:sigma_n_CW},
and \eqref{eq:signaCW}
invoke, for simplicity,  normally-distributed foreground counts.
We run Monte-Carlo Poisson simulations to test and correct for this non-Poissonian approximation.
Such simulations do not, however, capture structures and correlations in the actual catalogs.
We therefore study, in addition, the statistics of the actual catalog sources around a large number of control (or mock) cluster samples, in order to test for systematics and validate our method.

Our Monte-Carlo Poisson simulations involve drawing, from a Poisson distribution, $10^6$ random foreground source-counts for each $\tau$ bin in each cluster, based on the estimated foreground source-density $\FS(\tau,\Myc)$ of the cluster in each catalog.
These counts represent $\NS(\tau,\Myc)$ under the field-only null hypothesis.

The control sample analysis is based on
$\sim 10^6$
samples of control clusters for each catalog, each such sample including $\Nclust$ clusters with the same $\tefh$ values as in the real cluster sample, but placed in random positions within the allowed, \ie non-masked, part of the sky.
We stack the real catalog sources around each such control cluster, and calculate the corresponding distributions of foreground and excess source properties, just as in the real sample.

\begin{center}
\begin{figure}[b!]
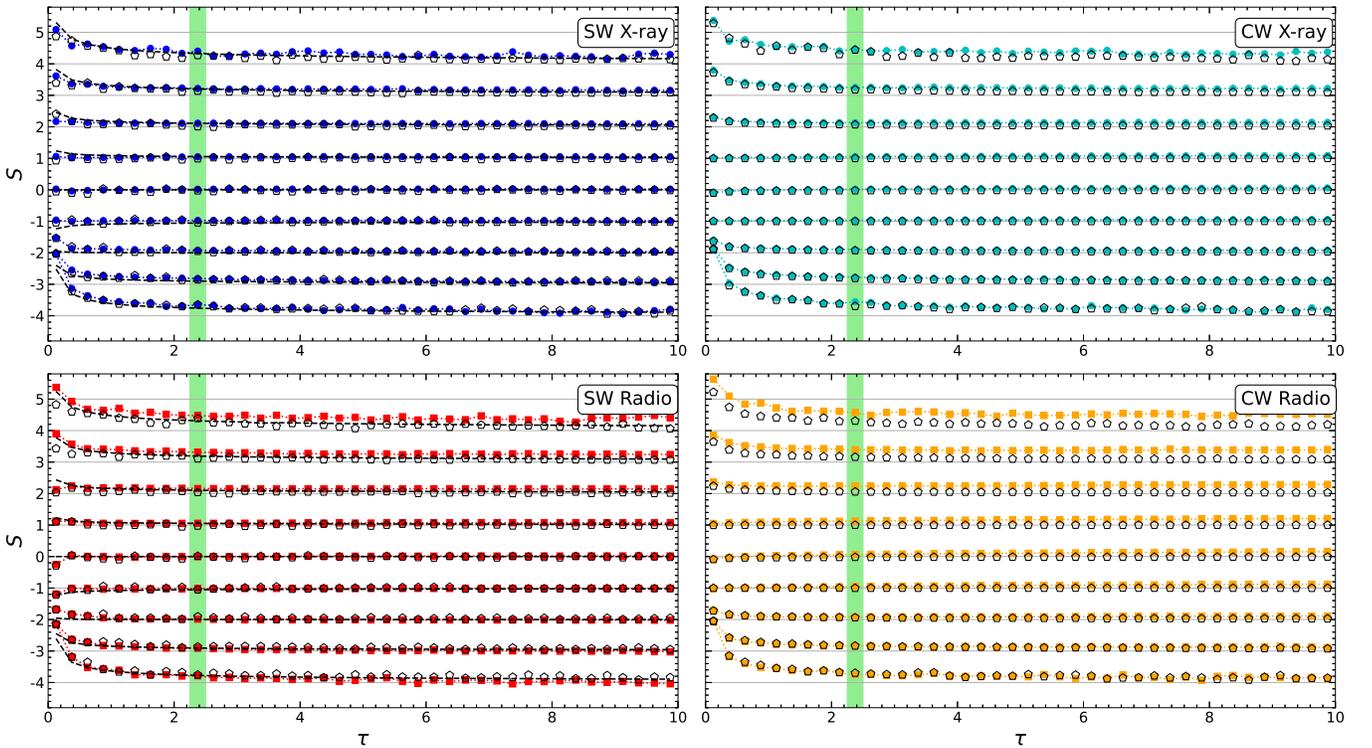

    \centerline{
    \DrawFig{\MYepsfbox{\myfig{mocks_norm_check_diff_Hou.eps}}}}
	\caption{
	\label{fig:mocks_norm_check_1}
    Statistical foreground analysis of SW (left panels) and CW (right) stacked X-ray (upper panels) and radio (bottom) sources.
    The
    $\pm1,\pm2,\pm3,\ldots$ CLs of control sample (filled colored symbols with dotted guides to the eye) and Poisson distribution estimates (black; empty symbols are Monte-Carlo simulations, dashed lines are the SW analytic
    approximation of \S\ref{app:Poisson})
    with respect to their median
    are plotted as a function of the normalized radius $\tau\equiv\te/\theta_{\rm 500}$ from the center of the cluster, against our nominal, normal-distribution approximation $S(\tau)$ (thin solid horizontal lines with left ticks). The anticipated VS region (the $2.25<\tau<2.5$ bin
    of previous \gama-ray \citep{reiss2018detection} and radio \citep{HouEtAl23} stacked detections) is highlighted (vertical green shading).
	}
\end{figure}
\end{center}

For both Monte-Carlo simulations and control-sample tests, we then study the resulting distributions of $S(\tau)$ values, each quantifying the significance of the putative excess above the foreground in a particular instance.
Figure \ref{fig:mocks_norm_check_1} shows the $\pm1\sigma, \pm2\sigma,\ldots$ CLs of $S$ around the median of its distribution, as inferred from the corresponding $68\%, 95\%, \ldots$ containment values.
These CLs are shown based on both Poisson simulations and control samples, for both X-ray and radio catalogs, using both \SW and \CW stacking methods.

As expected, the results approach Eqs.~\eqref{eq:sig_sw} and \eqref{eq:sig_cw} at large $\tau$, where $\FS\gg1$, but deviate substantially at small $\tau$.
The deviations are particularly large for $\NS<\FS$ source deficits, but we are mostly interested in the $S>0$ significance of a positive excess.
The approximate agreement of the $S=0,+1,+2 \ldots$ curves with the corresponding Poisson and control-sample containment fractions confirms that our nominal \S\ref{subsec:SW} and \S\ref{subsec:CW} estimates capture the significance of the excess reasonably well, with small errors $\Delta S$ in $S$.
Quantitatively, for the relevant, $\tau>1$ range, we find modest, $|\Delta S|/S \lesssim 10\%$ fractional errors, except for radio \CW stacking where $|\Delta S|/S \lesssim 15\%$.
While small, this $\Delta S$ should be accounted for
at the VS region.

Much of the $\Delta S$ difference is attributed to the inaccurate normal-distribution approximation of Poisson statistics.
Indeed, the Poisson and control-sample estimates are in much better agreement with each other, for both positive and negative $S$, and even for $\tau<1$.
The difference $\Delta S$ between them is very small in X-rays, where $|\Delta S|/S \simeq 4\%$ ($|\Delta S|/S \simeq 2\%$) for $S>0$ ($S<0$),
and is acceptable for radio \SW stacking, where $|\Delta S|/S \simeq 6\%$ ($|\Delta S|/S \simeq 2\%$).
For these cases, we henceforth apply an analytic correction (\S\ref{app:Poisson}),
correcting our nominal-distribution estimates for Poisson statistics; the relevant Eqs.~\eqref{eq:PoissonCL_corrected0}--\eqref{eq:PoissonCL_corrected1}  are shown to accurately follow the simulated Poisson results in Fig.~\ref{fig:mocks_norm_check_1}.
The overall agreement between Poisson and control-sample results indicates that systematic effects and correlations are weak, and our Poisson-statistics estimates are reliable.

For radio \CW stacking, however, Poisson and control-sample estimates differ by a more substantial $\Delta S/S \simeq 10\%$ ($\Delta S/S \simeq 3\%$) for $S>0$ ($S<0$), perhaps due to extended sources or sidelobes misidentified as multiple sources.
In this case, we henceforth interpolate the control-sample CLs in order to obtain a more accurate estimate of $S$.

\onecolumngrid
\twocolumngrid

\par\null\newpage
\clearpage

\subsection{Stacked source-excess models}
\label{subsec:ExcessModels}

For a significant $S>0$ excess, we use the TS-test to determine the CL at which the foreground-only null hypothesis is rejected in favor of one of
two
simple, effective models for the projected, dimensionless source-density $n(\tau)$:
(i) a planar model, of a 2D ring around the cluster oriented within the plane of the sky;
and
(ii) a shell model, obtained by projecting a spherical 3D shell around the cluster.

For the planar
model, we take $n(\tau)$ proportional to a 2D Gaussian of peak radius $\tau_{\rm sh}$ and width $\Delta\tau_{\rm sh}$, giving the three-parameter model
\begin{equation}
\label{eq:gauss_text}
    n_{\rm{planar}}\left(\tau \right) \equiv \frac{n_0 \tau_{\rm sh}}{\sqrt{8\pi\,\Delta\tau_{\rm sh}^2}} \exp\left[-\frac{\left( \tau-\tsh\right)^2}{2\,\Delta\tau_{\rm sh}^2}\right] \fin
\end{equation}
The prefactor is chosen such that in the $\Delta\tau_{\rm sh}\to 0$, thin-ring limit, a normalized solid-angle integration yields
\begin{equation}
\label{eq:SourcesInVS}
\int n(\tau) \, d\omega \simeq \pi \tau_{\rm sh}^2 n_0
\end{equation}
sources inside the VS. Here, $d\omega$ is the solid-angle element $d\Omega$ normalized by either $\bar{\te}_{500}^2$ (for SW) or
$\theta_{500,\Myc}^2$
(CW), and $n_0$ is a constant dimensionless source density.

For the shell
model, it suffices for our purposes to consider a zero-width 3D shell, giving upon projection the two-parameter model \citep[\eg][]{keshet2018evidence}
\begin{equation}
\label{eq:shell_text}
    n_{\rm shell}\left(\tau \right) \equiv  \frac{n_0\tau_{\rm sh}\Theta\left( \tau_{\rm sh}-\tau\right)}{2\sqrt{\tau_{\rm sh}^2-\tau^2}} \, ,
\end{equation}
where $\Theta$ is the Heaviside step function.
The normalization is chosen to again satisfy Eq.~(\ref{eq:SourcesInVS}),
to facilitate a meaningful comparison of $n_0$ in the two models.

\subsection{Model fitting and TS-Test}
\label{subsec:TS_sig}

We estimate the best-fit model parameters by a least-squares procedure. In general, one may minimize
\begin{equation}
\label{eq:stand_chi}
    \chi^2 \simeq \sum_{\tau,\Myc} \frac{\left[\NS(\tau,\Myc)-\FS(\tau,\Myc)-\MS(\tau,\Myc)\right]^2}{\FS(\tau,\Myc)+\MS(\tau,\Myc)} \coma
\end{equation}
where $\MS(\tau,\Myc)$ is the model excess at radial bin $\tau$ in cluster $\Myc$, and we exclude (henceforth) the innermost, $\tau<1.5$ bins, to avoid the noisy centers of clusters.
However, clusters typically have
some inner
bins with no sources at all, so
the above quantity is not well-approximated by
a $\chi^2$ distribution.
We therefore consider figures of merit
based on the stacked quantities.

For \SW stacking,
we consider $\NS(\tau)$, already co-added over clusters,
as the relevant random variable,
leading to
\begin{equation}
\label{eq:chi_sw}
    \chi^2_{\rm \SWt} \simeq \sum_{\tau} \frac{\left[\NS(\tau)-\FS(\tau)-\MS(\tau)\right]^2}{\FS(\tau)+\MS(\tau)}\coma
\end{equation}
where $\MS\left(\tau\right)=\sum_{\Myc} \MS(\tau,\Myc)$ is accordingly the model excess co-added over clusters.
For \CW stacking,
taking $S_{\rm CW}(\tau)$ as the relevant random variable yields
\begin{align}
\label{eq:chi_cw}
    \chi^2_{\rm \CWt}
    & \simeq \sum_{\tau}\frac{ \left[\sumclust \frac{\NS(\tau,\Myc)-\FS(\tau,\Myc)-\MS(\tau,\Myc)}{\sqrt{\FS(\tau,\Myc)}}\right]^2}{\sumclust\left[1+\frac{\MS(\tau,\Myc)}{\FS(\tau,\Myc)}\right]} \fin
\end{align}
In both methods, one may also sum over the two catalogs in order to fit both X-ray and radio signals simultaneously and estimate their joint significance.

Minimizing $\chi^2$ is equivalent in the normal-error approximation to finding the maximum $\mathcal{L}_{\rm max}$ of the likelihood
$\mathcal{L}(\chi^2) = \exp(- \chi^2/2)$.
The significance of the excess within the model is estimated from
the test statistic \citep{MattoxEtAl96_TS}
\begin{equation}\label{eq:DefTS}
    \mbox{TS} \equiv -2 \ln \frac{\mathcal{L}_{\rm max,-}}{\mathcal{L}_{\rm max,+}} = \chi^2_- - \chi^2_+ \, ,
\end{equation}
where a subscript `$-$' (`$+$') refers to the best-fitted model without (with) the modelled excess.
Then $\mbox{TS}$
has an approximate chi-squared distribution $\chi_\mathsf{n}^2$ of order $\mathsf{n}\equiv \mathsf{n}_+-\mathsf{n}_-$, \ie with the number of degrees of freedom added by the excess-source model \citep{Wilks1938}.
Our single-parameter $1\sigma$ confidence intervals are defined through $\mbox{TS}=1$ after projecting onto the given parameter.

For modelling, we nominally adopt a $\Delta\tau=1/6$ resolution, to better capture the radial profile; results for other resolutions are given in \S\ref{app:sensitivity_and_methods}.
We use Monte-Carlo simulations to verify that although the data is not normally distributed, $\chi^2$ in both Eqs.~\eqref{eq:chi_sw} and \eqref{eq:chi_cw} indeed follows the expected $\chi^2$ distribution (see \S\ref{app:Poisson}), and so can be used to estimate the TS-based CLs, denoted $\sigmaTS$.

\section{Virial excess and its implications}
\label{sec:results}

Our nominal stacking procedure yields the significance $S(\tau)$ profiles of the radially binned excess in Fig.~\ref{fig:nu_sig_plot_4bins},
with the SW significance $S(\tau_x,\tau_y)$ of the 2D excess illustrated in Fig.~\ref{fig:quad_joint}.
Fractional source overdensity $\delta(\tau)$ profiles are shown in Fig.~\ref{fig:FinalOverdensity}, along with their effective modelling summarized in Table \ref{tab:short_summary} and in more detail in Table \ref{tab:long_summary}.
Each panel in these figures shows results for both X-ray and radio catalogs, as well as a joint analysis of both catalogs.
Additional results and models are provided in \S\ref{app:sensitivity_and_methods}.

\subsection{Stacking}

\begin{center}
\begin{figure}[h!]
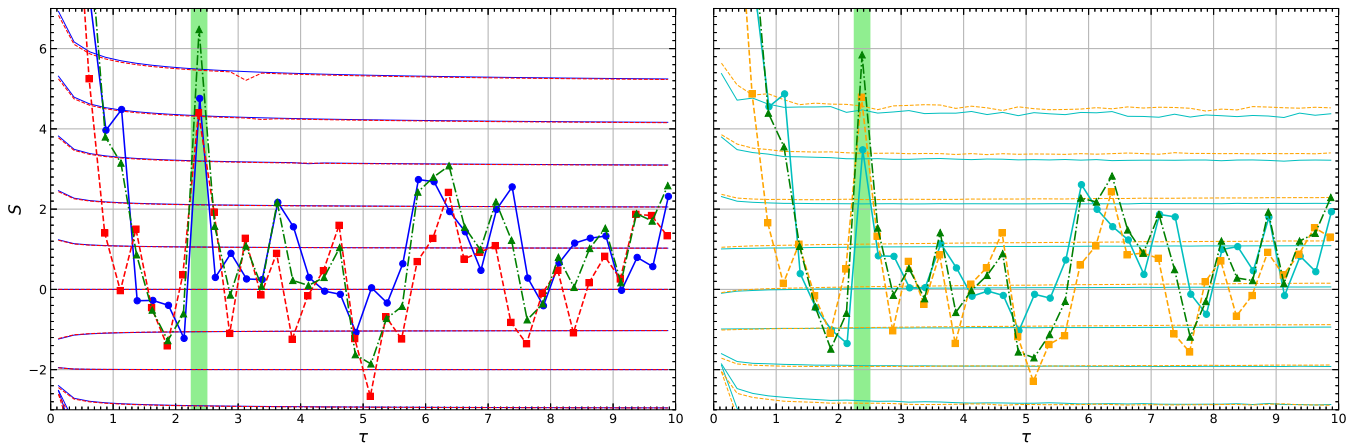

    \centerline{
    \DrawFig{\MYepsfbox{\myfig{joint_4bins_sig_Hou.eps}}}}
	\caption{
	\label{fig:nu_sig_plot_4bins}
        Significance $S(\tau)$ profiles of the SW/CW (left/right panel) co-added excess of X-ray (blue/cyan disks with solid guides to the eye), radio (red/orange squares with dashed guides), and co-added X-ray+radio (green triangles with dot-dashed guides) sources around our nominal MCXC cluster sample.
        Corrected confidence levels (thin, approximately horizontal curves) for X-ray (solid curves) and radio (dashed) are based on either a Poisson approximation (for SW, where they are accurate) or control clusters (for CW).
        The anticipated VS region is highlighted as in Fig.~\ref{fig:mocks_norm_check_1}.
	}
\end{figure}
\end{center}

\begin{center}
\begin{figure}[t!]
    \includegraphics[width=0.85\linewidth]{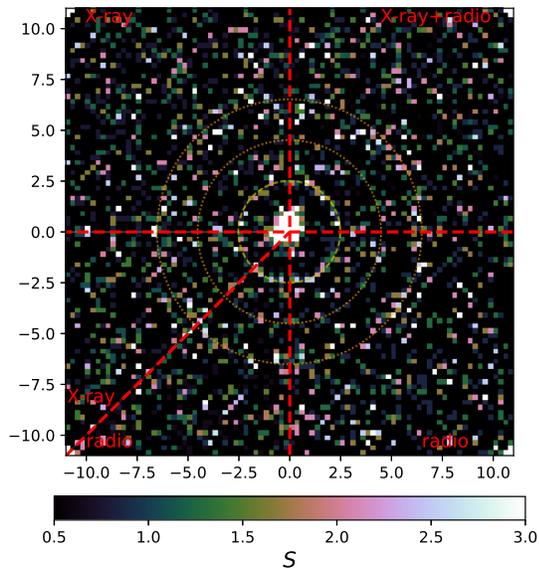}
	\caption{\label{fig:quad_joint}
    Excess-source significance $S$ in sky coordinates $\{\tau_x,\tau_y\}$ for SW-stacked and folded (see labels) X-ray (folded onto left quadrant and octant), radio (bottom quadrant and octant) and joint (upper right quadrant) catalogs.
    Dotted circles at $\tau=2.5$, $4.5$, and $6.5$ are guides to the eye.
    Stacked source images around individual clusters are rotated randomly around the center of the cluster before cluster co-addition.
    }
\end{figure}
\end{center}

\begin{center}
\begin{figure}[b!]
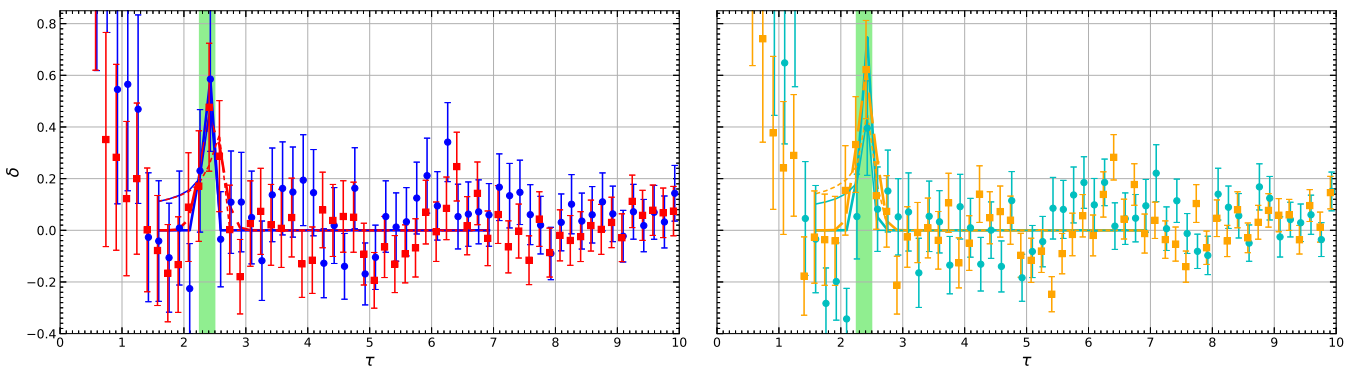

    \centerline{
    \DrawFig{\MYepsfbox{\myfig{joint_6_bins_Hou.eps}}}}
	\caption{ \label{fig:FinalOverdensity}
        Stacked fractional $\delta(\tau)$ overdensity (same notations as in Fig.~\ref{fig:nu_sig_plot_4bins}), along with the best-fitted planar (thick curves, three-parameter) and shell (thin, two-parameter) models for the X-ray (solid curves) and radio (dashed) sources.
    }	
\end{figure}
\end{center}

The stacking results are only weakly sensitive to the details of the analysis, as shown by the various sensitivity tests and method variations in \S\ref{app:sensitivity_and_methods}, as well as by the similar results obtained in Figs.~\ref{fig:nu_sig_plot_4bins} and \ref{fig:FinalOverdensity} using the \SW (left panels) vs. \CW (right panels) co-addition methods.
In Fig.~\ref{fig:quad_joint} and in the models of Table \ref{tab:short_summary} and \S\ref{subsec:Modeling},
we focus on the more conventional and well-calibrated \SW stacking; CW models are deferred to Table \ref{tab:long_summary}.

As expected, the centers of the stacked clusters show a strong excess of both X-ray and radio sources associated with the ICM.
This $\tau \lesssim 1.5$ excess reaches very high, $\delta\simeq 50$ ($\delta\simeq 20$) central overdensities in X-rays (radio), detected at extremely high, $S\simeq 100$ ($S\simeq 40$) CLs.
These and all subsequent CLs include the aforementioned correction for Poisson-statistics.

We are more interested here in cluster peripheries, where a nearly identical, highly significant, $\delta\simeq 0.5$ overdensity is identified in both X-ray ($4.4\sigma$) and radio ($4.0\sigma$) sources.
In both catalogs, this excess is highly localized, found in the same anticipated region (highlighted in Figs.~\ref{fig:mocks_norm_check_1}--\ref{fig:FinalOverdensity}), and peaked around the same $\tau\simeq 2.4$ normalized radius, identified previously as the VS in \gama-ray \citep{reiss2018detection} and radio \citep{HouEtAl23} continuum stacking.
Indeed, a joint-catalog analysis indicates a very high confidence, $5.9\sigma$ detection.
Below, we discuss this signal, henceforth referred to as the VS excess, in detail.

In addition to the central and VS signals, both catalogs show evidence for a very peripheral, $5.5\lesssim \tau \lesssim 7.5$ excess.
This weak, $\delta\simeq 0.2$ overdensity presents in both X-rays and radio at a low, $S\simeq 2$ significance in our nominal resolution scans.
However, the signal is very extended, and so presents with a more significant, $\sim 2.5\sigma$  detection at low resolution.
Furthermore, as the X-ray and radio signals overlap, a joint-catalog analysis implies a more significant, $\sim3\sigma$ overdensity.
Interestingly, a similar peripheral weak excess was previously identified independently, in continuum \gama-ray, radio, and SZ analyses, suggesting non-spherical, elongated VS morphologies; see discussion in Ref.~\citep{reiss2018detection}.

\subsection{Modeling}
\label{subsec:Modeling}

Consider first the VS excess in X-ray sources.
Fitting a planar (shell) model places the shock at $\tau_{\rm sh} = 2.44^{+0.06}_{-0.19}$ ($2.50^{+0.00}_{-0.02}$), detected at a
TS-significance of $3.8\sigma$ ($3.3\sigma$), as inferred from $\mbox{TS}\simeq20.6$ with $\mathsf{n}=3$ degrees of freedom ($\mbox{TS}\simeq14.2$ with $\mathsf{n}=2$).
The width of the shock in the planar model is
$\Delta\tau_{\rm sh} = 0.03^{+0.04}_{-0.03}$; for a cluster of median sample parameters
($z \simeq 0.077$ and $\theta_{500} \simeq 0\dgrdot15$), this corresponds to only $\sim 20\kpc$.
Given our prior knowledge of the VS radius, the effective significance of the detection is higher.
For instance, if one adopts as priors a shock radius $\tau_{\rm sh} = 2.4$ (based on \citep{reiss2018detection} and \citep{HouEtAl23}) and a sub-resolution shock width, then the TS-confidence rises to $4.5\sigma$ (for $\mathsf{n}=1$ remaining free parameter).

Modeling the radio-source excess yields very similar results.
Here, the planar (shell) model indicates a shock of similar radii $\tau_{\rm sh} = 2.44^{+0.04}_{-0.05}$ ($2.54^{+0.06}_{-0.03}$), detected at a $3.8\sigma$ ($3.3\sigma$) CL.
The shock width in the planar model, $\Delta\tau_{\rm sh} = 0.12^{+0.05}_{-0.04}$, is larger than inferred from X-rays, corresponding to $\sim 90\kpc$ for median (see Table \ref{tab:psc}) cluster parameters.
Here too, if one adopts as priors $\tau_{\rm sh} = 2.4$ and a sub-resolution shock width, the detection TS-confidence rises to $4.5\sigma$.

\onecolumngrid
\twocolumngrid

\par\null\newpage
\clearpage

\vspace{-1cm}
The X-ray and radio excess signals closely overlap, so a joint-catalog analysis produces similar results, at a higher CL.
Here, the planar (shell) model indicates a $\tau_{\rm sh} = 2.41^{+0.02}_{-0.02}$ ($\tau_{\rm sh} = 2.50^{+0.82}_{-0.77}$) shock, detected at $5.3\sigma$ ($4.4\sigma$).
The corresponding (planar model) shock width $\Delta \tau_{\rm sh} = 0.10^{+0.03}_{-0.02}$ translates to $\sim80\kpc$ for the median ($z \simeq 0.079$ and $\theta_{500} \simeq 0\dgrdot14$) parameters of the joint cluster sample.
Adopting as priors a $\tau_{\rm sh} = 2.4$ shock radius and a sub-resolution shock width would raise the TS-confidence to $5.8\sigma$.

\begin{table}[h!]
    \caption{
    Nominal VS modeling.
    \label{tab:short_summary}}
\centering
\begin{tabular}{ll@{\hskip 0.2cm}|@{\hskip 0.2cm}llllll}
Model & Cat. & $n_{0,X}^* $  & $n_{0,\nu}^* $  & $\tau_{\rm sh}$ & $\Delta \tau_{\rm sh}$& TS & $\sigmaTS$ \\
(1) & (2) & (3) & (4) & (5) & (6) & (7) & (8) \\
\hline
Shell  & 2RXS  & $8.2^{+1.4}_{-3.4}$ & --- & $2.50^{+0.00}_{-0.04}$& --- & $14.2$ & $3.3$ \\
Shell  & NVSS  & --- & $8.1^{+2.4}_{-2.3}$ & $2.54^{+0.06}_{-0.03}$& --- & $14.0$ & $3.3$ \\
Shell & Joint & $7.3^{+2.3}_{-2.1}$& $6.8^{+2.2}_{-2.1}$& $2.50^{+0.83}_{-0.78}$ & --- & $25.8$ & $4.4$ \\
Planar & 2RXS  & $5.5^{+1.6}_{-1.5}$ & --- & $2.35^{+0.05}_{-0.02}$& $0.03^{+0.04}_{-0.03}$ & $20.6$ & $3.8$ \\
Planar & NVSS  & --- & $6.5^{+2.1}_{-1.8}$ & $2.44^{+0.04}_{-0.05}$& $0.12^{+0.05}_{-0.04}$ & $20.4$ & $3.8$ \\
Planar & Joint & $5.8^{+1.7}_{-1.5}$& $5.9^{+1.9}_{-1.7}$& $2.41^{+0.02}_{-0.02}$ & $0.10^{+0.03}_{-0.02}$ & $38.1$ & $5.3$ \\
\end{tabular}
    \begin{tablenotes}[flushleft]
    \item
    Nominal (SW; $\Delta\tau=1/6$) best-fitted parameters; additional models are provided in Table \ref{tab:long_summary}.
    Table columns: (1) Effective VS model (planar or shell); (2) Stacked catalog (2RXS, NVSS, or the two catalogs jointly); (3) and (4) Normalized, $n_0^*\equiv 10^3 n_0$ X-ray and radio source densities in Eqs.~\eqref{eq:gauss_text} or \eqref{eq:shell_text};
    (5) Dimensionless shock radius; (6) Dimensionless signal width; (7) TS value; (8) TS-based significance.
    \end{tablenotes}
\end{table}

Curiously, we find similar source densities for the VS excess in the X-ray and radio catalogs.
With the planar model, for instance, $n_0\simeq 5\times10^{-3}$ (sources per $\theta_{500}^2$)
in X-rays and $n_0\simeq 7\times10^{-3}$ in radio; the shell model yields $n_0$ values similar but slightly higher due to the 3D projection.
The exact difference between X-ray and radio peak densities is not particularly meaningful, as $n_0$ is somewhat sensitive to details such as catalog parameters and cuts.
Despite their similar properties, in \S\ref{subsec:joint_xray_radio_sources} we show that X-ray and radio excess signals are not dominated by the same individual objects.

\onecolumngrid
\twocolumngrid

\par\null\newpage
\clearpage

As argued in \S\ref{sec:Intro}, an excess of AGN in a narrow radial range at the far outskirts of clusters would be highly unnatural.
Our above results establish systematically and quantitatively that both X-ray and radio sources emerge in a very narrow, $\Delta r/r\lesssim20$ radial range, at least in a large fraction of clusters.
This sharp excess is inconsistent with enhanced AGN triggering, given the
weak environmental gradients found at these large, $\gtrsim2\Mpc$ radii, and the low ambient pressure even behind the VS.
We conclude that the signals are
unrelated to AGN;
indeed, we find no significant localized signal when examining or stacking sources associated with identified AGN, as shown in \S\ref{app:AGN_identification}.
Moreover, the properties of the excess sources, discussed in \S\ref{sec:characteristics}, are inconsistent with AGN.

The robust spatial coincidence of the excess with the previously-determined VS location indicates that the excess is associated with the shock itself, and is most likely triggered by it.
While the shock provides the strong spatial localization necessary to account for the sharpness of the excess, its downstream pressure is still far too weak to penetrate into the very center of a galaxy and trigger an AGN, so it is necessary to explore alternative models.
Indeed, the constraints derived in \S\ref{sec:characteristics} on the properties of the excess sources indicate that they constitute
a different class of objects, explored in \S\ref{sec:origin}.

\section{Properties of VS excess sources}
\label{sec:characteristics}

The $\delta\simeq 0.5$ fractional overdensity of both X-ray and radio catalog sources within the narrow VS range indicates that roughly $\sim1/3$ of the sources therein are associated with the VS, while the remaining $\sim2/3$ majority are field sources.
The sources found in the VS regions, listed in \S\ref{app:SourceTables}, are mostly unidentified; note that these peripheral sources lie well outside the radii of known ICM radio sources, including halos, relics, and even mega-halos \citep{Keshet24}, and cannot be attributed to the smooth distribution of cluster sources or to AGN (\S\ref{app:AGN_identification}).
While we cannot attribute individual sources to the VS, the most likely properties of those VS-induced sources can be deduced from a differential statistical analysis that compares the properties of sources in the VS region to their field counterparts.

\begin{center}
\begin{figure}[t!]
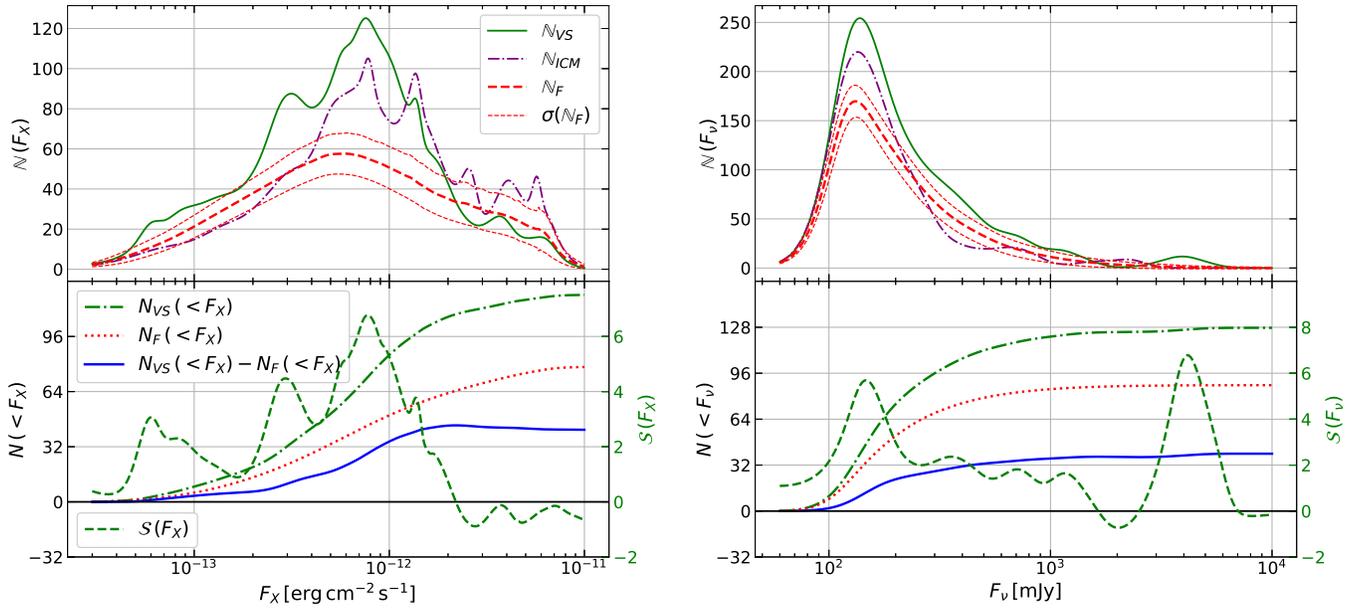

    \centering{
    \DrawFig{\MYepsfbox{\myfig{Pl_flux_pdf_new_Hou.eps}}}}
    \vspace{-0.7cm}
	\caption{
    Differential $\mathbb{N}(\myP)$ (top panels) and cumulative $N(<\myP)$ (bottom panels with left axes) distributions of logarithmic, $\myP=\log_{10}F$  energy flux in X-rays (left panels) and radio (right panels).
    Differential distributions are shown for the VS region (solid green) and comparable ICM (dot-dashed magenta) and field (red; thick dashed for mean value and thin dotted for $1\sigma$ dispersion) estimates; see text.
    Bottom panels show, in addition to the cumulative distributions in the VS region (dot-dashed green), field estimate (dotted red), and their difference (solid blue), also the local significance of the VS excess (dashed green with right axes).
    In radio, we add a $\sigma_{\rm smooth}^2 \equiv (0.2\,F_{\nu})^2$ smoothing variance to the $F_\nu$ uncertainty, for visibility.
    \label{fig:X_ray_typ_flux1}\label{fig:radio_typ_flux1} }
\end{figure}
\end{center}

\begin{center}
\begin{figure}[h!]
    \begin{tikzpicture}[
        every node/.style={anchor=south west,inner sep=0pt},
        x=1cm, y=1cm,
      ]
     \node (fig1) at (0,0)
       {\centerline{\DrawFig{\includegraphics[width=18cm]{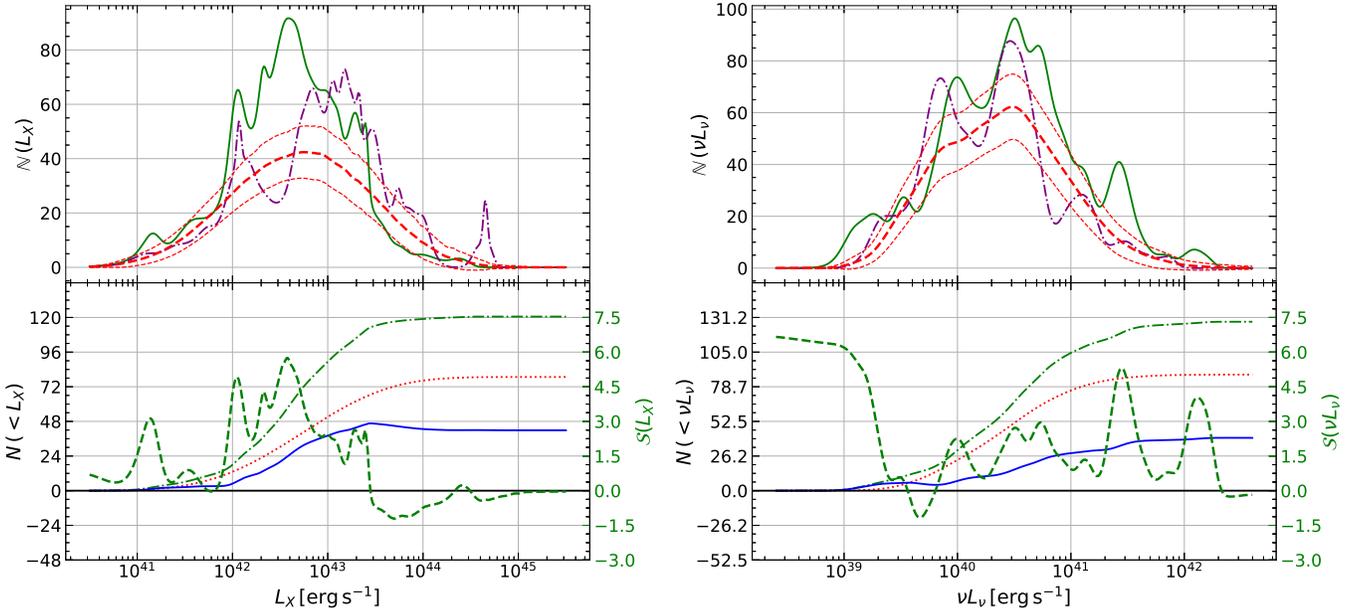}}}};
\end{tikzpicture}
	\caption{\label{fig:X_ray_typ_lum1}\label{fig:radio_typ_lum1}
    Distribution of luminosity, $\myP=\log_{10}L_X$ in X-rays (left) and $\myP=\log_{10}\nu L_\nu$
    with added $\sigma_{\rm smooth}^2(L_\nu) \equiv (0.2\,L_{\nu})^2$ variance (for visibility) in radio (right); notations are as in Fig.~\ref{fig:X_ray_typ_flux1}.
    }
\end{figure}
\end{center}

\begin{figure}[h]
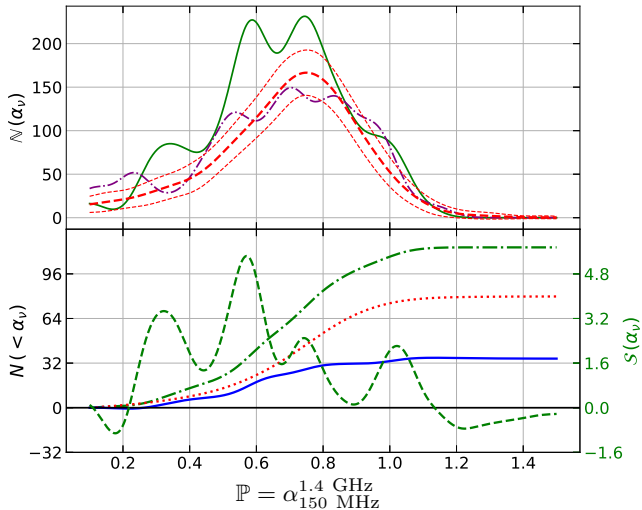

    \begin{overpic}[width=1\linewidth,trim={0cm 0cm 0cm 0cm},clip]
    {\myfig{radio_PL_ind_hist_spec_30sec_Hou.eps}}
        \put (35,1) {\colorbox{white}{\textcolor{black}{$\mathbb{P}=\alpha_{150\mbox{ \scriptsize MHz}}^{1.4\mbox{ \scriptsize GHz}}$}}}
    \end{overpic}
    \vspace{-0.2cm}
	\caption{\label{fig:radio_spec}
	Distribution of $\mathbb{P}=\alpha_{150\mbox{ \scriptsize MHz}}^{1.4\mbox{ \scriptsize GHz}}$ radio spectral indices among GMRT-NVSS catalog pairs; notations as in Fig.~\ref{fig:X_ray_typ_flux1}.
    }
\end{figure}

\begin{figure}[h!]
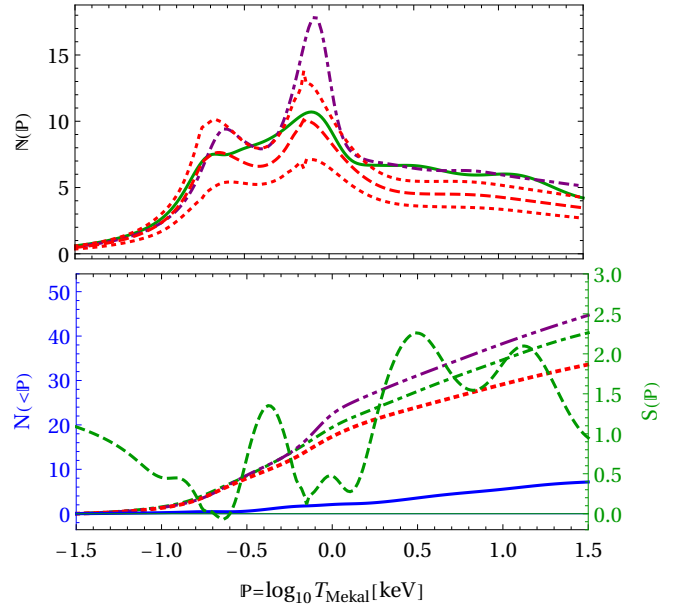

    \begin{center}
        \hspace*{-0.4cm}
        \includegraphics[width=0.96\linewidth,trim={0cm 0.9cm 0cm 0cm},clip]{\myfig{PropTMekalXrays1b.eps}}
    \end{center}
    \vspace{-0.75cm}
    \begin{center}
        \includegraphics[width=1\linewidth,trim={0cm 0cm 0cm 0cm},clip]{\myfig{PropTMekalXrays2b.eps}}
    \end{center}
    \vspace{-0.5cm}
    \caption{
    Distribution of the $\myP=\log_{10}k_B T$ Mekal-fitted temperature; notations as in Fig.~\ref{fig:X_ray_typ_flux1}.
    Here we show also the cumulative ICM-source distribution (double-dot magenta).
    \label{fig:PropTMekal} }
\end{figure}

\begin{figure}[h!]
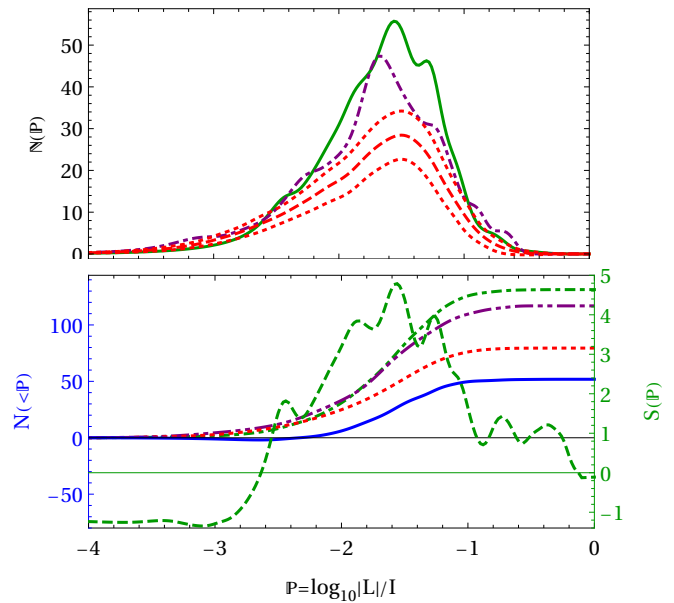

    \begin{center}
        \includegraphics[width=0.95\linewidth,trim={0cm 1.0cm 0cm 0cm},clip]{\myfig{PropPolFracRadio1b.eps}}
    \end{center}
    \vspace{-0.6cm}
    \begin{center}
        \includegraphics[width=1\linewidth,trim={0cm 0cm 0cm 0cm},clip]{\myfig{PropPolFracRadio2b.eps}}
    \end{center}
    \vspace{-0.5cm}
    \caption{
    Distribution of the $\myP=\log_{10}|L|/I$ linearly-polarized radio fraction; notations as in Fig.~\ref{fig:PropTMekal}.
    \label{fig:PropPolFrac2} }
\end{figure}

\begin{center}
\begin{figure}[b!]
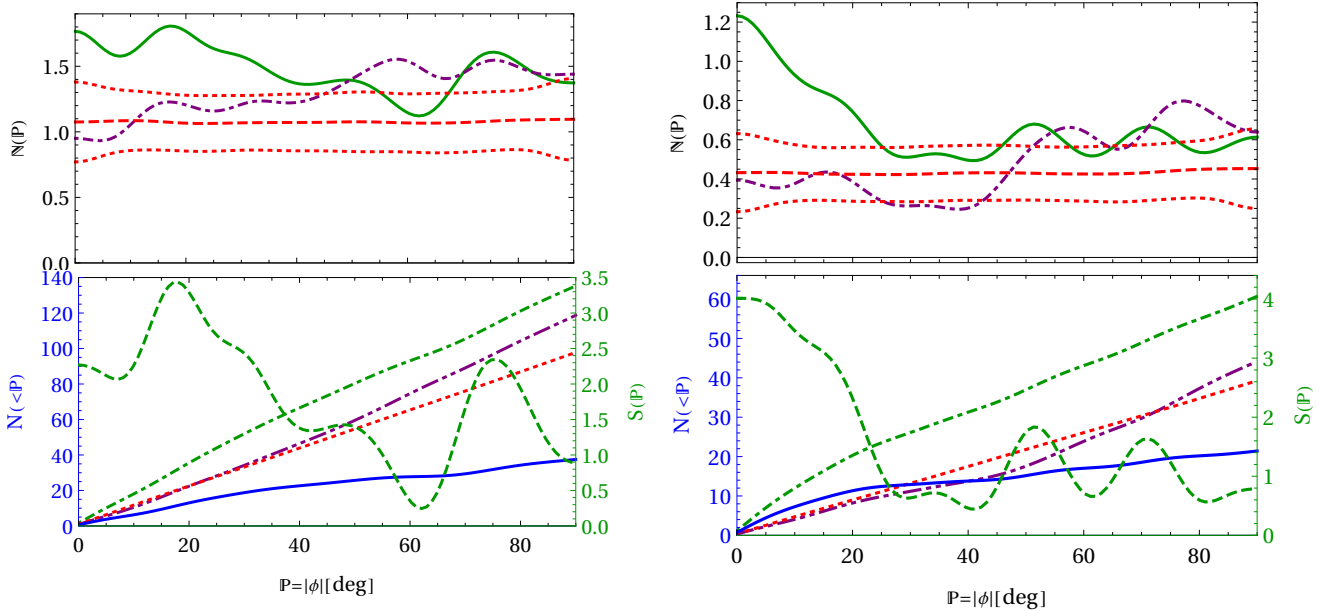

    \begin{center}
        \includegraphics[width=0.46\linewidth,trim={0cm 0.9cm 0cm 0cm},clip]{\myfig{PropPA1b.eps}}
        \hspace{0.3cm}
        \raisebox{0cm}{\includegraphics[width=0.48\linewidth,trim={0cm 0.9cm 0cm 0cm},clip]{\myfig{PropPA1d20b.eps}}}
    \end{center}
    \vspace{-0.4cm}
    \begin{center}
            \hspace*{-0.2cm}
            \includegraphics[width=0.476\linewidth,trim={0cm 0cm 0cm 0cm},clip]{\myfig{PropPA2b.eps}}
            \hspace{0.1cm}
            \raisebox{-0.17cm}{\includegraphics[width=0.476\linewidth,trim={0cm 0cm 0cm 0.0cm},clip]{\myfig{PropPA2d20b.eps}}}
    \end{center}
    \vspace{-0.5cm}
    \caption{
    Distribution of the $\mathbb{P}=|\phi|$ angle between polarization vector and cluster radius for all (left) and $\mbox{Dec}>20\dgr$ (right) sources; notations and sources as in Fig.~\ref{fig:PropPolFrac2}. A $\sigma_{\rm smooth}^2(\varphi)=(5\dgr)^2$ variance is added for visibility.
    \label{fig:PropPolPA} }
\end{figure}
\end{center}

\begin{center}
\begin{figure}[h!]
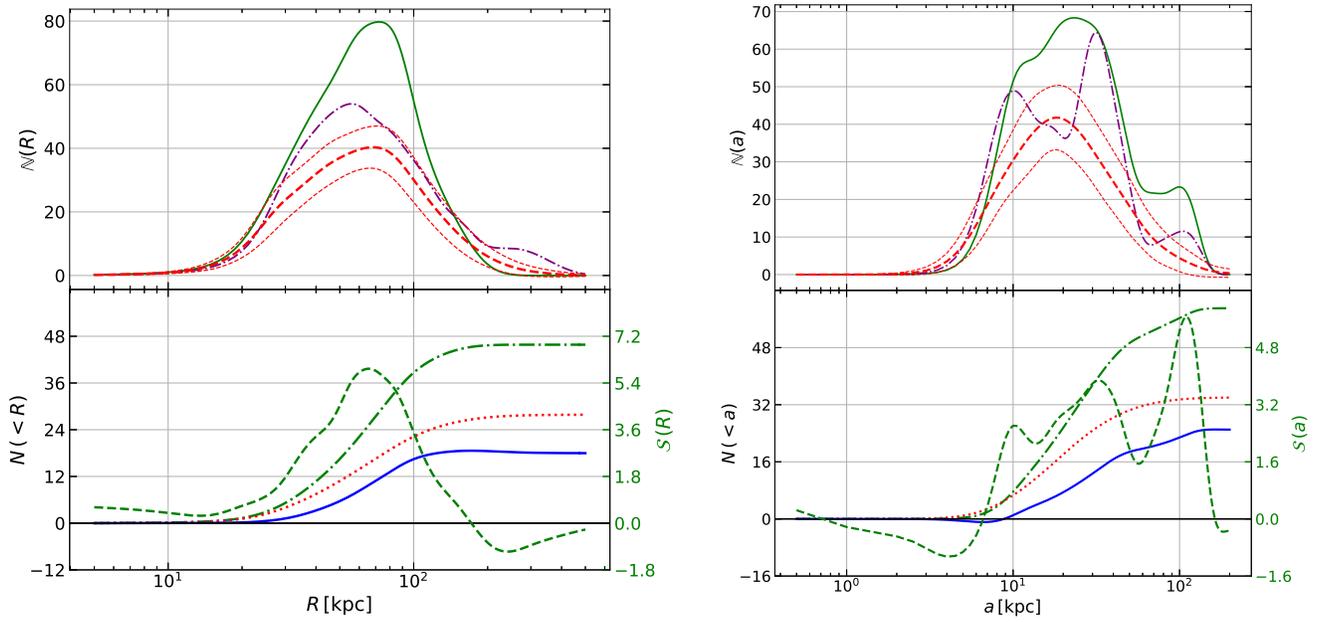

    \begin{center}
        \includegraphics[width=0.508\linewidth,trim={20cm 0cm 0cm 0cm},clip]{\myfig{Xray_size_Hou.eps}}
        \hspace{0.1cm}
        \includegraphics[width=0.452\linewidth,trim={20cm 20.5cm 0cm 0cm},clip]{\myfig{radio_size_kpc_pdf_spec_Hou.eps}}
    \end{center}
	\caption{\label{fig:PropSize}
    Distribution of source size: radius $\mathbb{P}=R$ in X-rays (left) and semi-major axis $\mathbb{P}=a$ with added $\sigma_{\rm smooth}^2(a) \equiv (0.2\,a)^2$ variance in radio (right) for visibility; notations are the same as in Fig.~\ref{fig:X_ray_typ_flux1}.
    Normalized (to $\theta_{500}$) and semi-minor axes are shown in Figs.~\ref{fig:X_ray_typ_phys1} and \ref{fig:radio_typ_ext1}.
	}
\end{figure}
\end{center}

\begin{center}
\begin{figure}[b!]
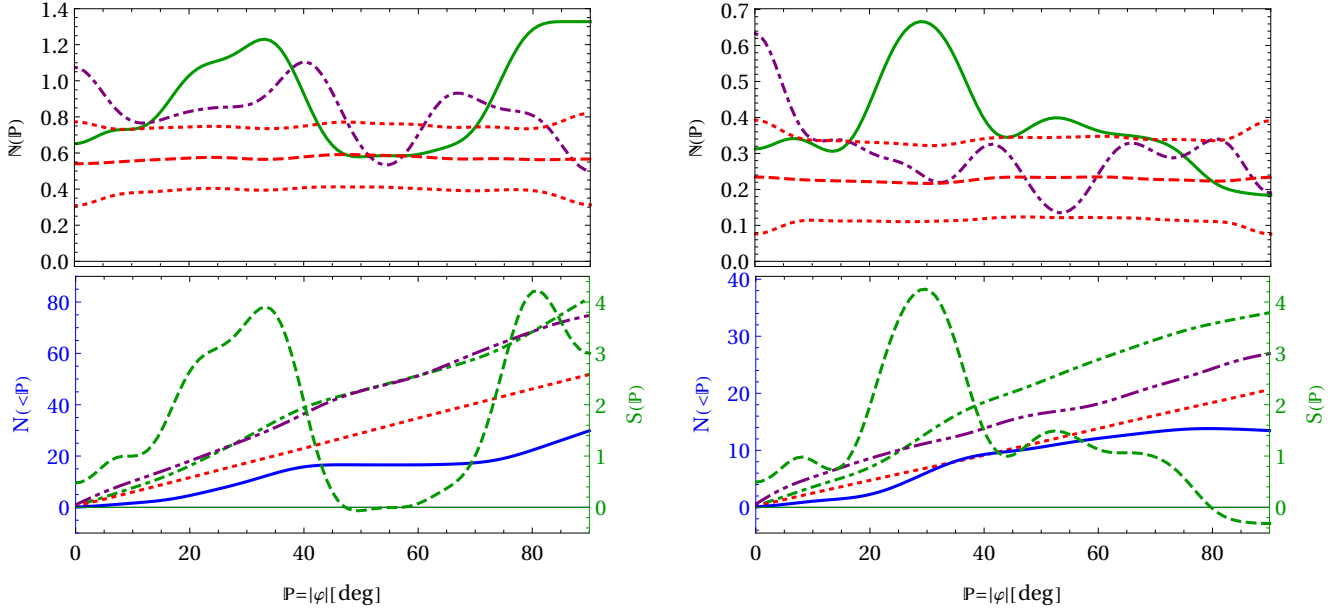

    \begin{center}
        \includegraphics[width=0.476\linewidth,trim={0cm 0.9cm 0cm 0cm},clip]{\myfig{PropAPA1b.eps}}
        \hspace{0.3cm}
        \raisebox{0cm}{\includegraphics[width=0.476\linewidth,trim={0cm 0.9cm 0cm 0cm},clip]{\myfig{PropAPA1d20b.eps}}}
    \end{center}
    \vspace{-0.4cm}
    \begin{center}
            \hspace*{0.cm}
            \includegraphics[width=0.47\linewidth,trim={0cm 0cm 0cm 0cm},clip]{\myfig{PropAPA2b.eps}}
            \hspace{0.4cm}
            \raisebox{0cm}{\includegraphics[width=0.47\linewidth,trim={0cm 0cm 0cm 0.0cm},clip]{\myfig{PropAPA2d20b.eps}}}
            \hspace{-0.3cm}
    \end{center}
    \vspace{-0.5cm}
    \caption{
    Distribution of the $\mathbb{P}=|\varphi|$ angle between source elongation and cluster radius for all (left) and $\mbox{Dec}>20\dgr$ (right) sources; notations and sources as in Fig.~\ref{fig:PropPolFrac2}. A $\sigma_{\rm smooth}^2(\varphi)=(5\dgr)^2$ variance is added for visibility.
    \label{fig:PropAPA} }
\end{figure}
\end{center}

Consider some source property $\myP$ of interest, such as flux, spectral index, or size.
Figures \ref{fig:X_ray_typ_flux1}--\ref{fig:PropAPA} compare, for various such properties, the distribution of $\myP$  among sources in the VS region ($2.25<\tau<2.5$) to $\myP$ among sources in the ICM region ($1.0<\tau<1.5$) and in the field.
Here, we adopt a normalized, $7<\tau<35$ radial field range, instead of the fixed angular scale invoked earlier for field removal, in order to obtain a better behaved VS-to-field source-number ratio.
In addition to the nominal analysis, we also consider different field regions ($10<\tau<30$) and cuts that do not exclude the FB and Loop-I regions, in order to improve the statistics.
For additional tests and source properties, see \S\ref{App:characteristics}.

When relevant, we show analogous X-ray (left panels) and radio (right panels) $\myP$ properties on the same figure.
Each figure presents both differential $\mathbb{N}(\myP)\equiv dN/d\myP$ (top panels) and cumulative $N(<\myP)$ (bottom panels) distributions in the VS region (solid and dot-dashed green), compared to equivalent field (thick dashed and thin dotted red) and ICM (dot-dashed and double dot-dashed magenta) distributions scaled to the same $\Omega$.
Bottom panels highlight also the cumulative, $N_{\mbox{\tiny VS}}(<\myP)-N_{\mbox{\tiny F}}(<\myP)$ distribution (solid blue) and the $S(\myP)=[\mathbb{N}_{\mbox{\tiny VS}}-\mu(\mathbb{N}_{\mbox{\tiny F}})]/\sigma(\mathbb{N}_{\mbox{\tiny F}})$ local significance (dashed green) of the VS excess, with respect to the field distribution $\mathbb{N}_{\mbox{\tiny F}}$ inferred from $>200$ random field samples.

Figure \ref{fig:X_ray_typ_flux1} shows the logarithmic, $\myP=\log_{10}F$ energy flux distribution.
In X-rays, the VS excess arises mostly in the $2\times 10^{-13}\erg\se^{-1}\cm^{-2}\lesssim F_X \lesssim2\times 10^{-12}\erg\se^{-1}\cm^{-2}$ flux range, dominating the cumulative distribution at local $\gtrsim3\sigma$ significance levels, and appears unaffected by our higher $F_X$ upper limit.
In radio, the excess arises mostly in the $10^2\mbox{ mJy}\lesssim F_\nu \lesssim 10^3\mbox{ mJy}$ range, with a local $\gtrsim2\sigma$ significance increasing towards lower $F_\nu$, suggesting additional excess sources below our lower $F_\nu$ cutoff.
The local excess of bright sources with $\sim 5$ Jy flux, seen in the figure, can probably be disregarded, as it is associated with only two sources and is not robust against changes in analysis details.

The cataloged sources do not have readily available associated redshifts.
Assigning sources to the redshift of the nearby MCXC cluster, as seen in projection, produces an intrinsic, albeit noisy, estimate of the typical luminosity of excess VS sources.
Figure \ref{fig:X_ray_typ_lum1} thus shows that the excess is associated with $10^{42}\erg\se^{-1}\lesssim L_X \lesssim 10^{43}\erg\se^{-1}$ 2RXS sources and with $10^{40}\erg\se^{-1}\lesssim \nu L_\nu \lesssim 10^{41}\erg\se^{-1}$ NVSS sources.
(The luminosity of excess GMRT sources is a factor $\sim 2$ lower than in NVSS; see Fig.~\ref{fig:GMRT_lum}.)
Again, local excess signals at lower or higher luminosities cannot be substantiated, as they are associated with few sources and are not robust against changes to the analysis details.

Consider the (unlikely) possibility that the VS excess signals in X-rays and in radio arise from the same power-law spectrum, extending over more than eight orders of magnitude in photon energy (although the implied synchrotron origin of the X-rays is ruled out in \S\ref{sec:origin}).
For the relevant VLA and \emph{ROSAT} channels, the associated photon spectral index would then be
\begin{equation}\label{eq:PurePowerLaw}
  \alpha \equiv -\frac{d\ln L}{d\ln \nu} \simeq 0.81\left(\frac{100\nu L_\nu}{L_X}\right)^{0.067} \,,
\end{equation}
providing a robust reference value for a spectral analysis.

\par\null\newpage
\clearpage

\onecolumngrid

\twocolumngrid

\par\null\newpage
\clearpage

The spectral index $\alpha_{150\mbox{ \scriptsize MHz}}^{1.4\mbox{ \scriptsize GHz}}$ is estimated by matching sources in the GMRT and NVSS catalogs through a projected proximity, $\theta<\theta_{\rm max}=30''$ angular separation criterion.
Most ($\sim90\%$) GMRT sources then show an NVSS counterpart, with only a small, $\pi \theta_{\rm max}^2 N_{\rm NVSS}N_{\rm GMRT}/(4\pi f_\Omega N_{\rm pairs})\sim 1\%$ fraction of the resulting $N_{\rm pairs}\simeq 5.6 \times 10^5$ pairs being false, chance associations.
Here, $N_{\rm NVSS}\simeq 1.6\times 10^6$ and $N_{\rm GMRT}\simeq 5.8 \times 10^5$ are respectively the numbers of NVSS and GMRT sources in the relevant, $f_\Omega=82\%$ fraction of the sky.
Defining $\mathbb{P}=\alpha_{150\mbox{ \scriptsize MHz}}^{1.4\mbox{ \scriptsize GHz}}$ for each pair, the radio VS excess may be inferred from Fig.~\ref{fig:radio_spec} to arise mostly from $0.5\lesssim \alpha_{150\mbox{ \scriptsize MHz}}^{1.4\mbox{ \scriptsize GHz}} \lesssim 0.8$ sources.
The considerable $\alpha_{150\mbox{ \scriptsize MHz}}^{1.4\mbox{ \scriptsize GHz}}<0.6$ excess suggests that this hard synchrotron power-law spectrum does not extend up to X-ray energies.

This estimate is broadly consistent but somewhat harder than in Eq.~\eqref{eq:PurePowerLaw}, and in the absence of NVSS-2RXS pairs (see \S\ref{subsec:joint_xray_radio_sources}), we cannot support such $\alpha_\nu$ values persisting out to X-rays.
Moreover, while we cannot localize any excess in the X-ray spectral index $\Gamma_X$ obtained from power-law fits, both Mekal and black-body fits suggest $k_B T\gtrsim 1\keV$ excess-source temperatures, as shown in Figs.~\ref{fig:PropTMekal} and \ref{fig:PropTmTb}, where $k_B$ is the Boltzmann constant.
The combined implications of hard power-law spectra in radio and thermal spectra in X-rays are discussed in \S\ref{sec:origin}.

The VS excess is associated in radio with sources of elevated linear polarization, in the $1$--$10\%$ range, as indicated by Fig.~\ref{fig:PropPolFrac2}.
Here we use the bias-corrected, linearly polarized flux density and its associated position angle, available for most sources in the NVSS catalog; see \citep{NVSS_paper} for details.
Again assigning sources to a nearby projected MCXC cluster, we may estimate the typical angle $\phi$ between the source polarization vector and the radius, \ie the line connecting it to the center of cluster.
As Fig.~\ref{fig:PropPolPA} shows, the polarization is preferentially radial ($\phi\simeq 0$), corresponding to magnetic fields aligned perpendicular to the cluster radius.

This effect is quite robust, strengthening at high Galactic latitudes or at high declinations \citep[where the typical Galactic Faraday rotation at NVSS frequencies drops below $\sim30\dgr$; see, \eg][]{TaylorEtAl09}, as illustrated by the figure.
Such a polarization supports the association of the sources not only with the cluster, but also specifically with its VS, suggesting that the emission arises from a gaseous medium shocked by the VS, resulting in enhanced tangential magnetic fields.
While the measured polarizations carry substantial uncertainties, the elevated polarized flux and the correlation between polarization vectors and radii among VS (but not field) sources suggest that the polarization effect is real.

Assigning again sources with nearby projected MCXC clusters, we may estimate the typical sizes of sources in the VS region.
Figure \ref{fig:PropSize} thus suggests that the X-ray excess is dominated by sources of $30\kpc\lesssim R\lesssim 100\kpc$ radii, whereas the radio excess arises from sources with semi-major axes in the range $10\kpc\lesssim a\lesssim 100\kpc$.
An analogous examination of the semi-minor axes $b$ of the radio sources suggests that the VS excess arises from elongated sources, of $6\kpc\lesssim b\lesssim 30\kpc$.
Similarly estimating the dimensionless (normalized to $R_{500}$) scales yields $0.05\lesssim \tau_R \lesssim 0.15$, $0.02\lesssim \tau_a \lesssim0.2$, and $0.01\lesssim \tau_b \lesssim0.15$, associated respectively with the X-ray radius and the radio semi-major and semi-minor axes (see \S\ref{App:characteristics}).
These scales are broadly consistent with the width $\Delta\tau_{\rm sh}$ of the stacked VS signal, as derived in \S\ref{sec:results}.

The distribution of the angle $\varphi$ between the semi-major axis of a radio source and the line connecting it to the center of the cluster, presented in Fig.~\ref{fig:PropAPA}, appears bimodal.
A significant excess is found for sources elongated both around $\varphi\simeq 30\dgr$, and perpendicular ($\varphi\simeq 90\dgr$) to the cluster radius; again, any well established $\varphi$ excess would support the association of the virial excess with the cluster, and specifically with the VS.
The second, $\varphi\simeq 90\dgr$ excess (see also Fig.~\ref{fig:PropPA}) is associated with the more extended sources, which appear to be stretched along the VS; however, we cannot substantiate this excess, as it vanishes at high declinations as shown in the figure.

\begin{center}
\begin{figure}[h]
    \begin{center}
        \hspace*{-0.4cm}
        \includegraphics[width=0.95\linewidth]{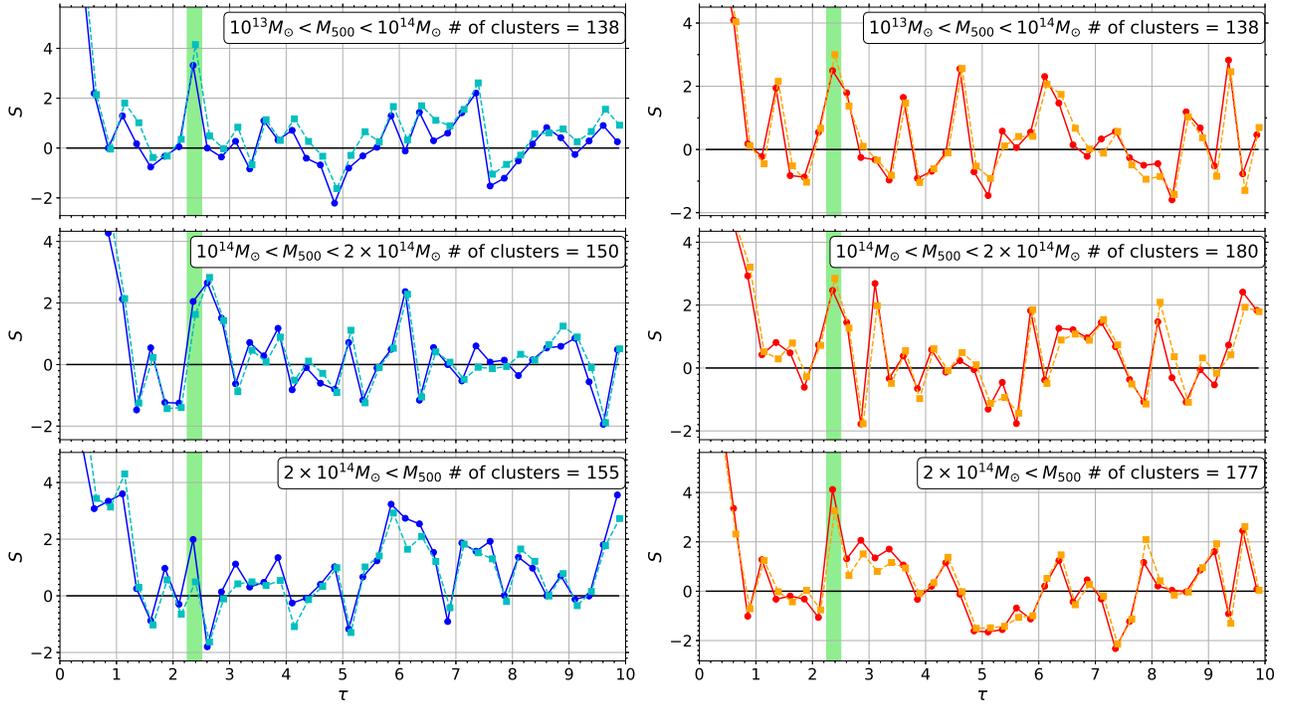}
    \end{center}
	\caption{
	\label{fig:M_slices}
        Significance $S(\tau)$ profile of excess X-ray (left panel) and radio (right) sources in different $M_{500}$ bins (see legend);
        other notations are the same as in Fig.~\ref{fig:nu_sig_plot_4bins}.
    }
\end{figure}
\end{center}

We examine the dependence of the VS excess upon cluster mass by crudely splitting the cluster sample into three $M_{500}$ bins, as shown in Fig.~\ref{fig:M_slices}.
The bins are adjusted to include a roughly fixed, $\sim150$ number of clusters per bin, to provide comparable statistics in each bin; this is achieved only approximately, as we adopt the same mass bins for both X-ray and radio analyses.
The figure suggests that the mass dependence of the VS excess is weak, unlike the strong mass dependence expected \citep{LoebWaxman00, TotaniKitayama00, WaxmanLoeb00}, simulated \citep{KeshetEtAl03, KeshetEtAl04_EGRB}, and indicated observationally \citep{HouEtAl23} for smooth accretion.

It is interesting to examine the properties of the ICM-region, $1<\tau<1.5$ excess, featuring sources clearly associated with the cluster, but residing away from its center and probably not directly related to its VS.
These sources are somewhat brighter than the VS sources in X-rays, extending out to $\sim10^{44}\erg\se^{-1}$, but not in radio, where a deficit is seen above $\sim 10^{40.5}\erg\se^{-1}$.
The source spectra indicate $\sim$keV temperatures in X-rays; the radio spectra cannot be determined, as their distribution does not significantly differ from the field.
The sources are comparable to the VS sources in terms of size, $30\kpc\lesssim R\lesssim 70\kpc$ in X-rays, and $5\kpc\lesssim a\lesssim 50\kpc$ in radio.
Finally, while these sources show a linear polarization fraction somewhat smaller than in the field, there is a low-significance indication of preferred polarization in the radial ($\varphi\simeq 90\dgr$), and not the tangential, direction.

\onecolumngrid
\twocolumngrid

\par\null\newpage
\clearpage

\onecolumngrid
\twocolumngrid


\onecolumngrid
\twocolumngrid

\null\newpage
\clearpage

\onecolumngrid
\twocolumngrid


\section{Origin of the VS sources}
\label{sec:origin}

The above results show a highly significant excess of X-ray and radio catalog sources narrowly peaked near the same, $\tau\simeq 2.4$ normalized radius where previously stacked \gama-ray \citep{reiss2018detection} and radio \citep{HouEtAl23} continuum signals were attributed to VS-induced emission.
This spatial coincidence indicates that the excess sources are directly associated with the VS of their host cluster, especially considering the very narrow, $\Delta\tau\simeq 0.1$ width of their stacked excess.
Indeed, it is difficult to envision how a well-localized spike in source density at this particular radius could arise without involving the VS.
The conclusion is further supported by anecdotal evidence, such as the well-defined angles of radio source polarization and possibly also elongation with respect to the host cluster.
Here, we consider the physical mechanism by which the VS may induce the excess sources, as implied by the stacked signal (\S\ref{sec:results}) and statistical source properties (\S\ref{sec:characteristics}).

The comparable and narrow widths of the X-ray and radio signals suggests that the sources radiate, or, more likely, brighten significantly, only during VS crossing or soon thereafter.
For a body of length $l=100l_{100}\kpc$, such brightening would span a normalized $\Delta\tau\simeq0.1l_{100}$ radial extent during the $t\simeq l/v_s \simeq 0.1l_{100}T_1^{-1/2}\Gyr$ passage of the body into the $k_B T_d\simeq T_1 \keV$ ICM, approximating its infall through the VS as fixed at the $v_s\simeq (16k_B T/3\bar{m})^{1/2}$ shock velocity.
This conclusion is supported by the $0.02\lesssim\{\tau_R,\tau_a\}\lesssim 0.2$ extent of the individual sources comprising the excess, being comparable and not much smaller than the signal width.

The $\Delta P\lesssim 0.02\eV\cm^{-3}$ pressure change induced by the VS \citep[\eg][]{ArnaudEtAl10,PlanckV13,HurierEtAl19, keshet20coincident} is much too weak to penetrate down to the vicinity of compact bodies, certainly those found in the centers of galaxies. As anticipated in \S\ref{sec:Intro}, our results thus challenge previous claims that a well-localized excess of X-ray sources near the virial radius could be attributed to AGN.
Indeed, sources already associated with identified AGN show no significant localized excess; see \S\ref{app:AGN_identification}.
Instead, sources triggered by the VS are likely to be extended and gaseous.
Indeed, this conclusion is directly corroborated by the inferred, $10$--$100$ kpc extent of the excess sources, as well as by the polarization angle of the radio emission.

Specifically, the extent, spectrum, and polarization of the radio sources identify them as synchrotron emission from an extended distribution of CREs which have not fully cooled, gyrating in the predominantly tangential magnetic fields $B$ amplified by the VS.
For the $\gamma_\nu\simeq 2\times 10^4 B_1^{-1/2}$ Lorentz factor of the $1.4\GHz$ radio-emitting CREs, the cooling length
\begin{equation}\label{eq:lCool}
  l_{\mbox{\tiny cool}} \simeq v_s t_{\mbox{\tiny cool}}
  \simeq v_s \frac{6\pi m_e c f_B}{\sigma_T \gamma_\nu B^2}
  \simeq 110\frac{\sqrt{B_1T_1}}{1+0.1B_1^2}\kpc
\end{equation}
is comparable to the width of the signal at the $B_1\equiv B/1\muG\simeq 1.9$ maximum of the cooling time $t_{\mbox{\tiny cool}}(B)$.
Here, $c$ is the speed of light, $\sigma_T$ is the Thomson cross-section, $m_e$ is the electron mass, and we defined the fractions $f_\epsilon\equiv 1-f_B$ and
\begin{equation}\label{eq:fB}
  f_B\equiv \frac{B^2/8\pi}{u_{cmb}+B^2/8\pi}\simeq \frac{0.1B_1^2}{1+0.1B_1^2}
\end{equation}
of IC and synchrotron CRE emission, $u_{cmb}$ being the CMB energy density.

For the $\sim$few $0.1\mu$G fields anticipated over $>10\kpc$ scales, $l_{\mbox{\tiny cool}}\propto \nu^{-1/2}$ is shorter than our nominal $\Delta \tau$ resolution by a factor of a few at $1.4\GHz$, but not at low frequencies.
Interestingly, some possible $74\MHz$ broadening is found at low-significance by comparing the stacked NVSS, GMRT, and VLSSr profiles (see Fig.~\ref{fig:BroadbandRadio}).

In contrast to radio, the $\epsilon=\epsilon_1\keV$ X-ray sources are unlikely to arise from non-thermal, neither IC nor synchrotron, emission.
Indeed, IC emission over the relevant scales, dominated by $\gamma_\epsilon\simeq 10^3\epsilon_1^{1/2}$ CREs scattering CMB photons, would extend over a far too long, $l_{\mbox{\tiny cool}} \simeq 2 (T_1/\epsilon_1)^{1/2}\Mpc$ cooling length, greatly exceeding the width of the signal; IC emission is also too faint to explain the sources, as shown below.
The opposite problem of $l_{\mbox{\tiny cool}} \lesssim 9 (T_1B_1/\epsilon_1)^{1/2}\pc$ cooling and $l_{\mbox{\tiny acc}}<cl_{\mbox{\tiny cool}}/v_s$ acceleration lengths much too short to explain the signal would arise if X-rays were synchrotron emission, also requiring a radio spectrum similar to Eq.~\eqref{eq:PurePowerLaw}, which is softer than inferred in \S\ref{sec:characteristics}, as well as an ongoing acceleration of excessively energetic, $\gamma\gtrsim 10^{8.5}(\epsilon_1/B_1)^{1/2}$ CREs.
In either case, if CREs were responsible for the X-ray sources, we would expect to find at least a few 2RXS--NVSS pairs, absent in the present data.

Therefore, the X-ray excess is likely dominated by thermal emission from hot plasma.
Indeed, the X-ray spectral fits are marginally more consistent with excess sources having high, $k_BT\gtrsim 1\keV$ temperatures, than with sources of power-law spectra (see Figs.~\ref{fig:PropTMekal}, \ref{fig:PropGamma2}, and \ref{fig:PropTmTb}).
In addition, the evidence suggests (albeit at low confidence) that the radio signal has a radial extent $\sim 2.5$ times wider than in X-rays (see Table \ref{tab:short_summary}) and that the radio $a$ and $b$ distributions extend to scales somewhat larger than the X-ray radius $R$ (see Figs.~\ref{fig:PropSize}, \ref{fig:X_ray_typ_ext1}, and \ref{fig:radio_typ_ext1}), consistent with a thermal X-ray component being more compact than its relativistic, radio counterpart.

X-ray emission from the hot plasma halo around a galaxy of mass $M_{500}$ is thought to follow the same $L_{0.5\mbox{ \scriptsize keV}}^{2.0\mbox{ \scriptsize keV}}\simeq 10^{43} (M_{500}/10^{14}M_\odot)^{1.85}\erg\se^{-1}$ scaling that extends up to galaxy-cluster scales \citep[][for a mean $C_{\rm bolo}\simeq 1.1$ bolometric correction factor]{AndersonEtAl15}.
Such a halo around an infalling galaxy would substantially brighten if a fair fraction $f_s$ of its mass is shocked, with $f_s$ being plausibly large if the density profile is shallow (\eg $n\propto r^{-0.9}$ in Ref.~\citep{FaermanEtAl20}).
For instance, the $[0.1,2.4]\keV$ luminosity $L_X\simeq2L_{0.5\mbox{ \scriptsize keV}}^{2.0\mbox{ \scriptsize keV}}$ of $k_BT=0.1\keV$ plasma, compressed (factor $\sim10$ brightening) and heated (factor $\sim 2$ brightening for $Z=0.2$ metallicity) by a (typical: see below) Mach $\mach\simeq 3.3$ shock, could account for at least the fainter of the $L_X\simeq 10^{42\mbox{\scriptsize{--}}43}\erg\se^{-1}$ sources.
The shocked halos of the more massive galaxies or galaxy aggregates could account for sources of higher luminosities and larger, $R\simeq 100\kpc$ radii.
Note that a shock is driven also into the ICM, but is weak: $\mach\simeq 3/\sqrt{5}\simeq 1.3$.

Infalling galactic halos processed by the VS could also explain the excess radio sources, through the acceleration or reacceleration of CREs and the magnetization of the plasma.
When CRE scattering is sufficiently isotropic \citep{keshet2020diffusive}, diffusive shock acceleration \citep{AxfordEtAl77, Krymskii77, Bell78, BlandfordOstriker78} injects a CRE power-law spectrum of index
$p\equiv -d\ln{N}/d\ln{E}\simeq 2(\mach^2+1)/(\mach^2-1)$.
The $\alpha_{150\mbox{ \scriptsize MHz}}^{1.4\mbox{ \scriptsize GHz}}\simeq 0.55\mbox{--}0.8$ radio indices can thus be attributed to non-cooled CREs accelerated by $\mach\simeq [1+2/(\alpha-1/2)]^{1/2}\simeq 2.8\mbox{--}6.4$ shocks, which compress the gas and the tangential magnetic fields by a factor $1+3/(2\alpha)\simeq 2.9\mbox{--}3.7$, thus accounting for the radial polarization.
Such shocks require plausible
$16(5\mach^2-1)^{-1}k_BT_d/(1+3\mach^{-2})\simeq (0.1\mbox{--}0.3)T_1\keV$
pre-shock galactic-halo temperatures.
Considering the main, $\alpha\simeq0.55$ excess in Fig.~\ref{fig:radio_spec} and accounting for partial CRE cooling renders the high, $\mach\simeq 6$ regime more likely.

Such a shock, heating a fraction $f_s$ of the gas mass $M_g$ and depositing a fraction $\xi_e$ of the resulting thermal energy in CREs, leads to a synchrotron luminosity
\begin{eqnarray}\label{eq:nuLnu}
  \nu L_\nu & \simeq & \frac{2\alpha-1}{2}\gamma_\nu^{-(2\alpha-1)}\xi_e f_B \frac{v_s}{\Delta r(\nu)}\frac{f_s M_g}{m_p}k_BT  \\
  & \simeq & 10^{40} \xi_{1} f_B  \left( \frac{f_sM_g}{10^{11}M_\odot} \right) \frac{T_1^{3/2}B_1^{0.05}}{\Delta r_{100}} \erg\se^{-1} \, , \nonumber
\end{eqnarray}
where we defined $\xi_{1}\equiv \xi_e/1\%\simeq 1$ based on detected continuum VS signals \citep{KeshetEtAl17, reiss2018detection, keshet2018evidence, keshet20coincident, HouEtAl23}, and $\Delta r_{100}\equiv \Delta r/100\kpc\simeq 1$ based on our NVSS stacking.
Equation \eqref{eq:nuLnu} can explain at least the fainter of the $\nu L_\nu\simeq 10^{40\mbox{\scriptsize{--}}41}\erg\se^{-1}$ sources, even without accounting for a compressed or re-accelerated pre-existing CRE population.
As mentioned above, the counterpart IC signal (replacing $f_B\to f_\epsilon$ in Eq.~\ref{eq:nuLnu}) is much too faint to explain $L_X$ for plausible, $B\gtrsim 0.1\muG$ fields.

An interesting possibility is that the excess sources, or a subset of these sources, are associated with remnant lobes produced by galactic outflows from previous AGN or star formation bursts.
Such lobes could provide seed CREs and magnetic fields, more easily explaining the excess radio sources even at their high-luminosity end.
For example, a VS-induced, factor $\sim10$ brightening of faint lobes could account for the entire radio excess, if they comprise $\sim10\%$ of the rapidly declining (without the $F_\nu^*$ cut) $\mathbb{N}(L_\nu)$ distribution of field sources.
Remnant galactic outflows could better explain why in the radio, the radial width of the stacked excess and the high end of the source-size distribution are somewhat larger than in X-rays, and may possibly account for the elongation angle $\varphi$ distribution in Fig.~\ref{fig:PropAPA}  if the lobes are, \eg longer than $l_{\mbox{\tiny cool}}$ or are stronger along a large-scale filament feeding the cluster.
While the X-ray sources should still have a thermal origin, they could be strengthened if such outflows enrich their host halos.

\section{Summary and discussion}
\label{sec:discussion}

Following reports of excess X-ray sources in galaxy-cluster peripheries, we stack X-ray (2RXS) and radio (NVSS, GMRT, VLSSr) catalog sources around MCXC galaxy clusters.
We impose cuts on the MCXC to match source-catalog limitations (Table \ref{tab:psc}), and an upper (lower specific) flux cutoff on the X-ray (radio) catalog, to avoid the dominant bright (faint) sources in the foreground (background); see Fig.~\ref{fig:all_typ_flux}.
Our results are not, however, sensitive to any of our cuts or analysis details (see \S\ref{app:sensitivity_and_methods}).

The radially binned and stacked projected source-density profiles are very similar to each other in X-rays and in radio (see Fig.~\ref{fig:FinalOverdensity}, the fit results in Table \ref{tab:short_summary}, and their counterparts in \S\ref{app:sensitivity_and_methods}), presenting a pronounced peak near the same, $\tau\simeq 2.4$ normalized radius where source-masked \gama-ray \citep{reiss2018detection} and radio \citep{HouEtAl23} continuum signals were previously detected and attributed to emission from the VS.
Both X-ray and radio catalog signals are highly localized ($\Delta\tau\lesssim 0.1$), indicating that the excess sources are directly driven by the VS.

These excess X-ray ($\sim 4\sigma$ CL for $443$ clusters), radio ($\sim4 \sigma$ for $485$ clusters), and joint ($>5\sigma$) source signals are significant enough (even without priors; see \S\ref{sec:results} and \S\ref{app:sensitivity_and_methods}) to see directly in the stacked images even without full radial binning (Figs.~\ref{fig:quad_joint} and \ref{fig:quad_joint_n}).
The signals present robustly, independently, and with similar parameters in the four (2RXS, NVSS, GMRT, and VLSSr) catalogs, and are not sensitive to variations in our analysis methods and parameters (see \S\ref{app:sensitivity_and_methods}).
The signals and their CLs are verified using control cluster samples, showing that the stacked excess closely follows the expected Poisson statistics (Fig.~\ref{fig:mocks_norm_check_1} and \S\ref{app:Poisson}) away from the real clusters.

The properties of the excess VS sources can be deduced only statistically and separately in each catalog, because the individual sources comprising the excess cannot be isolated from the $\sim$ twice as large number of coincident field sources, and we cannot demonstrate a VS excess of X-ray--radio source pairs (see \S\ref{subsec:joint_xray_radio_sources}) with present statistics.
We thus characterize the excess distributions of the flux (Fig.~\ref{fig:X_ray_typ_flux1}), luminosity (Fig.~\ref{fig:X_ray_typ_lum1}), and size (Fig.~\ref{fig:PropSize}) of the sources, the temperature (Fig.~\ref{fig:PropTMekal}) of the X-ray sources, and the spectral index (Fig.~\ref{fig:radio_spec}) and elongation angle (Fig.~\ref{fig:PropAPA}) of the radio sources, as well as their linear polarization fraction (Fig.~\ref{fig:PropPolFrac2}) and angle (Fig.~\ref{fig:PropPolPA}). Additional source properties are provided in \S\ref{App:characteristics}.

The inferred properties of the excess X-ray and radio sources support their association with the VS, and constrain the physical mechanism responsible for their emergence (see \S\ref{sec:origin}).
In particular, the radial, $\phi\simeq 0$ polarization of the radio sources, and their elongation angle distribution with its tentative tangential, $\varphi\simeq 90\dgr$ component, relate these sources to the VS orientation.
The hard, $0.55\lesssim\alpha_{150\mbox{ \scriptsize MHz}}^{1.4\mbox{ \scriptsize GHz}}\lesssim0.8$, partially polarized radio signal is identified as synchrotron emission from CREs energized by the VS, gyrating in predominately tangential magnetic fields amplified by the VS, whereas the X-ray emission is most probably thermal, arising from an extended plasma distribution, compressed and heated to $k_BT\gtrsim 1\keV$ temperatures by the VS.
The excess sources are thus identified as heated gas clumps accreted through the VS, probably the hot halos of galaxies or galaxy aggregates, possibly enriched by dormant lobes from galactic outflows.

The narrowness of the excess-source signals and their coincidence with each other and with the VS also corroborate previous claimed detections of VS signals based on continuum emission \citep{KeshetEtAl17, reiss2018detection, keshet2018evidence, keshet20coincident, HurierEtAl19, HouEtAl23}, which were more complicated due to smaller numbers of relevant clusters, as well as challenges associated for example with data reduction, point-source removal, and interferometric sidelobes.
A comparison with these studies indicates that in all cases, planar models fit the stacked $\tau\sim 2.4$ signals better than their projected-shell counterparts, and that a secondary weaker, broader, and farther, $5\lesssim\tau\lesssim 7$ signal emerges (\emph{cf.} Ref.~\citep{reiss2018detection}, \citep{HouEtAl23}, and the present Figs.~\ref{fig:nu_sig_plot_4bins}--\ref{fig:FinalOverdensity}).
Both effects indicate that a spherical, isolated model of the VS is oversimplified.
The secondary signal would imply, if established at high confidence, that projected VS are elongated, with a mean $\sim2.5$ aspect ratio, as indicated in Coma \cite{KeshetEtAl17, keshet2018evidence}, while the better fit of the planar model may also reflect quenching of the background by the VS; for a discussion, see Ref.~\citep{HouEtAl23}.
For the present signals, comprised of discrete excess sources, a planar model may also be favored by catalog selection effects preferring VS-compressed sources seen edge-on, but we are unable to study this possibility or the nature of the $\tau\sim 6$ excess sources with the available statistics.

Better statistics and new constraints are expected with future source and cluster catalogs, and are already accessible in part with catalogs recently made available.
Following this work, an analysis \citep{IlaniEtAl24} of the \emph{eROSITA}-DE Early Data Release (EDR) catalogs \citep{BrunnerEtAl22eRosita, LiuEtAl22eRosita, BaharEtAl22eRosita} established a significant, $3\sigma$--$4\sigma$ excess of X-ray sources at $2.0<\tau<2.25$ radii around EDR clusters.
In radio, the deep LOFAR Two-metre Sky Survey \citep[LoTSS;][]{Shimwell19} and Rapid ASKAP Continuum Survey \citep[RACS;][]{McConnellEtAl20} can be used to test our results, supplement them with better statistics, stronger spectral constraints, and additional polarization information, and possibly extend them in the future to higher redshifts.

A study of individual VS-driven source candidates is beyond the scope of the present work, but should become feasible with future dedicated studies and upcoming catalogs.
In particular, excess VS-driven radio sources, which should become clearer and more radially elongated at very low frequencies due to the longer $l_{\mbox{\tiny cool}}\propto \nu^{-1/2}$, may have already been observed, for example as the ``accretion relic'' or the NGC 4849 signature in Coma \citep{BonafedeEtAl22}, both extending out to the IC-traced elliptical VS \citep{KeshetEtAl17, keshet2018evidence}.
In the absence of concrete examples of VS-driven sources or an excess of X-ray--radio pairs, our statistical determination of the excess-source properties relies on a fairly small number of sources, and may be confused by superimposing different types of sources.
Thus, the physical origins of the X-ray and radio sources may differ more significantly from each other than inferred above, and tentative bimodality signatures (\eg in $a$, $b$, $\varphi$, and $T$) may indicate multiple source types within each catalog.

The excess radio sources should emit an IC signal at hard X-ray and \gama-ray energies $\epsilon$.
In the strong, $p\to2$ shock limit, the anticipated IC luminosity of such a source is $\epsilon L_\epsilon^{\mbox{\tiny(IC)}} \simeq (f_{\epsilon}/f_B)\nu L_\nu \simeq  10^{40\mbox{\scriptsize--}41}B_1^{-2}\erg\se^{-1}$ (assuming $l_{\mbox{\tiny cool}}\simeq \Delta r$ for NVSS sources, as inferred in \S\ref{sec:origin}).
Such values are comparable, for $B\simeq 1\muG$ fields within the source, to the luminosity of the entire VS ring, as inferred from \gama-rays in stacked \citep{reiss2018detection} and individual \citep{KeshetEtAl17, keshet2018evidence, keshet20coincident} clusters.
Hence, a few such sources, activated by some VS, may outshine the entire IC signal from smooth accretion through this VS, especially at the lower energies $\epsilon$ accessible by the $p>2$ source CREs.
Similarly, a $\nu L_\nu\simeq 10^{40\mbox{\scriptsize{--}}41}\erg\se^{-1}$ radio source outshines the $\nu L_\nu\simeq 10^{39}\erg\se^{-1}$ synchrotron ring from smooth accretion (extrapolating the Ref.~\citep{HouEtAl23} signal to higher frequencies with their median parameters).

Nevertheless, the radiative VS signatures due to excess sources vs. smooth accretion are easily distinguished from each other by their markedly different properties, in particular brightness, extent, spectra, and cluster-mass dependence.
For instance, in contrast to the strong $M_{500}$ dependence anticipated and measured \citep{HouEtAl23} for smooth accretion, we find no significant mass dependence in the excess-source properties, as demonstrated in Fig.~\ref{fig:M_slices}.
Moreover, localized sources were vigorously removed and masked in previous \emph{Fermi}-LAT and LWA analyses.
In contrast, a discrete source component cannot be ruled out in the somewhat high \citep{keshet20coincident} VS flux inferred in Coma from VERITAS data \citep{KeshetEtAl17}.

The link established here between the VS and a coincident spike in discrete thermal (X-ray) and non-thermal (radio, and possibly also X-ray and \gama-ray) sources opens new pathways to tracing the VS, mapping its relation to the cosmic web, and uncovering the underlying physical processes.
If the multiple, bright peripheral X-ray sources per cluster reported previously at $z\sim 1$ are indeed identified with the VS, then the redshift evolution
of VS sources is substantial, and similar analyses could shed light on the cosmological evolution of clusters, galaxies, galactic outflows, or other infalling objects.
The strong, naively $L_X\propto \rho^2 \propto (1+z)^6$ redshift dependence of thermal emission from shocked objects would render them useful VS tracers at high redshifts.
If the weak dependence of the signal upon host cluster, hinted by Fig.~\ref{fig:M_slices}, is confirmed, then a radial spike in discrete sources around an object could map its VS even on small, sub-cluster scales.

\section*{Acknowledgements}
We thank T. Boller, I. Rofeh, D. Kushnir, I. Gurwich, I. Reiss, Y. Lyubarsky, M. Gedalin, and Y. Faerman, for helpful discussions and support.
This research has received funding from the Israel Science Foundation (ISF grants No. 1769/15 and 2126/22), from the IAEC-UPBC joint research foundation (grant No. 300/18), and from the Ministry of Science, Technology \& Space, Israel. \\
May Gideon Ilani's memory be a blessing.

\putbib
\end{bibunit}

\clearpage

\onecolumngrid
\begin{center}
{\large Supplementary Material}
\vspace{0.3cm}
\end{center}
\twocolumngrid

\setcounter{page}{1}
\appendix

\begin{bibunit}[apsrev4-2-author-truncate]

\renewcommand\thesubsection{\roman{subsection}}
\renewcommand\thefigure{\thesection.\arabic{figure}}
\counterwithin{figure}{section}

\section{Spectral modeling}
\label{app:X_ray_emission_models}

\begin{center}
\begin{figure}[h]
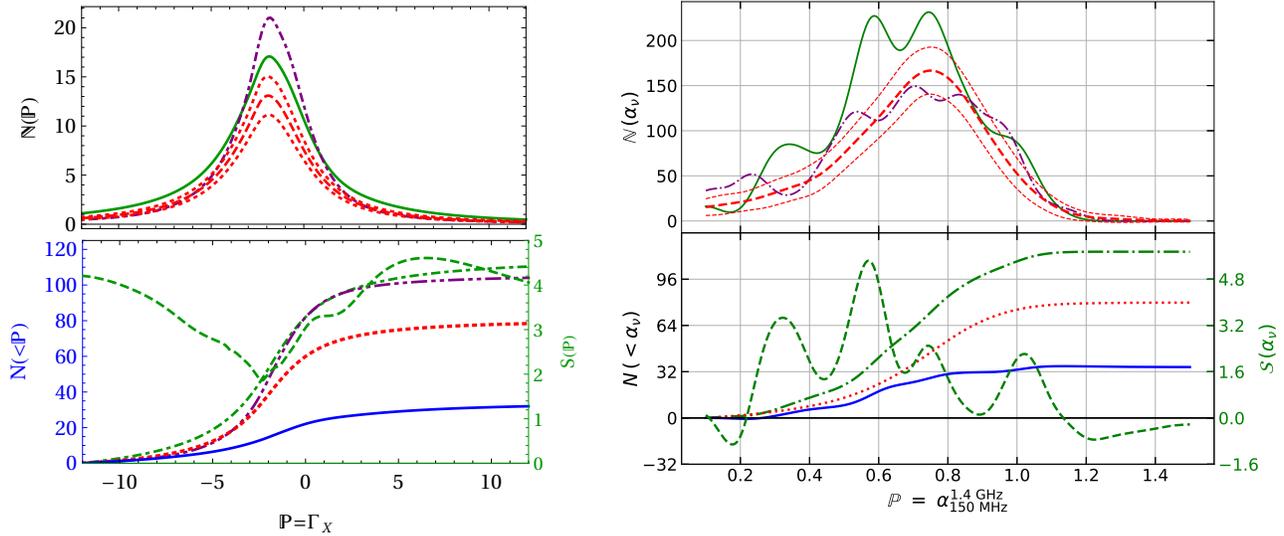

    \begin{center}
        \includegraphics[width=0.41\linewidth,trim={0cm 0.95cm 0cm 0cm},clip]{\myfig{PropGammaXRays1b.eps}}
        \hspace{0.3cm}
        \raisebox{-3.8cm}{\includegraphics[width=0.50\linewidth,trim={0cm 0cm 0cm 0cm},clip]{\myfig{radio_PL_ind_hist_spec_30sec_Hou.eps}}
        }
    \end{center}
    \vspace{-4.2cm}
    \begin{center}
            \hspace*{-9.8cm}
            \includegraphics[width=0.42\linewidth,trim={0cm 0cm 0cm 0cm},clip]{\myfig{PropGammaXRays2b.eps}}
    \end{center}
    \vspace{-0.4cm}
    \caption{
    Distributions of power-law spectral indices $\myP=\Gamma_X$ in X-rays (left panel) and $\mathbb{P}=\alpha_{150\mbox{\scriptsize MHz}}^{1.4\mbox{\scriptsize GHz}}$ in radio (right); notations are the same as in Fig.~\ref{fig:X_ray_typ_flux1}.
    \label{fig:PropGamma2} }
\end{figure}
\end{center}

\begin{center}
\begin{figure}[h!]
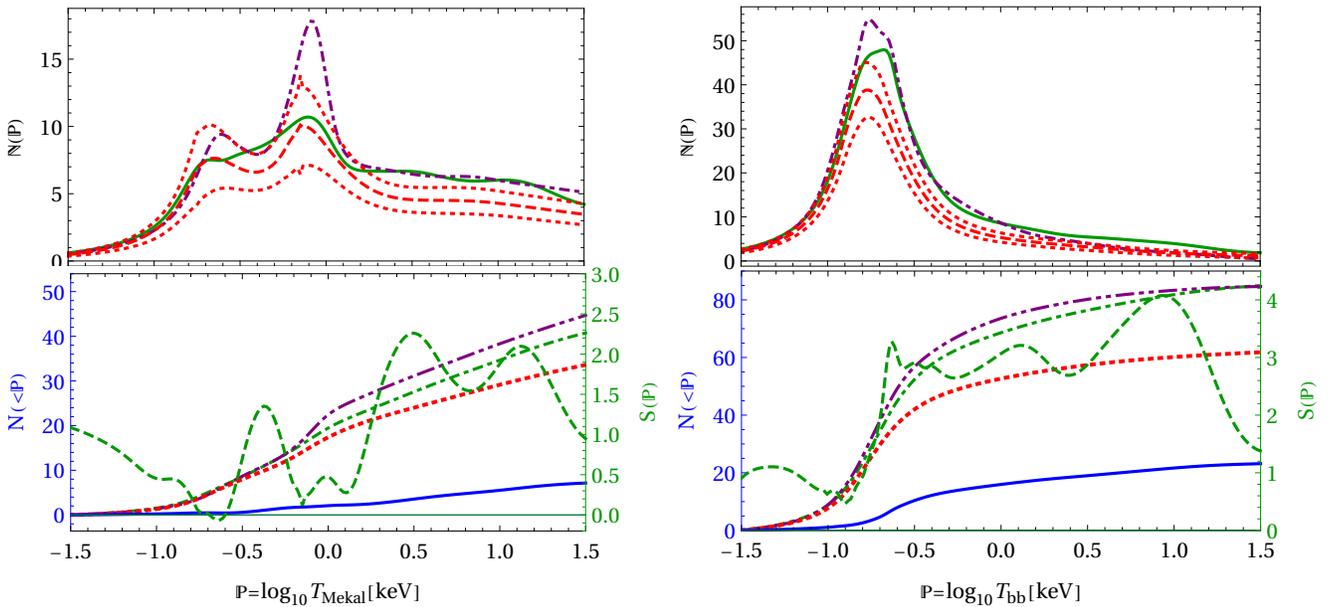

    \begin{center}
        \includegraphics[width=0.47\linewidth,trim={0cm 1.0cm 0cm 0cm},clip]{\myfig{PropTMekalXRays1b.eps}}
        \hspace{0.3cm}
        \raisebox{-0.1cm}{\includegraphics[width=0.473\linewidth,trim={0cm 0.9cm 0cm 0cm},clip]{\myfig{PropTbbXRays1b.eps}}}
    \end{center}
    \vspace{-0.5cm}
    \begin{center}
            \hspace*{-0.1cm}
            \includegraphics[width=0.485\linewidth,trim={0cm 0cm 0cm 0cm},clip]{\myfig{PropTMekalXRays2b.eps}}
            \hspace{0.0cm}
            \raisebox{0.cm}{\includegraphics[width=0.475\linewidth,trim={0cm 0cm 0cm 0.0cm},clip]{\myfig{PropTbbXRays2b.eps}}}
    \end{center}
    \caption{
    Distribution of X-ray Mekal (left) and black-body (right) fitted temperature $\mathbb{P}=\log_{10}T$.
    Notations are as in Fig.~\ref{fig:PropPolFrac2}.
    \label{fig:PropTmTb} }
\end{figure}
\end{center}

\vspace{-0.3cm}
Figure \ref{fig:PropGamma2} shows that while the radio spectral index has a significant $\sim 5\sigma$ excess of $\alpha_{150\mbox{ \scriptsize MHz}}^{1.4\mbox{ \scriptsize GHz}}\simeq 0.55$ indices associated with $\sim 10$ sources, and, in addition, indications of $\sim 10$ more $0.6\lesssim\alpha_{150\mbox{ \scriptsize MHz}}^{1.4\mbox{ \scriptsize GHz}}\lesssim0.8$ excess sources (right panel), there is no consistent signal in the $\Gamma_X$ distribution of X-ray spectral indices in the power-law fit (left panel). The temperatures of both Mekal and black-body X-ray spectral fits in Fig.~\ref{fig:PropTmTb} show a more localized excess, consistent with about half of the excess sources having $T\gtrsim 1\keV$ temperatures.

\onecolumngrid

\twocolumngrid

\par\null\newpage
\clearpage

\section{Catalog source sizes}
\label{app:cat_properties}

The 2RXS catalog provides an estimate of the angular extent of each source, expressed (T. Boller, private communications 2020) as the radius of the source minus the radius of the point spread function (PSF). For simplicity, we approximate the source radius as the e-fold drop in photon distribution \citep[][]{boese2001maximum}.
The PSF varies with the location of the source across the PSPC field of view, so we adopt an averaged, approximate PSF value.
For this purpose, we take the e-fold radius of the RASS PSF model \citep[][]{boese2000rosat}, providing an estimated radius $r_{psf}\sim 40''$.
We verify that this result is sensible, by examining individual sources.
Our conclusions are not sensitive to the precise value of the actual PSF.

The NVSS catalog provides the FWHM major axes $2a$ and minor axes $2b$ of sources, constrained in the range $14''<2b<2a<286''$.
Source localizations have an angular precision of $\gtrsim 7''$ for sources fainter than $15 \mJy$, but the brighter sources comprising our sample have a high, $\gtrsim 1''$ precision.

\vspace{-0.15cm}
\section{Confidence levels for Poisson statistics}
\label{app:Poisson}

Monte-Carlo simulations were used to numerically estimate the CLs $S_p$ for Poisson, rather than normal, statistics, demonstrated in Fig.~\ref{fig:mocks_norm_check_1} (the empty black symbols show the $S$ values corresponding to $S_p=\pm1,\pm2\ldots$), but it is useful to derive an approximate expression for $S$ given $S_p$.
For a Poisson distribution of mean $\lambda$, where measurement $k$ has probability $p(k)= e^{-\lambda} \lambda^k/k!$, we may analytically sum the probabilities of equal or larger (smaller) $k$ values, in order to quantify the significance of an excess (deficit) in units of Gaussian standard errors,
\begin{equation}
\label{eq:PoissonCL_corrected0}
\!\!\!\!\!\!S_p(k;\lambda) =
\begin{cases}
\vspace{0.1cm}
\sqrt{2}\,\mathrm{erfc}^{-1}\left[\frac{\Gamma(k)-\Gamma(k,\lambda)}{\Gamma(k)/2}\right]>0 & \mbox{if } k>\lambda\,; \\
\vspace{0.1cm}
-\sqrt{2}\,\mathrm{erfc}^{-1}\left[\frac{\Gamma(1+k,\lambda)}{\Gamma(1+k)/2}\right]<0 & \mbox{if } k<\lambda\,; \\
0 & \mbox{if } k=\lambda\,,
\end{cases}
\end{equation}
where $\mathrm{erfc}(x)$ and $\Gamma(x)$ are respectively the complementary error and Euler Gamma functions.
For SW co-addition, where $S_{\rm \SWt}$ is the function of $\NS$ and $\FS$ given in Eq.~\eqref{eq:sig_sw}, one can thus relate $S_{\rm \SWt}$ to $S_p$ for any given $\FS$ by solving the equation
\begin{equation}\label{eq:PoissonCL_corrected1}
  S_p = S_p(\NS=\FS+S_{\rm \SWt}\sqrt{\FS};\FS) \fin
\end{equation}
The resulting $S_{\rm \SWt}$ profiles are shown for integer $S_p$ values in the SW figures (as dashed black curves).

For modelling, we Monte-Carlo simulate our $\chi^2_{\rm \SWt}$ and $\chi^2_{\rm \CWt}$ expressions under the field-only null hypothesis and Poisson statistics. We verify that the results follow the expected $\chi^2$ distributions, thus justifying the parameter fit and TS significance estimates.
In particular, when fitting models to the stacked excess, we estimate the relative goodness of fit using the TS test, following Eqs.~\eqref{eq:chi_sw}--\eqref{eq:DefTS} and Wilks' theorem \citep{Wilks1938}, which are based on underlying normal statistics. To verify that the results do not change significantly under Poisson statistics, we use Monte-Carlo simulations, with the null hypothesis of Poisson-distributed field sources only, to estimate $\chi^2$ and TS. The resulting distributions are found to closely follow the anticipated $\chi_{\mathsf{n}}^2$ distribution with the correct effective number $\mathsf{n}$ of degrees of freedom.

\vspace{-0.15cm}
\section{Convergence and sensitivity tests}
\label{app:sensitivity_and_methods}

Our results are qualitatively insensitive to variations in the selection criteria (of clusters, sources, or field sources) and analysis methods (of field removal and source binning, stacking, or modeling).
This robustness is demonstrated in Figs.~\ref{fig:galac_cut_sensi}--\ref{fig:quad_joint_n} below, each varying some analysis parameter with little qualitative impact on the outcome with respect to the nominal results (which are shown for reference as a solid red curve in most figures).

In particular, the results are shown to be robust to variations in the $|b|$ cut (Fig.~\ref{fig:galac_cut_sensi}), the masking of clusters in different sky regions and the minimal projected cluster interdistance (Fig.~\ref{fig:binary_cut_sensi}), the order of the polynomial fit to the field source distribution $\FS$ around each cluster (Fig.~\ref{fig:on_const_fg_Xray}), the radial $\Delta\tau$ resolution (Fig.~\ref{fig:StacikingDifferentDTau}), the X-ray and radio flux cutoffs $F^*$ (Fig.~\ref{fig:flux_cut_sensi}), the likelihood $\mathcal{L}$ cutoff on the X-ray sources (Fig.~\ref{fig:likelihood_cut_test}), and the presentation and folding of stacked sources in 2D sky coordinates (\eg varying the resolution and smoothing of the map; Fig.~\ref{fig:quad_joint_n}).
When adopting the nominal, constant density $\FS$ of field sources (\ie a polynomial of order zero) around each cluster, we vary the region used to measure $\FS$, with inner radii spanning the range $1\dgr\till5\dgr$ and outer radii in the range $6\dgr\till10\dgr$, resulting in only negligible changes to the $S(\tau)$ profile.

\begin{center}
\begin{figure}[h]
    \begin{center}
        \hspace*{-0.4cm}
        \includegraphics[width=0.85\linewidth]{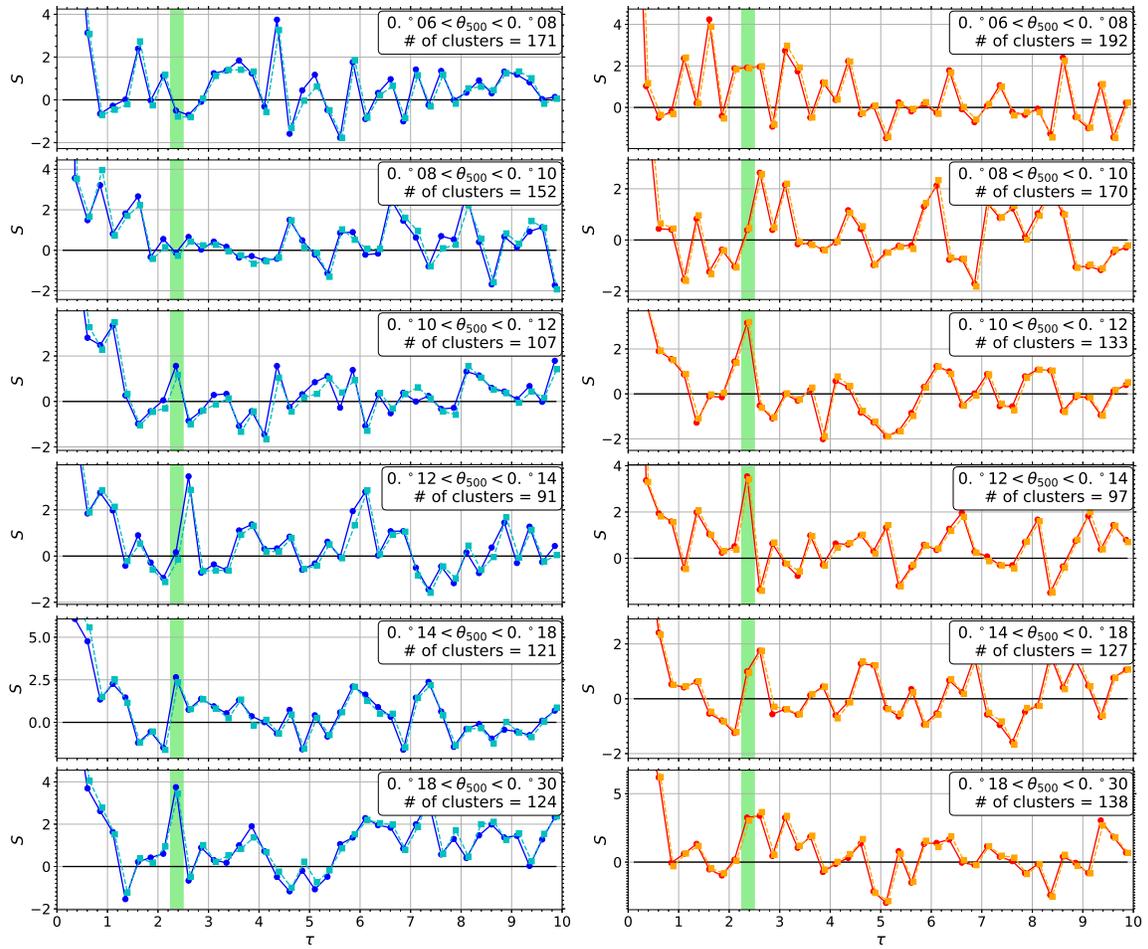}
    \end{center}
	\caption{
	\label{fig:theta_slices}
        Same as Fig.~\ref{fig:M_slices}, but for different $\tefh$ bins (see legend).
    }
\end{figure}
\end{center}

The same procedure used to test for a mass dependence in Fig.~\ref{fig:M_slices} is used in Fig.~\ref{fig:theta_slices} to examine the dependence of the excess upon $\theta_{500}$, here with six bins of $\sim150$ clusters each, extending beyond the nominal $\theta_{500}$ range.
For sufficiently extended clusters, the $\theta_{500}$ dependence appears to be weak.
However, the VS excess entirely vanishes for compact, $\theta_{500}\lesssim 0\dgrdot1$ clusters, at least in X-rays.
The high resolutions of the catalogs suggest that this disappearance is association with the intrinsic widths of the signals.
Indeed, this cutoff scale corresponds to $d_A(\bar{z})\theta_{500}\Delta\tau\simeq 140\kpc$, broadly consistent with the signal widths inferred in \S\ref{sec:results}.
Here, $d_A$ is the angular diameter distance, and we adopted the median $\bar{z}\simeq 0.08$ redshift in the sample.

Variations in the modelling of the VS excess are demonstrated in Table \ref{tab:long_summary}, expanding on the nominal models presented in Table \ref{tab:short_summary}.
Here, we also examine the use of priors on the model (numbers in {\bf bold} are frozen parameters).
As the table shows, the determination of the position and width of the VS excess is robust, and the significance of the detection increases if prior information, in particular regarding the shock position, is used.

\clearpage
\newpage

\onecolumngrid
\begin{center}
\begin{figure}[h]
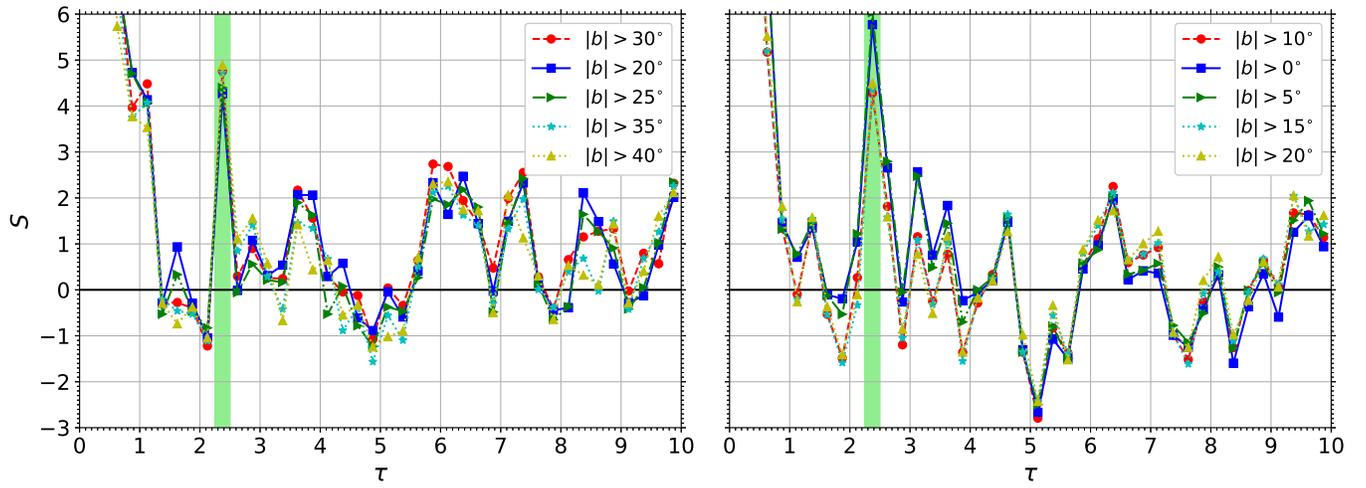

    \hspace{-0.2cm}
    \centerline{
    \DrawFig{\MYepsfbox{\myfig{Galactic_cut_test.eps}}}}
    \vspace{-0.5cm}
	\caption{\label{fig:galac_cut_sensi}
     Significance $S(\tau)$ profiles of the SW-stacked X-ray (left panel) and radio (right panel) source excess under variations in the Galactic latitude cutoff $|b|$ (see legend); the nominal analysis is shown, for reference, as a solid red curve.
     The anticipated VS region is highlighted as in Fig.~\ref{fig:mocks_norm_check_1}.
     Notice the contribution of $|b|<10\dgr$ radio sources, neglected in the nominal analysis.
	}
\end{figure}
\end{center}

\begin{center}
\begin{figure}[h]
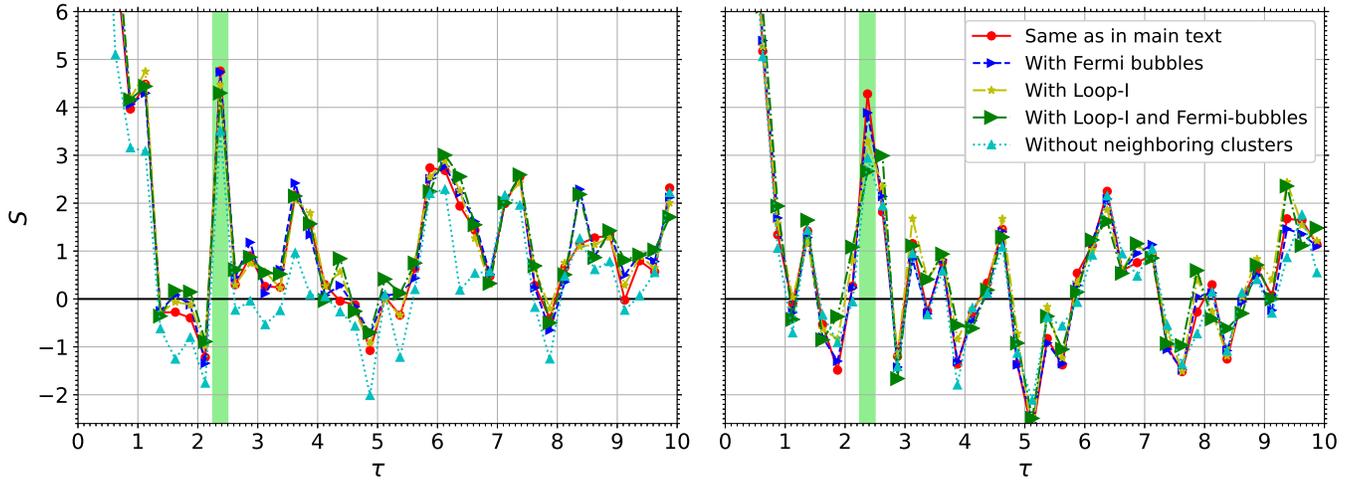

    \hspace{-0.2cm}
    \begin{overpic}[width=1\linewidth]{\myfig{Binary_cut_test.eps}}
        \put (33.9,35.8) {
        {}
        }
        \put (82.8,35.8) {
        {}
        }
    \end{overpic}
    \vspace{-0.5cm}
	\caption{\label{fig:binary_cut_sensi}
    Variations in cluster masking in different regions (see legend), with neighboring clusters defined as pairs within $5\tefh$ in projection;
    notations are the same as in Fig.~\ref{fig:galac_cut_sensi}.
	}
\end{figure}
\end{center}
\twocolumngrid

\clearpage
\newpage

\onecolumngrid
\begin{center}
\begin{figure}[h]
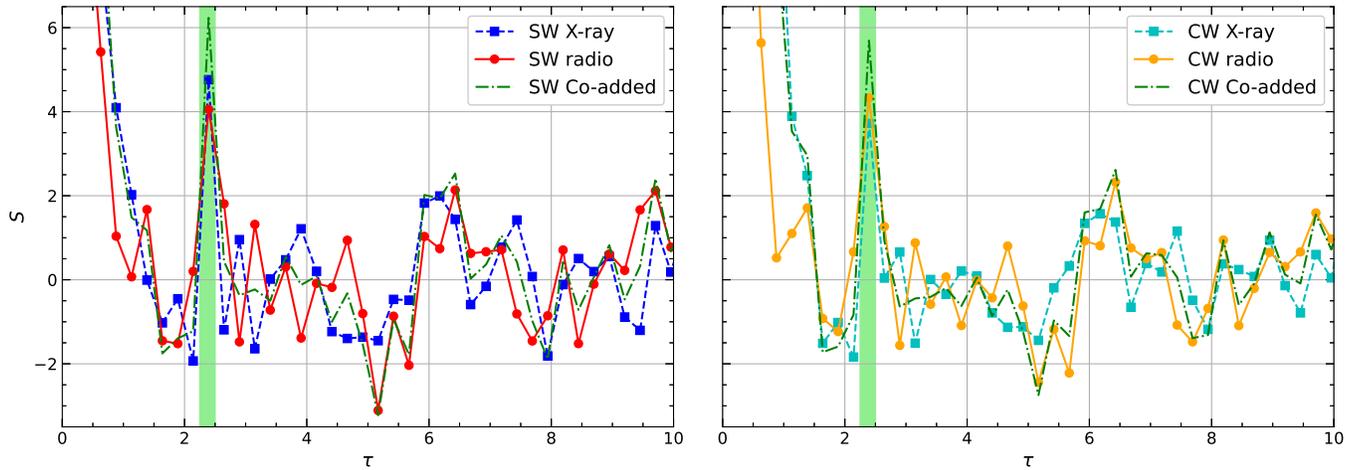

    \begin{center}
    \DrawFig{\MYepsfbox{\myfig{Non_const_fg_R15_Np4_Hou.eps}}}
    \end{center}
	\caption{\label{fig:on_const_fg_Xray}
    Same as the nominal Fig.~\ref{fig:nu_sig_plot_4bins}, but with field source densities $\FS$ inferred from a fourth-order polynomial fit to their sky distribution within $1\le \tau \le15$ around each cluster.
	}
\end{figure}
\end{center}
\twocolumngrid

\onecolumngrid
\begin{center}
\begin{figure}[h]
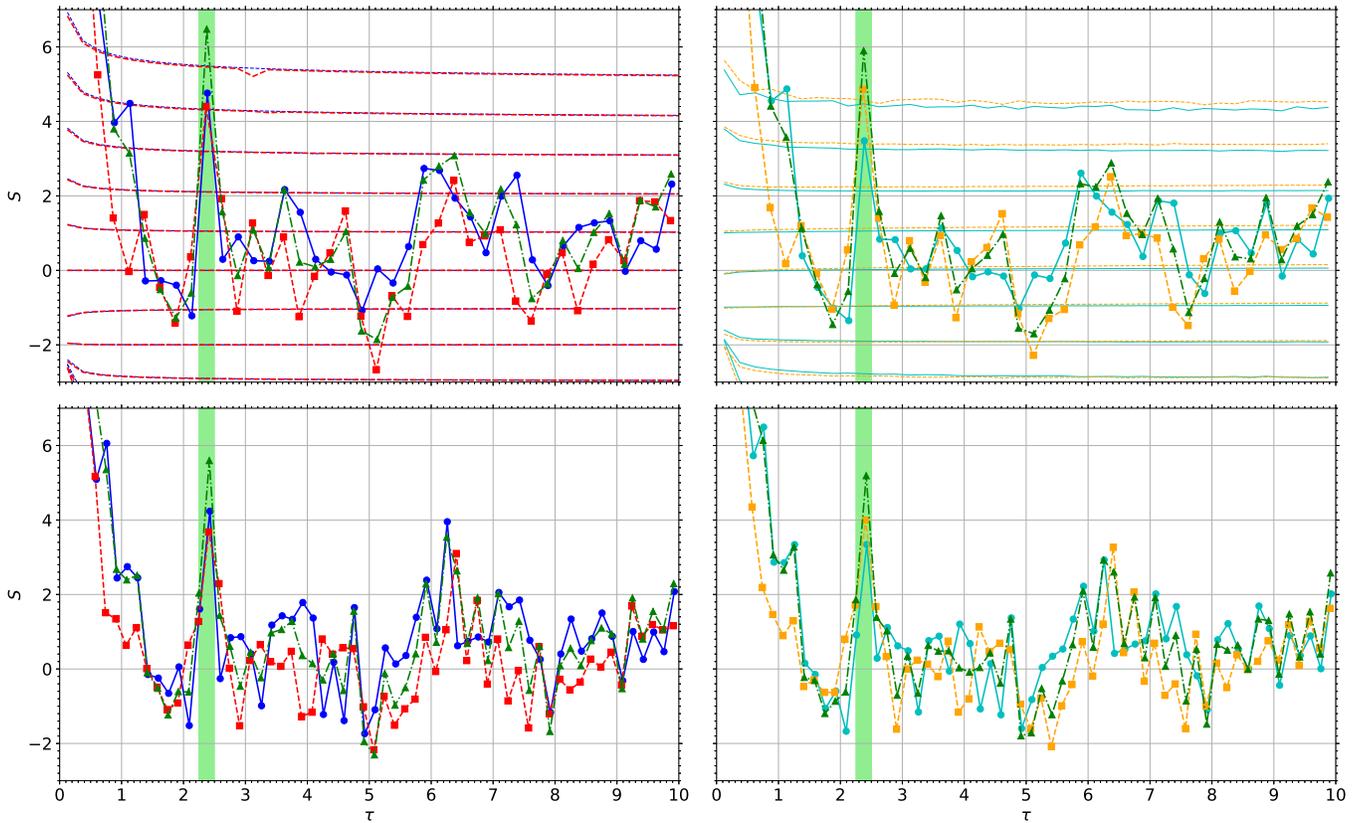

    \begin{center}
    \DrawFig{\MYepsfbox{\myfig{joint_all_sig_Hou.eps}}}
    \end{center}
	\caption{
        \label{fig:StacikingDifferentDTau}
        Same as the nominal Fig.~\ref{fig:nu_sig_plot_4bins}, comparing $\Delta\tau=1/4$ (top panels) and $\Delta\tau=1/6$ (bottom) resolutions.
		}
\end{figure}
\end{center}
\twocolumngrid

\vspace{-5cm}

\clearpage
\newpage

\onecolumngrid
\begin{center}
\begin{figure}[h]
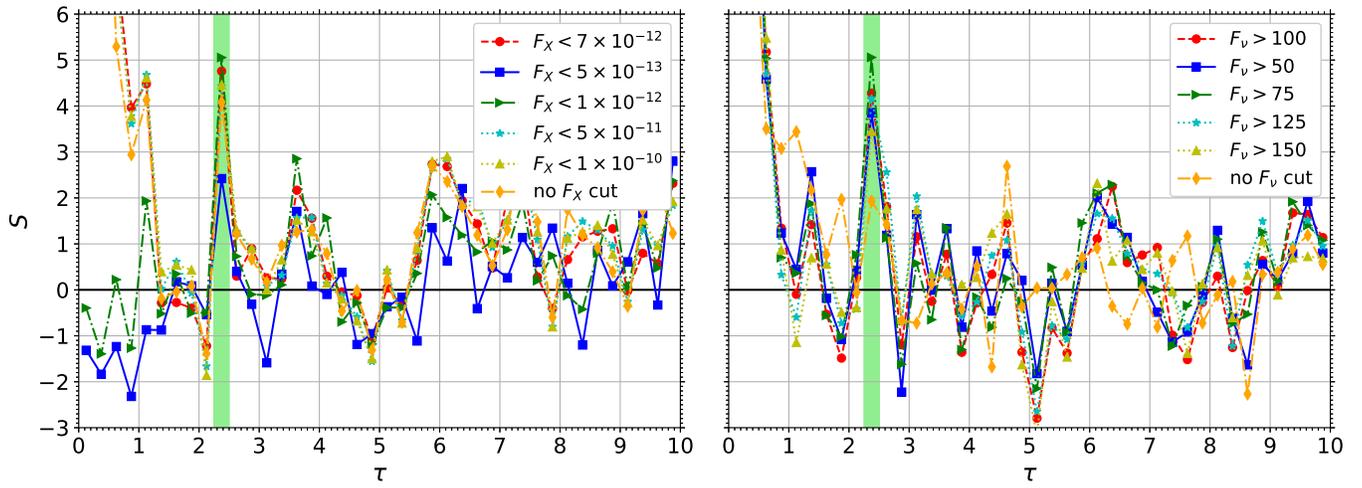

    \hspace{-0.2cm}
    \centerline{
    \DrawFig{\MYepsfbox{\myfig{flux_cut_test_Hou_v1.eps}}}}
	\caption{\label{fig:flux_cut_sensi}
    Variations in flux cutoffs (see legend), including the case of no flux cut at all (orange diamonds); notations are the same as in Fig.~\ref{fig:galac_cut_sensi}.
	}
\end{figure}
\end{center}
\twocolumngrid

\begin{figure}[t]
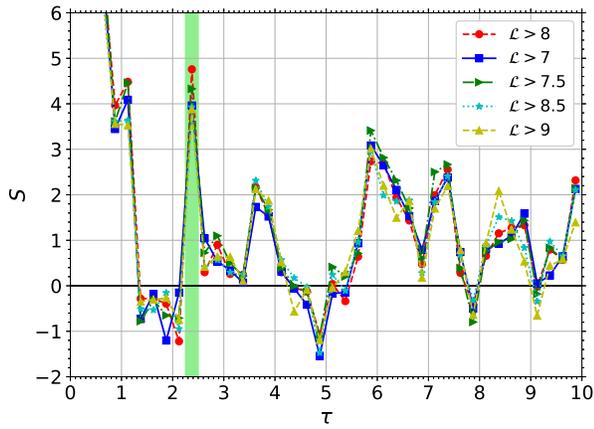

    \hspace{-0.2cm}
    \centerline{
    \DrawFig{\MYSepsfbox{\myfig{likelihood_cut_test.eps}}}}
	\caption{\label{fig:likelihood_cut_test}
    Variations in X-ray source likelihood (see legend); notations are the same as in the left panel of Fig.~\ref{fig:galac_cut_sensi}.
	}
\end{figure}

\onecolumngrid
\begin{center}
\begin{figure}[h]
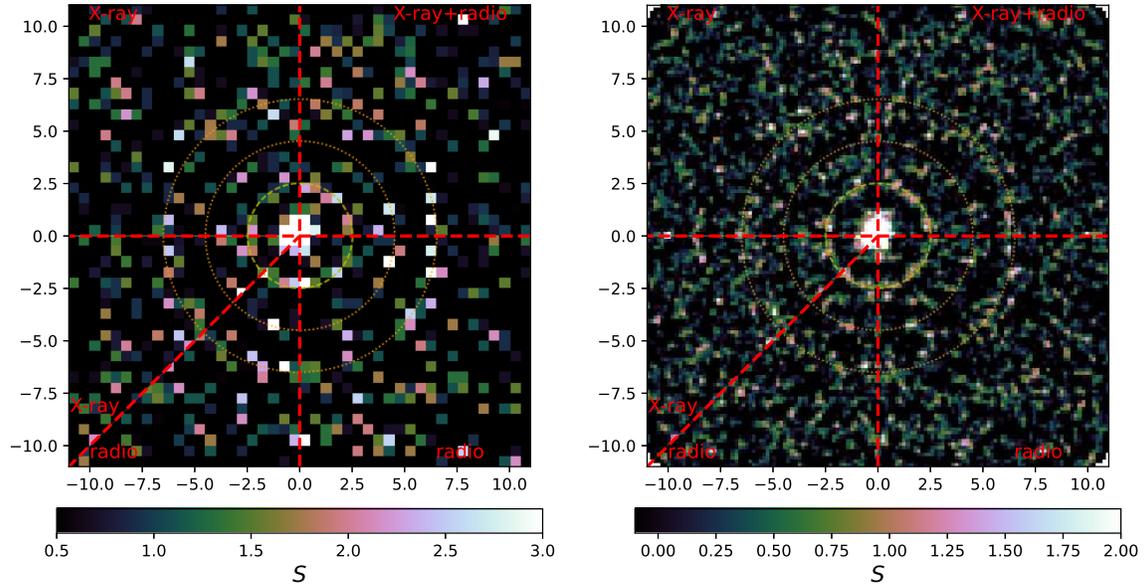

    \begin{center}
        \includegraphics[height=0.45\linewidth]{\myfig{comp_fig_significance_nonsmoothed_Hou.eps}}
        \includegraphics[height=0.45
\linewidth]{\myfig{comp_fig_significance_smoothed3pixs_Hou.eps}}
    \end{center}
    \caption{
    \label{fig:quad_joint_n}
    Same as Fig.~\ref{fig:quad_joint}, but for $\Delta\tau=1/2$ resolution (left panel) or after convolution with a Gaussian matrix of $0.5\theta_{500}$
    standard deviation  (right panel).
    }
\end{figure}
\end{center}
\twocolumngrid

\onecolumngrid
\twocolumngrid

\onecolumngrid
\begin{center}
\begin{table}
 \caption{
    Best-fitted parameters for different variants of the virial-excess model.
 }
 \vspace{0.2cm}
\begin{tabular}{l@{\hskip 0.2cm}l@{\hskip 0.4cm}ll@{\hskip 0.4cm}c@{\hskip 0.2cm}|@{\hskip 0.2cm}l@{\hskip 0.4cm}l@{\hskip 0.4cm}l@{\hskip 0.4cm}l@{\hskip 0.4cm}l@{\hskip 0.4cm}l}
Method & $\Delta\tau$ & Catalog & Model& $\mathsf{n}$ & $n_{0,x}$ & $n_{0,r} $ & $\tau_{sh}$ & $\Delta \tau_{sh}$& TS & $\sigmaTS$ \\
 (1) & (2) & (3) & (4) & (5) & (6) & (7) & (8) & (9) & (10) & (11) \\
\hline
SW                   & $1/4$ & 2RXS & Shell& 2 & $7.0^{+2.2}_{-2.0}$ & --- &$2.50^{+0.00}_{-0.02}$ & --- &  $14.0$ & $3.32$ \\
                     &   &       & Planar& 1 & $5.4^{+1.5}_{1.5}$ & ---  & $\mathbf{2.40}$& $\mathbf{0.01}$ &  $22.8$ & $4.77$ \\
                     &   &       &     & 2 & $5.4^{+1.6}_{-1.1}$ & --- & $\mathbf{2.40}$ & $0.05^{+0.04}_{-0.05}$ & $22.8$ & $4.39$ \\
                     &   &       &     & 3 & $5.4^{+2.1}_{-1.7}$ & --- & $2.44^{+0.06}_{-0.19}$ & $0.04^{+0.05}_{-0.04}$ & $22.8$ & $4.08$ \\
                     &   & NVSS & Shell & 2 & --- & $8.4^{+2.5}_{-2.4}$ &$2.52^{+0.05}_{-0.02}$ & --- &  $14.7$ & $3.42$ \\
                     &   &       & Planar& 1 & --- & $6.9^{+1.9}_{-1.7}$  & $\mathbf{2.40}$& $\mathbf{0.01}$ &  $21.3$ & $4.61$ \\
                     &   &       &     & 2 & --- & $6.9^{+1.4}_{-1.4}$ & $\mathbf{2.40}$ & $0.12^{+0.04}_{-0.04}$ & $21.3$ & $4.23$ \\
                     &   &       &     & 3 & --- & $6.6^{+2.1}_{-2.0}$ & $2.44^{+0.06}_{-0.06}$ & $0.10^{+0.05}_{-0.10}$ & $21.7$ & $3.96$ \\
                     &   & Joint & Shell      & 3 &$7.0^{+2.2}_{-2.0}$& $6.9^{+0.7}_{-0.7}$ & $2.50^{+0.24}_{-0.25}$ & --- & $27.8$ & $4.61$ \\
                     &   &       & Planar & 2 &$6.1^{+1.2}_{-1.2}$& $5.9^{+1.2}_{-1.2}$ & $\mathbf{2.40}$ & $\mathbf{0.01}$  & $42.6$ & $6.20$ \\
                     &   &       &     & 3 &$6.1^{+1.2}_{-1.2}$& $6.9^{+0.7}_{-0.7}$ & $\mathbf{2.40}$ & $0.09^{+0.03}_{-0.03}$ & $42.6$ & $5.93$ \\
                     &   &       &     & 4 &$5.9^{+1.7}_{-1.5}$& $5.7^{+1.7}_{-1.6}$& $2.47^{+0.02}_{-0.02}$ & $0.04^{+0.02}_{-0.02}$& $43.3$ & $5.75$ \\
                     & $1/6$ & 2RXS & Shell& 2 & $8.2^{+1.4}_{-3.4}$ & --- &$2.50^{+0.00}_{-0.04}$ & --- &  $14.2$ & $3.34$ \\
                     &   &       & Planar& 1 & $5.3^{+1.5}_{-1.5}$ & ---  & $\mathbf{2.40}$& $\mathbf{0.01}$ &  $19.8$ & $4.45$ \\
                     &   &       &     & 2 & $5.3^{+1.6}_{-1.6}$ & --- & $\mathbf{2.40}$ & $0.07^{+0.03}_{-0.03}$ & $19.8$ & $4.06$ \\
                     &   &       &     & 3 & $5.5^{+1.6}_{-1.5}$ & --- & $2.35^{+0.05}_{-0.02}$ & $0.03^{+0.04}_{-0.03}$ & $20.6$ & $3.83$ \\
                     &   & NVSS & Shell & 2 & --- & $8.1^{+2.4}_{-2.3}$ &$2.54^{+0.06}_{-0.03}$ & --- &  $14.0$ & $3.32$ \\
                     &   &       & Planar& 1 & --- & $6.9^{+1.9}_{-1.8}$  & $\mathbf{2.40}$& $\mathbf{0.01}$ &  $19.8$ & $4.45$ \\
                     &   &       &     & 2 & --- & $6.9^{+2.1}_{-1.4}$ & $\mathbf{2.40}$ & $0.14^{+0.04}_{-0.04}$ & $19.8$ & $4.06$ \\
                     &   &       &     & 3 & --- & $6.5^{+2.1}_{-1.8}$ & $2.44^{+0.04}_{-0.05}$ & $0.12^{+0.05}_{-0.04}$ & $20.4$ & $3.81$ \\
                     &   & Joint & Shell & 3 &$7.3^{+2.3}_{-2.1}$& $6.8^{+2.2}_{-2.1}$ & $2.50^{+0.83}_{-0.78}$ & --- & $25.8$ & $4.41$ \\
                     &   &       & Planar & 2 &$5.8^{+1.2}_{-1.2}$& $5.9^{+1.2}_{-1.2}$ & $\mathbf{2.40}$ & $\mathbf{0.01}$  & $38.0$ & $5.83$ \\
                     &   &       &     & 3 &$5.8^{+1.2}_{-1.2}$& $5.9^{+1.8}_{-1.2}$& $0.10^{+0.03}_{-0.02}$& $\mathbf{2.40}$& $38.0$ & $5.55$ \\
                     &   &       &     & 4 &$5.8^{+1.7}_{-1.5}$& $5.9^{+1.9}_{-1.7}$& $2.41^{+0.02}_{-0.02}$ & $0.10^{+0.03}_{-0.02}$& $38.1$ & $5.31$ \\
CW                   & $1/4$ & 2RXS & Shell& 2 & $4.2^{+2.1}_{-1.9}$ & --- &$2.50^{+0.02}_{-0.05}$ & --- &  $5.9$ & $1.94$ \\
                     &   &       & Planar& 1 & $3.7^{+1.4}_{-1.1}$ & ---  & $\mathbf{2.40}$& $\mathbf{0.01}$ &  $12.1$ & $3.48$ \\
                     &   &       &     & 2 & $3.7^{+1.5}_{-1.1}$ & --- & $\mathbf{2.40}$ & $0.01^{+0.03}_{-0.01}$ & $12.1$ & $3.04$ \\
                     &   &       &     & 3 & $4.3^{+1.7}_{-1.5}$ & --- & $2.47^{+0.03}_{-0.21}$ & $0.04^{+0.05}_{-0.04}$ & $12.8$ & $2.80$ \\
                     &   & NVSS & Shell & 2 & --- & $7.6^{+2.4}_{-2.2}$ &$2.50^{+0.03}_{-0.03}$ & --- &  $18.9$ & $3.95$ \\
                     &   &       & Planar& 1 & --- & $6.3^{+1.7}_{-1.5}$  & $\mathbf{2.40}$& $\mathbf{0.01}$ &  $24.9$ & $4.99$ \\
                     &   &       &     & 2 & --- & $6.3^{+1.9}_{-1.3}$ & $\mathbf{2.40}$ & $0.11^{+0.03}_{-0.03}$ & $24.9$ & $4.62$ \\
                     &   &       &     & 3 & --- & $6.2^{+1.9}_{-2.0}$ & $2.41^{+0.08}_{-0.06}$ & $0.11^{+0.04}_{-0.10}$ & $24.9$ & $4.32$ \\
                     &   & Joint & Shell & 3 &$4.9^{+2.4}_{-2.1}$& $7.7^{+0.7}_{-0.8}$ & $2.50^{+0.24}_{-0.25}$ & --- & $27.5$ & $4.58$ \\
                     &   &       & Planar & 2 &$3.8^{+1.2}_{-1.2}$& $5.9^{+1.8}_{-1.2}$ & $\mathbf{2.40}$ & $\mathbf{0.01}$  & $36.9$ & $5.74$ \\
                     &   &       &     & 3 &$3.8^{+1.2}_{-1.2}$& $7.7^{+0.7}_{-0.8}$ & $\mathbf{2.40}$ & $0.10^{+0.03}_{-0.03}$ & $36.9$ & $5.46$ \\
                     &   &       &     & 4 &$3.8^{+1.4}_{-1.2}$& $5.5^{+2.1}_{-1.6}$& $2.47^{+0.02}_{-0.02}$ & $0.04^{+0.02}_{-0.02}$& $37.5$ & $5.26$ \\
                     & $1/6$ & 2RXS & Shell& 2 & $4.3^{+2.0}_{-1.8}$ & --- &$2.50^{+0.01}_{-0.04}$ & --- &  $6.7$ & $2.10$ \\
                     &   &       & Planar& 1 & $3.8^{+1.5}_{-1.2}$ & ---  & $\mathbf{2.40}$& $\mathbf{0.01}$ &  $12.0$ & $3.47$ \\
                     &   &       &     & 2 & $3.8^{+1.5}_{-1.2}$ & --- & $\mathbf{2.40}$ & $0.07^{+0.04}_{-0.07}$ & $12.0$ & $3.03$ \\
                     &   &       &     & 3 & $3.9^{+1.7}_{-1.5}$ & --- & $2.39^{+0.10}_{-0.06}$ & $0.07^{+0.04}_{-0.07}$ & $12.1$ & $2.69$ \\
                     &   & NVSS & Shell & 2 & --- & $7.6^{+2.3}_{-2.2}$ &$2.51^{+0.03}_{-0.01}$ & --- &  $16.7$ & $3.68$ \\
                     &   &       & Planar& 1 & --- & $6.3^{+1.7}_{-1.6}$  & $\mathbf{2.40}$& $\mathbf{0.01}$ &  $22.0$ & $4.69$ \\
                     &   &       &     & 2 & --- & $6.3^{+1.9}_{-1.3}$ & $\mathbf{2.40}$ & $0.14^{+0.05}_{-0.04}$ & $22.0$ & $4.30$ \\
                     &   &       &     & 3 & --- & $6.2^{+2.1}_{-1.8}$ & $2.42^{+0.05}_{-0.05}$ & $0.13^{+0.05}_{-0.04}$ & $22.1$ & $4.00$ \\
                     &   & Joint & Shell & 3 &$4.8^{+2.4}_{-2.1}$& $7.2^{+0.7}_{-0.7}$ & $2.51^{+0.24}_{-0.25}$ & --- & $25.0$ & $4.33$ \\
                     &   &       & Planar & 2 &$3.8^{+1.1}_{-1.1}$& $5.6^{+1.1}_{-1.1}$ & $\mathbf{2.40}$ & $\mathbf{0.01}$  & $33.2$ & $5.42$ \\
                     &   &       &     & 3 &$3.8^{+1.1}_{-1.1}$& $7.2^{+0.7}_{-0.7}$ & $\mathbf{2.40}$ & $0.10^{+0.04}_{-0.02}$ & $33.2$ & $5.13$ \\
                     &   &       &     & 4 &$3.7^{+1.5}_{-1.3}$& $5.6^{+1.8}_{-1.6}$& $2.41^{+0.02}_{-0.02}$ & $0.10^{+0.04}_{-0.02}$& $33.3$ & $4.88$ \\
\hline
\end{tabular}
    \begin{tablenotes}
    \item
    Table columns: (1) Co-addition method (SW or CW); (2) Dimensionless (in $\theta_{500}$ units) binning resolution; (3) Stacked catalog (2RXS, NVSS, or joint catalogs); (4) Effective VS model (planar or shell); (5) Effective number of free parameters; (6) and (7) Normalized, $n_0^*\equiv 10^3 n_0$ X-ray and radio source densities in Eqs.~\eqref{eq:gauss_text} or \eqref{eq:shell_text};
    (8) Dimensionless shock radius; (9) Dimensionless signal width; (10) TS value; (11) TS-based significance.
    Numbers in {\bf bold} are imposed constraints.
    \end{tablenotes}
    \label{tab:long_summary}
    \end{table}
    \end{center}
\twocolumngrid

\clearpage
\newpage

\section{Additional VS excess-source properties}
\label{App:characteristics}

\begin{center}
\begin{figure}[ht]
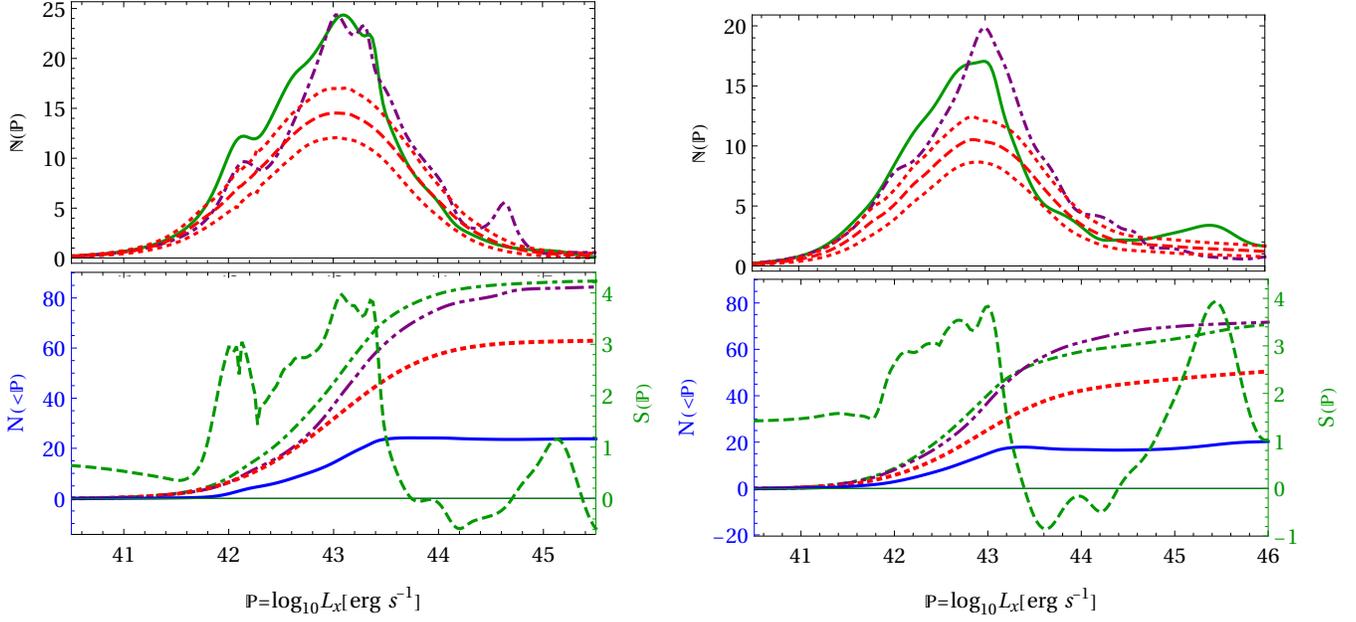

    \begin{center}
        \includegraphics[width=0.477\linewidth,trim={0cm 0.9cm 0cm 0cm},clip]{\myfig{PropLumXRays1b.eps}}
        \hspace{0.3cm}
        \raisebox{-0.05cm}{\includegraphics[width=0.477\linewidth,trim={0cm 0.95cm 0cm 0cm},clip]{\myfig{PropBBLumXrays1b.eps}}}
    \end{center}
    \vspace{-0.5cm}
    \begin{center}
            \hspace*{-0.1cm}
            \includegraphics[width=0.48\linewidth,trim={0cm 0cm 0cm 0cm},clip]{\myfig{PropLumXRays2b.eps}}
            \hspace{0.1cm}
            \raisebox{0.cm}{\includegraphics[width=0.49\linewidth,trim={0cm 0cm 0cm 0.0cm},clip]{\myfig{PropBBLumXrays2b.eps}}}
    \end{center}
    \caption{
    Distribution of X-ray luminosity $\mathbb{P}=\log_{10}L_X$ according to power-law (left panel) and black-body (right) spectral fits.
    Notations are the same as in Fig.~\ref{fig:PropPolFrac2}.
    \label{fig:PropLum2} }
\end{figure}
\end{center}

\begin{center}
\begin{figure}[h]
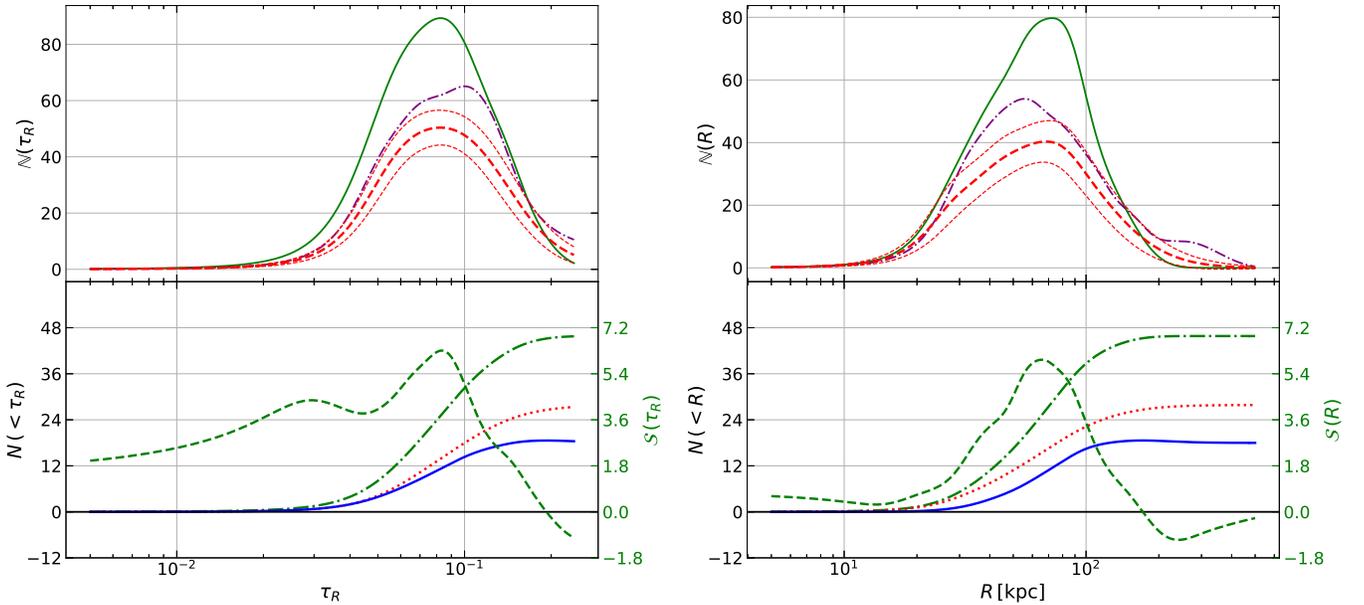

    \centerline{
    \DrawFig{\MYepsfbox{\myfig{Xray_size_Hou.eps}}}}
	\caption{\label{fig:X_ray_typ_ext1}\label{fig:X_ray_typ_phys1}
    Distributions of normalized (to $\theta_{500}$), dimensionless $\mathbb{P}=\tau_R$ (left panel) and non-normalized $\mathbb{P}=R$ (right) 2RXS source radii.
    Notations are the same as in Fig.~\ref{fig:X_ray_typ_flux1}.
	}
\end{figure}
\end{center}

\begin{center}
\begin{figure}[h]
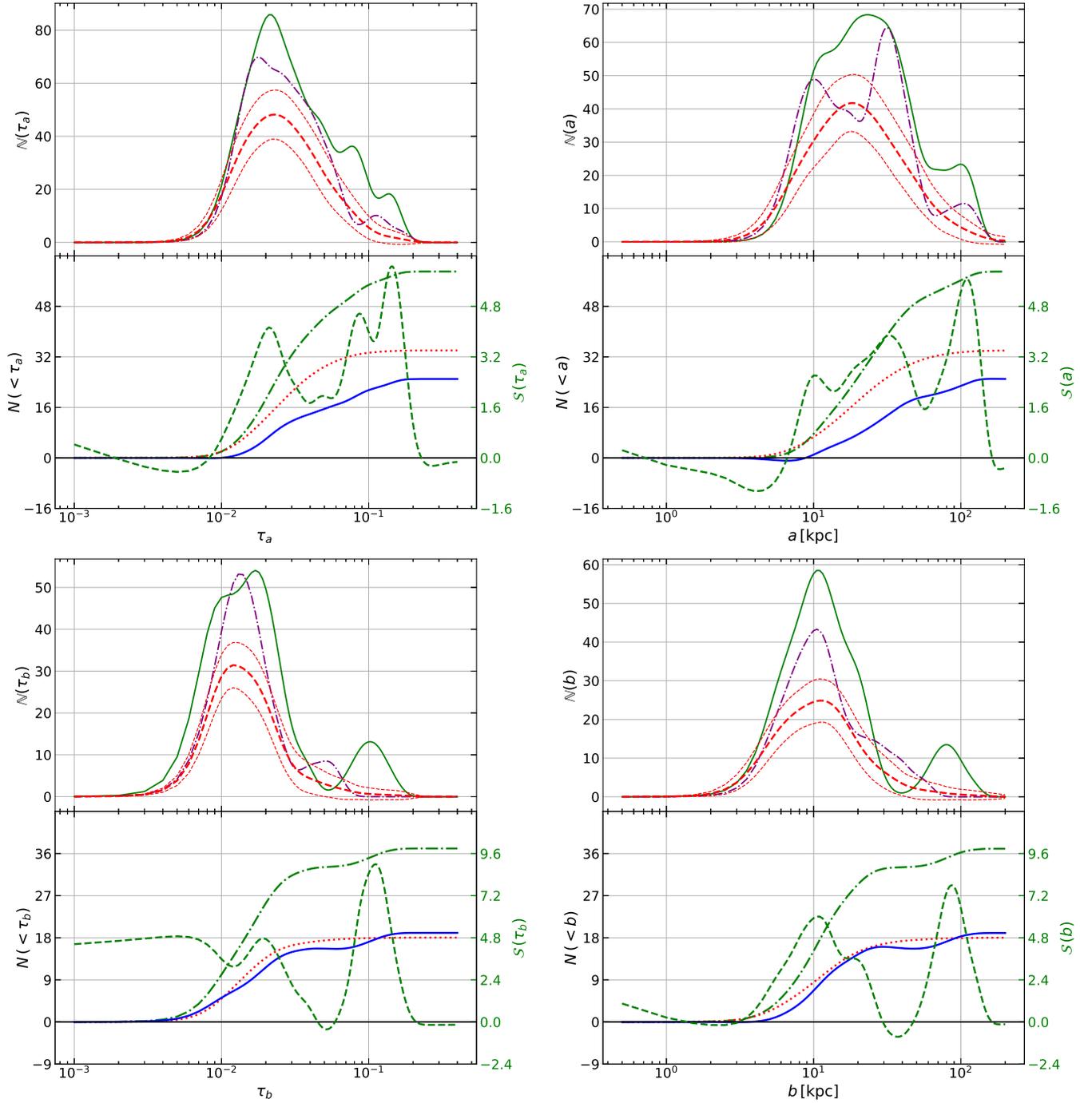

    \centerline{
    \DrawFig{\MYepsfbox{\myfig{radio_size_kpc_pdf_spec_Hou.eps}}}}
	\caption{\label{fig:radio_typ_ext1}\label{fig:radio_typ_phys1}\label{fig:radio_typ_ext2}\label{fig:radio_typ_phys_min1}
	The same as Fig.~\ref{fig:X_ray_typ_phys1}, but for the semi-major ($\mathbb{P} = a$, upper panels) and semi-minor ($\mathbb{P} = b$, lower panels) axes of radio sources. A variance $\sigma_{\rm smooth}^2(x)=(0.2x)^2$ is added to each scale $x$, for visibility.
}
\end{figure}
\end{center}

\begin{figure}[h!]
   \includegraphics[width=0.95\linewidth]{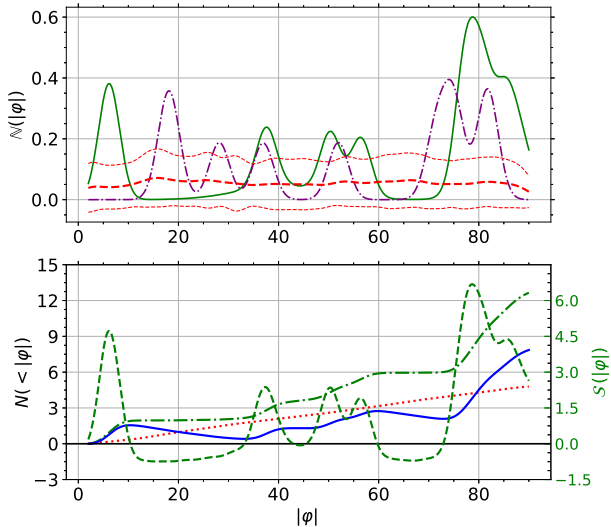}
	\caption{\label{fig:PropPA}
	Distribution of the angle $\mathbb{P}=|\varphi|$ between source elongation and cluster radius, as in Fig.~\ref{fig:PropAPA}, but for extended, $\te_a>60\arcsec$ sources.
    }
\end{figure}

\begin{figure}[h]
    \includegraphics[width=9cm]{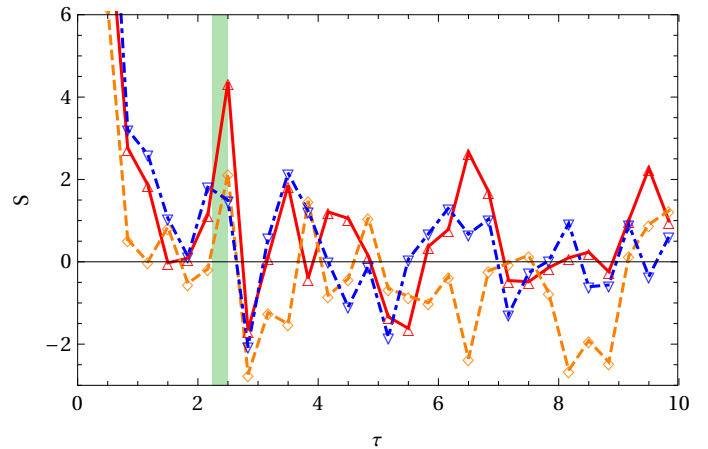}
	\caption{\label{fig:BroadbandRadio}
       Significance $S(\tau)$ profiles of SW-stacked NVSS (red up triangles), GMRT (orange diamonds), and VLSSr (blue down triangles) sources around the same clusters used for the NVSS sample. A low, $\Delta\tau=1/3$ resolution is used here, due to the smaller low-frequency catalogs.
	}
\end{figure}

\begin{figure}[h]
    \includegraphics[width=9cm]{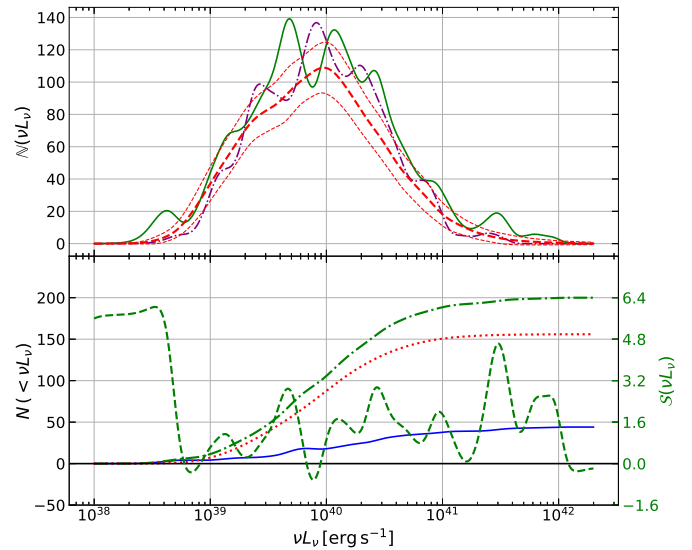}
	\caption{\label{fig:GMRT_lum}
   Same luminosity $\myP=\log_{10}\nu L_\nu$ distribution as in Fig.~\ref{fig:radio_typ_lum1}, but for GMRT sources, around the same clusters as in the NVSS sample.
	}
\end{figure}

Figures \ref{fig:PropLum2}--\ref{fig:GMRT_lum} demonstrate additional properties of the VS-excess sources, inferred in the same statistical method used in \S\ref{sec:characteristics}.

Figure \ref{fig:PropLum2} shows that the black-body X-ray fit indicates similar, $L_X\simeq 10^{42\mbox{\scriptsize{--}}43}\erg\se^{-1}$ luminosities as those inferred from the power-law fit. Next, a comparison of normalized and non-normalized sizes is shown for X-ray (Fig.~\ref{fig:X_ray_typ_ext1}) and radio (Fig.~\ref{fig:radio_typ_ext1}) sources.

\onecolumngrid
\twocolumngrid

\onecolumngrid
\twocolumngrid

\onecolumngrid
\twocolumngrid

\par\null\newpage
\clearpage

Figure \ref{fig:PropPA} shows that extended radio sources may preferentially be oriented with their long axis perpendicular to the cluster radius, \ie $|\varphi|\simeq \pi/2$, but this result is of low significance, based on only several low-declination (see Fig.~\ref{fig:PropAPA}) sources.

Finally, we examine the excess radio sources at lower frequencies, using the GMRT and VLSSr catalogs.
Comparing the stacked profiles of NVSS, GMRT, and VLSSr sources in Fig.~\ref{fig:BroadbandRadio} suggests that the signal may be broadened at $74\MHz$, as expected due to the enhanced cooling length, but the effect is of low significance due to the poor statistics of stacking the smaller, low-frequency catalogs.
Similarly, Fig.~\ref{fig:GMRT_lum} suggests that the luminosity of the excess GMRT sources is $\nu L_\nu(150\MHz)\simeq 5\times 10^{39\mbox{\scriptsize{--}}40}\erg\se^{-1}$, as expected based on the NVSS luminosities and the GMRT--NVSS spectra, but again, the low-frequency statistics are poor.

\par\null\newpage
\clearpage

\section{No significant 2RXS-NVSS pair excess}
\label{subsec:joint_xray_radio_sources}

\begin{figure}[h!]
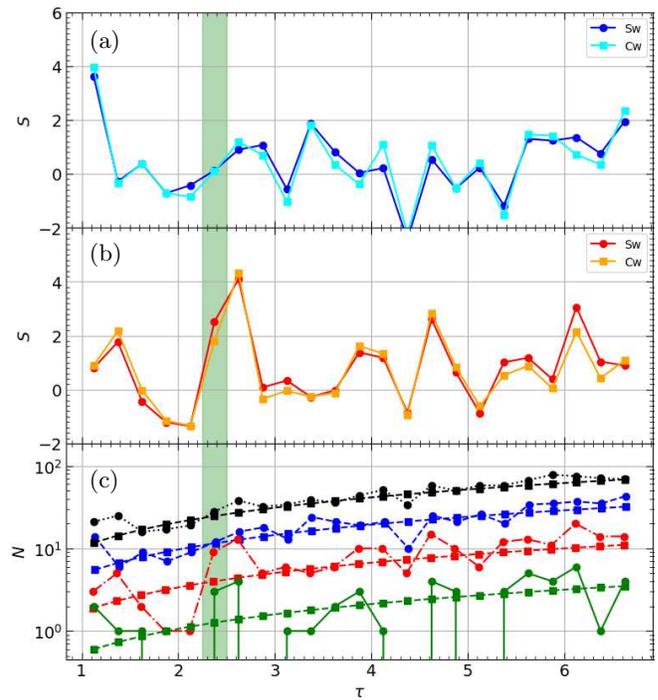

    \begin{center}
        \hspace*{-0.4cm}
       \begin{overpic}[width=1\linewidth,trim={0cm 0cm 17.9cm 0cm},clip]{\myfig{joint_N_1arcmin.eps}}
        \put (12,93) {\textcolor{black}{(a)}}
        \put (12,63) {\textcolor{black}{(b)}}
        \put (12,31) {\textcolor{black}{(c)}}
       \end{overpic}
   \end{center}
   	\caption{
    \label{fig:Pairs}
    Excess significance $S(\tau)$ (panels a and b; see legends) and total number $N(\tau)$ (panel c; real pairs as disks with guidelines to the eye; expected field pairs as squares) profiles of 2RXS-NVSS pairs.
    Imposing the nominal flux cuts (solid green lines in panel c) leaves insufficient pairs.
    Relaxing $F_\nu^*$ (panel a and dashed blue in panel c) yields no signal.
    Relaxing $F_X^*$ (panel b and dot-dashed red in panel c) or both $F_\nu^*$ and $F_X^*$ (dotted black in panel c) leaves strong fluctuations, with a $\tau>2.5$ signal we cannot substantiate.
    }
\end{figure}

In search of excess VS sources that radiate both in X-rays and in radio, we construct a catalog of 2RXS--NVSS pairs with an angular separation $<1'$, chosen such that $<30\%$ of these pairs are likely to be coincidental projected overlaps.
Stacking these pairs around our MCXC cluster sample does not produce a robust VS signal, as demonstrated in Fig.~\ref{fig:Pairs}.
We conclude that the X-ray and radio excess VS signals are not dominated by the same individual objects.

\section{Excess sources are not identified AGN}
\label{app:AGN_identification}

We examine the possible association of excess VS sources with previously-identified AGN.
For X-rays, 2RXS already provides associations with an AGN catalog \citep{Veron_QSOs}, whereas in radio, we cross-correlate NVSS with the same catalog using a proximity criterion of $<3'$ angular separation.
Almost all AGN associated with sources located in projection around an MCXC cluster are found to have redshifts inconsistent with the cluster.
In addition, we find no significant difference in the results of stacking sources with vs. without an AGN association.
Hence, we can rule out the possibility that a significant fraction of the sources comprising the excess VS signal is associated with these identified AGN.

\newpage

\onecolumngrid

\section{VS-region sources}\label{app:SourceTables}

Tables \ref{tab:VS_2RXS_list} and \ref{tab:VS_NVSS_list} list respectively the 2RXS and NVSS sources found in the VS region, naming the cluster tentatively associated with each source.
Note that statistically, only about $\sim1/3$ of these sources are associated with the localized VS excess, the rest attributed to field (foreground or background) sources.

\begin{center}
\begin{longtable*}{lll|lll}
\caption{2RXS sources in the VS region}
\label{tab:VS_2RXS_list} \\
\multicolumn{1}{c}{Source} & \multicolumn{1}{c}{R.A.} & \multicolumn{1}{c|}{Dec} & \multicolumn{1}{c}{Cluster} & \multicolumn{1}{c}{Alt. name} & \multicolumn{1}{c}{$z$} \\
\multicolumn{1}{c}{(1)} & \multicolumn{1}{c}{(2)} & \multicolumn{1}{c|}{(3)} & \multicolumn{1}{c}{(4)} & \multicolumn{1}{c}{(5)} & \multicolumn{1}{c}{(6)} \\ \hline \endfirsthead
\multicolumn{6}{c}%
{{\bfseries \tablename\ \thetable{} -- continued from previous page}} \\

\multicolumn{1}{c}{Source} & \multicolumn{1}{c}{R.A.} & \multicolumn{1}{c|}{Dec} & \multicolumn{1}{c}{Cluster} & \multicolumn{1}{c}{Alt. name} & \multicolumn{1}{c}{$z$} \\ \hline
\endhead

\hline \multicolumn{6}{r}{{Continued on next page}} \\
\endfoot

\hline
\endlastfoot
2RXS J213734.6-720725 & 324.3927 & -72.1233 & MCXC J2133.4-7156 &                  & 0.0559 \\
2RXS J043118.7-620535 & 67.8281 & -62.0940 & MCXC J0431.4-6126 & A3266            & 0.0589 \\
2RXS J034914.6-544813 & 57.3109 & -54.8043 & MCXC J0352.3-5453 & RBS 0485         & 0.0447 \\
2RXS J034324.8-530041 & 55.8534 & -53.0122 & MCXC J0342.8-5338 & A3158            & 0.0590 \\
2RXS J034517.9-530544 & 56.3248 & -53.0963 & MCXC J0342.8-5338 & A3158            & 0.0590 \\
2RXS J033852.2-532427 & 54.7174 & -53.4082 & MCXC J0342.8-5338 & A3158            & 0.0590 \\
2RXS J034710.8-533449 & 56.7951 & -53.5811 & MCXC J0342.8-5338 & A3158            & 0.0590 \\
2RXS J034329.7-541259 & 55.8740 & -54.2171 & MCXC J0342.8-5338 & A3158            & 0.0590 \\
2RXS J224403.1-525611 & 341.0127 & -52.9362 & MCXC J2246.3-5243 & A3911            & 0.0965 \\
2RXS J032345.2-513226 & 50.9385 & -51.5412 & MCXC J0321.9-5119 & A3120            & 0.0696 \\
2RXS J032133.8-514003 & 50.3907 & -51.6683 & MCXC J0321.9-5119 & A3120            & 0.0696 \\
2RXS J024638.7-461700 & 41.6615 & -46.2840 & MCXC J0245.2-4627 & A3047            & 0.0868 \\
2RXS J034211.4-455724 & 55.5477 & -45.9575 & MCXC J0340.8-4542 & RBS 0459         & 0.0698 \\
2RXS J224909.0-453551 & 342.2874 & -45.5975 & MCXC J2250.8-4521 & S1067            & 0.0511 \\
2RXS J232040.0-424706 & 350.1665 & -42.7850 & MCXC J2319.1-4206 & ACOS1111         & 0.0450 \\
2RXS J225144.6-371317 & 342.9362 & -37.2214 & MCXC J2249.2-3727 & S1065            & 0.0289 \\
2RXS J225159.4-373447 & 342.9978 & -37.5797 & MCXC J2249.2-3727 & S1065            & 0.0289 \\
2RXS J053523.7-362736 & 83.8495 & -36.4610 & MCXC J0533.3-3619 & S0535            & 0.0479 \\
2RXS J235847.4-340950 & 359.6976 & -34.1640 & MCXC J2357.0-3445 & A4059            & 0.0475 \\
2RXS J000606.1-350150 & 1.5255 & -35.0306 & MCXC J0006.0-3443 & A2721            & 0.1147 \\
2RXS J002301.4-330721 & 5.7563 & -33.1225 & MCXC J0025.5-3302 & S0041            & 0.0491 \\
2RXS J052401.7-320833 & 81.0078 & -32.1435 & MCXC J0525.5-3135 & A3341            & 0.0380 \\
2RXS J040615.2-304956 & 61.5638 & -30.8330 & MCXC J0408.2-3053 & A3223            & 0.0600 \\
2RXS J001131.5-282046 & 2.8817 & -28.3462 & MCXC J0011.3-2851 & A2734            & 0.0620 \\
2RXS J001247.7-291301 & 3.1991 & -29.2171 & MCXC J0011.3-2851 & A2734            & 0.0620 \\
2RXS J005049.8-284533 & 12.7079 & -28.7594 & MCXC J0051.3-2831 & A2829            & 0.1125 \\
2RXS J024421.9-262500 & 41.0917 & -26.4172 & MCXC J0244.1-2611 &                  & 0.1362 \\
2RXS J001358.9-233820 & 3.4958 & -23.6391 & MCXC J0015.4-2350 & A14              & 0.0645 \\
2RXS J010013.3-215003 & 15.0559 & -21.8344 & MCXC J0102.7-2152 & A0133            & 0.0569 \\
2RXS J005002.6-212709 & 12.5111 & -21.4528 & MCXC J0048.6-2114 & A2824            & 0.0581 \\
2RXS J215644.7-201430 & 329.1866 & -20.2414 & MCXC J2158.3-2006 & A2401            & 0.0570 \\
2RXS J045714.9-181757 & 74.3127 & -18.3002 & MCXC J0454.8-1806 & CID 28           & 0.0335 \\
2RXS J011459.1-153909 & 18.7467 & -15.6530 & MCXC J0116.1-1555 & SCG 16           & 0.0448 \\
2RXS J010832.7-155511 & 17.1369 & -15.9202 & MCXC J0108.9-1537 & A0151S           & 0.0970 \\
2RXS J010905.1-145729 & 17.2717 & -14.9584 & MCXC J0108.8-1524 & A0151N           & 0.0533 \\
2RXS J011029.5-151015 & 17.6233 & -15.1712 & MCXC J0108.8-1524 & A0151N           & 0.0533 \\
2RXS J005725.1-142056 & 14.3552 & -14.3492 & MCXC J0058.4-1425 & A123             & 0.0957 \\
2RXS J103914.9-090431 & 159.8131 & -9.0757 & MCXC J1039.7-0841 & A1069            & 0.0650 \\
2RXS J101848.3-062729 & 154.7021 & -6.4585 & MCXC J1020.4-0631 & A0978            & 0.0540 \\
2RXS J234648.0-014221 & 356.7006 & -1.7059 & MCXC J2347.4-0218 & HCG 97           & 0.0223 \\
2RXS J003606.6-021227 & 9.0281 & -2.2078 & MCXC J0034.6-0208 &                  & 0.0812 \\
2RXS J003545.5-020703 & 8.9403 & -2.1178 & MCXC J0034.2-0204 & SH518            & 0.0822 \\
2RXS J011731.8+002126 & 19.3833 & 0.3570 & MCXC J0115.2+0019 & A0168            & 0.0450 \\
2RXS J011326.7+000035 & 18.3620 & 0.0095 & MCXC J0115.2+0019 & A0168            & 0.0450 \\
2RXS J011703.9+000025 & 19.2671 & 0.0069 & MCXC J0115.2+0019 & A0168            & 0.0450 \\
2RXS J213637.9+004203 & 324.1585 & 0.7016 & MCXC J2137.1+0026 & VMF98 202        & 0.0509 \\
2RXS J013243.2+004109 & 23.1806 & 0.6856 & MCXC J0131.7+0033 & A208             & 0.0798 \\
2RXS J013038.7+004006 & 22.6622 & 0.6683 & MCXC J0131.7+0033 & A208             & 0.0798 \\
2RXS J010620.9+022017 & 16.5877 & 2.3380 & MCXC J0108.1+0210 & A0147            & 0.0436 \\
2RXS J121929.8+034850 & 184.8748 & 3.8140 & MCXC J1217.6+0339 & ZW 1215.1+0400   & 0.0766 \\
2RXS J122710.3+090646 & 186.7935 & 9.1131 & MCXC J1227.4+0849 & A1541            & 0.0896 \\
2RXS J025806.4+093216 & 44.5277 & 9.5373 & MCXC J0255.8+0918 & IC 1867          & 0.0258 \\
2RXS J110233.0+105722 & 165.6382 & 10.9561 & MCXC J1100.8+1033 & A1142            & 0.0353 \\
2RXS J235542.9+114259 & 358.9297 & 11.7166 & MCXC J2355.6+1120 & A2675            & 0.0726 \\
2RXS J224939.6+110016 & 342.4157 & 11.0048 & MCXC J2250.0+1137 & RX J2250.0+1137  & 0.0255 \\
2RXS J114410.5+151430 & 176.0442 & 15.2417 & MCXC J1145.3+1529 & A1371            & 0.0670 \\
2RXS J011459.8+152854 & 18.7499 & 15.4816 & MCXC J0113.0+1531 & A0160            & 0.0442 \\
2RXS J145502.0+164416 & 223.7590 & 16.7386 & MCXC J1454.4+1622 &                  & 0.0456 \\
2RXS J145502.0+164416 & 223.7590 & 16.7386 & MCXC J1452.9+1641 & A1983            & 0.0444 \\
2RXS J145427.1+162242 & 223.6136 & 16.3790 & MCXC J1452.9+1641 & A1983            & 0.0444 \\
2RXS J160624.0+181522 & 241.6005 & 18.2569 & MCXC J1604.5+1743 & A2151A           & 0.0370 \\
2RXS J160726.2+175017 & 241.8596 & 17.8390 & MCXC J1604.5+1743 & A2151A           & 0.0370 \\
2RXS J160624.0+181522 & 241.6005 & 18.2569 & MCXC J1606.8+1746 & A2151B           & 0.0321 \\
2RXS J124142.2+181103 & 190.4265 & 18.1845 & MCXC J1241.3+1834 & A1589            & 0.0730 \\
2RXS J232047.5+182931 & 350.1989 & 18.4924 & MCXC J2318.5+1842 & A2572            & 0.0403 \\
2RXS J232047.5+182931 & 350.1989 & 18.4924 & MCXC J2318.4+1843 & A2572B           & 0.0389 \\
2RXS J111124.7+212406 & 167.8532 & 21.4016 & MCXC J1109.7+2145 & A1177            & 0.0319 \\
2RXS J112204.0+245206 & 170.5169 & 24.8684 & MCXC J1122.3+2419 & HCG 51           & 0.0258 \\
2RXS J083659.6+243500 & 129.2488 & 24.5828 & MCXC J0838.1+2506 & CGCG120-014      & 0.0286 \\
2RXS J142324.0+264159 & 215.8503 & 26.7003 & MCXC J1423.1+2615 & RX J1423.1+2616  & 0.0375 \\
2RXS J134738.5+271451 & 206.9107 & 27.2479 & MCXC J1348.8+2635 & A1795            & 0.0622 \\
2RXS J135006.7+271253 & 207.5284 & 27.2153 & MCXC J1348.8+2635 & A1795            & 0.0622 \\
2RXS J134755.3+255544 & 206.9809 & 25.9292 & MCXC J1348.8+2635 & A1795            & 0.0622 \\
2RXS J233319.1+271429 & 353.3308 & 27.2417 & MCXC J2335.0+2722 & A2622            & 0.0613 \\
2RXS J152419.8+275742 & 231.0828 & 27.9625 & MCXC J1522.4+2742 & A2065            & 0.0723 \\
2RXS J120914.3+280712 & 182.3100 & 28.1201 & MCXC J1206.6+2811 & NGC 4104         & 0.0283 \\
2RXS J120745.5+273652 & 181.9399 & 27.6145 & MCXC J1206.6+2811 & NGC 4104         & 0.0283 \\
2RXS J091750.4+341607 & 139.4603 & 34.2682 & MCXC J0919.8+3345 & A779             & 0.0230 \\
2RXS J093429.7+333438 & 143.6241 & 33.5770 & MCXC J0933.4+3403 & UGC 05088        & 0.0269 \\
2RXS J133712.0+344230 & 204.3001 & 34.7087 & MCXC J1334.3+3441 & NGC 5223         & 0.0240 \\
2RXS J102045.2+381116 & 155.1885 & 38.1876 & MCXC J1022.0+3830 & RX J1022.1+3830  & 0.0491 \\
2RXS J102128.8+380655 & 155.3702 & 38.1150 & MCXC J1022.0+3830 & RX J1022.1+3830  & 0.0491 \\
2RXS J131156.8+385139 & 197.9870 & 38.8612 & MCXC J1311.1+3913 & A1691            & 0.0720 \\
2RXS J120722.0+385213 & 181.8417 & 38.8705 & MCXC J1205.2+3920 & RX J1205.1+3920  & 0.0381 \\
2RXS J171026.5+400332 & 257.6108 & 40.0598 & MCXC J1711.0+3941 & A2250            & 0.0647 \\
2RXS J163031.2+410137 & 247.6304 & 41.0277 & MCXC J1627.6+4055 & A2197            & 0.0301 \\
2RXS J162959.4+423008 & 247.4980 & 42.5030 & MCXC J1627.3+4240 & A2192            & 0.0317 \\
2RXS J141348.9+440015 & 213.4539 & 44.0047 & MCXC J1413.7+4339 & A1885            & 0.0890 \\
2RXS J141229.3+435534 & 213.1223 & 43.9267 & MCXC J1413.7+4339 & A1885            & 0.0890 \\
2RXS J171124.1+434903 & 257.8508 & 43.8184 & MCXC J1714.3+4341 & NGC 6329         & 0.0276 \\
2RXS J152115.3+485847 & 230.3139 & 48.9806 & MCXC J1520.9+4840 & A2064            & 0.1076 \\
2RXS J113803.7+490500 & 174.5151 & 49.0835 & MCXC J1134.8+4903 & A1314            & 0.0341 \\
2RXS J113408.7+483414 & 173.5360 & 48.5705 & MCXC J1134.8+4903 & A1314            & 0.0341 \\
2RXS J161157.2+490138 & 242.9884 & 49.0280 & MCXC J1614.1+4909 & A2169            & 0.0579 \\
2RXS J142404.6+493154 & 216.0190 & 49.5323 & MCXC J1421.5+4933 & ZW 1420.2+4952   & 0.0716 \\
2RXS J142354.7+492828 & 215.9781 & 49.4749 & MCXC J1421.5+4933 & ZW 1420.2+4952   & 0.0716 \\
2RXS J141945.7+491822 & 214.9405 & 49.3067 & MCXC J1421.5+4933 & ZW 1420.2+4952   & 0.0716 \\
2RXS J102406.0+500302 & 156.0250 & 50.0504 & MCXC J1022.5+5006 & A980             & 0.1580 \\
2RXS J100837.2+543735 & 152.1549 & 54.6262 & MCXC J1010.2+5430 & VMF98 84         & 0.0450 \\
2RXS J105939.2+563211 & 164.9127 & 56.5363 & MCXC J1058.4+5647 & A1132            & 0.1369 \\
2RXS J143027.2+565003 & 217.6129 & 56.8346 & MCXC J1428.4+5652 & A1925            & 0.1051 \\
2RXS J162948.3+584616 & 247.4515 & 58.7720 & MCXC J1629.7+5831 & A2208            & 0.1329 \\
2RXS J133520.4+584501 & 203.8343 & 58.7508 & MCXC J1336.1+5912 & A1767            & 0.0701 \\
2RXS J170902.9+641730 & 257.2623 & 64.2925 & MCXC J1712.7+6403 & A2255            & 0.0809 \\
2RXS J171640.0+641048 & 259.1670 & 64.1808 & MCXC J1712.7+6403 & A2255            & 0.0809 \\
2RXS J171708.3+640146 & 259.2850 & 64.0304 & MCXC J1712.7+6403 & A2255            & 0.0809 \\
2RXS J175341.7+654242 & 268.4244 & 65.7126 & MCXC J1751.1+6531 & NGC6505          & 0.0428 \\
2RXS J174833.2+652351 & 267.1389 & 65.3985 & MCXC J1751.1+6531 & NGC6505          & 0.0428 \\
2RXS J113146.0+660731 & 172.9404 & 66.1254 & MCXC J1133.2+6622 & A1302            & 0.1160 \\
2RXS J174532.3+674811 & 266.3849 & 67.8039 & MCXC J1742.7+6735 &                  & 0.0420 \\
2RXS J174603.5+672706 & 266.5151 & 67.4525 & MCXC J1742.7+6735 &                  & 0.0420 \\
2RXS J175614.0+680708 & 269.0588 & 68.1198 & MCXC J1755.7+6752 &                  & 0.0833 \\
2RXS J173039.0+681130 & 262.6628 & 68.1925 & MCXC J1736.3+6803 & ZW 1745.6+6703   & 0.0248 \\
2RXS J174241.6+680013 & 265.6739 & 68.0045 & MCXC J1736.3+6803 & ZW 1745.6+6703  & 0.0248 \\
2RXS J170234.2+681129 & 255.6427 & 68.1922 & MCXC J1659.6+6826 &                  & 0.0504 \\
2RXS J170102.2+680513 & 255.2593 & 68.0878 & MCXC J1659.6+6826 &                  & 0.0504 \\
2RXS J172207.7+694303 & 260.5323 & 69.7183 & MCXC J1724.2+6956 &                  & 0.0386 \\
2RXS J172400.1+694026 & 261.0006 & 69.6749 & MCXC J1724.2+6956 &                  & 0.0386 \\
2RXS J115509.4+734312 & 178.7870 & 73.7202 & MCXC J1156.0+7325 & A1412            & 0.0830 \\
\end{longtable*}
    \begin{tablenotes}
    \item
    Table columns: (1) Source name; (2) Right ascension (deg); (3) Declination (deg); (4) Name of tentatively-associated cluster; (5) Alternative cluster name; (6) Cluster redshift.
    Note that each of the two sources, 2RXS J145502.0+164416 and 2RXS J232047.5+182931, appears twice, as it falls in the VS radii of two clusters. Such double counting affects field sources, too, and has a negligible effect on our results.
    \end{tablenotes}
\end{center}

\begin{center}
\begin{longtable*}{lll|lll}
\caption{NVSS sources in the VS region}
\label{tab:VS_NVSS_list} \\
\multicolumn{1}{c}{Source} & \multicolumn{1}{c}{R.A.} & \multicolumn{1}{c|}{Dec} & \multicolumn{1}{c}{Cluster} & \multicolumn{1}{c}{Alt. name} & \multicolumn{1}{c}{$z$} \\
\multicolumn{1}{c}{(1)} & \multicolumn{1}{c}{(2)} & \multicolumn{1}{c|}{(3)} & \multicolumn{1}{c}{(4)} & \multicolumn{1}{c}{(5)} & \multicolumn{1}{c}{(6)} \\ \hline \endfirsthead

\multicolumn{6}{c}%
{{\bfseries \tablename\ \thetable{} -- continued from previous page}} \\
\multicolumn{1}{c}{Source} & \multicolumn{1}{c}{R.A.} & \multicolumn{1}{c|}{Dec} & \multicolumn{1}{c}{Cluster} & \multicolumn{1}{c}{Alt. name} & \multicolumn{1}{c}{$z$} \\ \hline
\endhead

\hline \multicolumn{6}{r}{{Continued on next page}} \\
\endfoot

\hline
\endlastfoot
NVSS J000029-345224  & 0.1227 & -34.8734 & MCXC J2357.0-3445 & A4059           & 0.0475 \\
NVSS J011413-321800  & 18.5563 & -32.3002 & MCXC J0113.9-3145 & S141A           & 0.0191 \\
NVSS J011448-321951  & 18.7037 & -32.3310 & MCXC J0113.9-3145 & S141A           & 0.0191 \\
NVSS J011632-314001  & 19.1349 & -31.6671 & MCXC J0113.9-3145 & S141A           & 0.0191 \\
NVSS J004839-294718  & 12.1645 & -29.7885 & MCXC J0049.4-2931 & S0084           & 0.1084 \\
NVSS J005037-292128  & 12.6577 & -29.3578 & MCXC J0049.4-2931 & S0084           & 0.1084 \\
NVSS J000956-282923  & 2.4846 & -28.4898 & MCXC J0011.3-2851 & A2734           & 0.0620 \\
NVSS J001001-282940  & 2.5052 & -28.4944 & MCXC J0011.3-2851 & A2734           & 0.0620 \\
NVSS J054438-253222  & 86.1613 & -25.5395 & MCXC J0545.4-2556 & A0548W          & 0.0424 \\
NVSS J054916-260421  & 87.3176 & -26.0727 & MCXC J0548.6-2527 & A0548E          & 0.0420 \\
NVSS J055016-255502  & 87.5699 & -25.9173 & MCXC J0548.6-2527 & A0548E          & 0.0420 \\
NVSS J001403-233834  & 3.5152 & -23.6430 & MCXC J0015.4-2350 & A14             & 0.0645 \\
NVSS J234326-215422  & 355.8610 & -21.9062 & MCXC J2344.4-2153 & A2655           & 0.1122 \\
NVSS J220245-214333  & 330.6908 & -21.7259 & MCXC J2203.8-2130 &                 & 0.0732 \\
NVSS J055158-211948  & 87.9928 & -21.3302 & MCXC J0552.8-2103 & A0550           & 0.0989 \\
NVSS J082717-202622  & 126.8221 & -20.4397 & MCXC J0826.7-2007 & S611            & 0.0876 \\
NVSS J215707-202441  & 329.2812 & -20.4116 & MCXC J2158.3-2006 & A2401           & 0.0570 \\
NVSS J023743-193235  & 39.4324 & -19.5432 & MCXC J0236.6-1923 & A0367           & 0.0907 \\
NVSS J221145-122639  & 332.9387 & -12.4442 & MCXC J2210.3-1210 & A2420           & 0.0846 \\
NVSS J091541-114807  & 138.9248 & -11.8022 & MCXC J0918.1-1205 & A0780           & 0.0539 \\
NVSS J104108-114139  & 160.2867 & -11.6944 & MCXC J1041.5-1123 &                 & 0.0839 \\
NVSS J221519-090004  & 333.8319 & -9.0011 & MCXC J2216.2-0920 & A2428           & 0.0825 \\
NVSS J094446-082722  & 146.1937 & -8.4562 & MCXC J0945.4-0839 & A0868           & 0.1535 \\
NVSS J205703-075258  & 314.2643 & -7.8830 & MCXC J2058.2-0745 & A2331           & 0.0793 \\
NVSS J234515-035930  & 356.3143 & -3.9918 & MCXC J2344.2-0422 &                 & 0.0786 \\
NVSS J033834-022200  & 54.6449 & -2.3668 & MCXC J0340.6-0239 &                 & 0.0352 \\
NVSS J003438-014309  & 8.6595 & -1.7192 & MCXC J0034.2-0204 & SH518           & 0.0822 \\
NVSS J222347-020135  & 335.9484 & -2.0265 & MCXC J2223.8-0138 & A2440           & 0.0906 \\
NVSS J101444-000417  & 153.6845 & -0.0715 & MCXC J1013.7-0006 & A0954           & 0.0927 \\
NVSS J234057+000447  & 355.2402 & 0.0797 & MCXC J2341.1+0018 &                 & 0.1100 \\
NVSS J011546+005204  & 18.9441 & 0.8679 & MCXC J0115.2+0019 & A0168           & 0.0450 \\
NVSS J213638+004154  & 324.1607 & 0.6985 & MCXC J2137.1+0026 & VMF98 202       & 0.0509 \\
NVSS J101135+005747  & 152.8969 & 0.9631 & MCXC J1013.6-0054 & A0957           & 0.0445 \\
NVSS J091800+003450  & 139.5004 & 0.5808 & MCXC J0920.0+0102 & MKW 1S          & 0.0175 \\
NVSS J092035+002330  & 140.1490 & 0.3918 & MCXC J0920.0+0102 & MKW 1S          & 0.0175 \\
NVSS J092124+013834  & 140.3506 & 1.6428 & MCXC J0920.0+0102 & MKW 1S          & 0.0175 \\
NVSS J023101+014958  & 37.7542 & 1.8330 & MCXC J0231.9+0114 & RCS145          & 0.0221 \\
NVSS J102710+033927  & 156.7941 & 3.6577 & MCXC J1028.3+0346 & A1024           & 0.0730 \\
NVSS J235254+060033  & 358.2277 & 6.0092 & MCXC J2350.8+0609 & A2665           & 0.0562 \\
NVSS J003808+071223  & 9.5341 & 7.2066 & MCXC J0040.0+0649 & A76             & 0.0395 \\
NVSS J082152+080430  & 125.4679 & 8.0751 & MCXC J0821.0+0751 & RX J0820.9+0751 & 0.1100 \\
NVSS J122631+090037  & 186.6315 & 9.0105 & MCXC J1227.4+0849 & A1541           & 0.0896 \\
NVSS J041528+103317  & 63.8691 & 10.5548 & MCXC J0413.4+1028 & A0478           & 0.0882 \\
NVSS J114049+105754  & 175.2069 & 10.9651 & MCXC J1141.2+1044 & A1345           & 0.1090 \\
NVSS J122901+120012  & 187.2546 & 12.0034 & MCXC J1229.9+1147 & A1552           & 0.0852 \\
NVSS J122935+120548  & 187.3960 & 12.0968 & MCXC J1229.9+1147 & A1552           & 0.0852 \\
NVSS J123056+115933  & 187.7336 & 11.9927 & MCXC J1229.9+1147 & A1552           & 0.0852 \\
NVSS J025829+133328  & 44.6227 & 13.5579 & MCXC J0257.8+1302 & A0399           & 0.0722 \\
NVSS J025831+133417  & 44.6329 & 13.5715 & MCXC J0257.8+1302 & A0399           & 0.0722 \\
NVSS J025914+132735  & 44.8122 & 13.4598 & MCXC J0257.8+1302 & A0399           & 0.0722 \\
NVSS J025634+133435  & 44.1457 & 13.5765 & MCXC J0258.9+1334 & A0401           & 0.0739 \\
NVSS J232309+140444  & 350.7902 & 14.0789 & MCXC J2324.3+1439 & A2593           & 0.0428 \\
NVSS J091204+161829  & 138.0167 & 16.3083 & MCXC J0912.4+1556 & A763            & 0.0851 \\
NVSS J145602+162702  & 224.0084 & 16.4508 & MCXC J1454.4+1622 &                 & 0.0456 \\
NVSS J145605+162652  & 224.0235 & 16.4480 & MCXC J1454.4+1622 &                 & 0.0456 \\
NVSS J145608+162646  & 224.0339 & 16.4462 & MCXC J1454.4+1622 &                 & 0.0456 \\
NVSS J123535+164732  & 188.8974 & 16.7924 & MCXC J1236.4+1631 & A1569           & 0.0780 \\
NVSS J145420+162055  & 223.5846 & 16.3488 & MCXC J1452.9+1641 & A1983           & 0.0444 \\
NVSS J145430+162250  & 223.6253 & 16.3806 & MCXC J1452.9+1641 & A1983           & 0.0444 \\
NVSS J232609+171745  & 351.5393 & 17.2960 & MCXC J2323.8+1648 & A2589           & 0.0416 \\
NVSS J160148+175401  & 240.4511 & 17.9004 & MCXC J1604.5+1743 & A2151A          & 0.0370 \\
NVSS J160616+181459  & 241.5669 & 18.2498 & MCXC J1604.5+1743 & A2151A          & 0.0370 \\
NVSS J160616+181459  & 241.5669 & 18.2498 & MCXC J1606.8+1746 & A2151B          & 0.0321 \\
NVSS J075000+182311  & 117.5014 & 18.3866 & MCXC J0748.1+1832 &                 & 0.0400 \\
NVSS J231822+191452  & 349.5954 & 19.2478 & MCXC J2317.1+1841 & A2572A          & 0.0422 \\
NVSS J232046+182925  & 350.1948 & 18.4905 & MCXC J2318.5+1842 & A2572           & 0.0403 \\
NVSS J232046+182925  & 350.1948 & 18.4905 & MCXC J2318.4+1843 & A2572B          & 0.0389 \\
NVSS J003912+213405  & 9.8002 & 21.5681 & MCXC J0039.6+2114 & A0075           & 0.0619 \\
NVSS J104702+221033  & 161.7617 & 22.1761 & MCXC J1048.7+2214 & A1100           & 0.0450 \\
NVSS J115427+233804  & 178.6144 & 23.6346 & MCXC J1155.3+2324 & A1413           & 0.1427 \\
NVSS J160212+241010  & 240.5525 & 24.1695 & MCXC J1604.9+2355 & AWM 4           & 0.0326 \\
NVSS J115752+240748  & 179.4701 & 24.1302 & MCXC J1156.9+2415 & ZWCL4673        & 0.1392 \\
NVSS J112145+245201  & 170.4414 & 24.8672 & MCXC J1122.3+2419 & HCG 51          & 0.0258 \\
NVSS J112331+235047  & 170.8832 & 23.8465 & MCXC J1122.3+2419 & HCG 51          & 0.0258 \\
NVSS J005122+242232  & 12.8430 & 24.3756 & MCXC J0049.8+2426 & A104            & 0.0815 \\
NVSS J204130+244822  & 310.3752 & 24.8062 & MCXC J2042.1+2426 &                 & 0.1019 \\
NVSS J204140+244924  & 310.4190 & 24.8234 & MCXC J2042.1+2426 &                 & 0.1019 \\
NVSS J005655+263125  & 14.2296 & 26.5236 & MCXC J0055.9+2622 & A115            & 0.1971 \\
NVSS J134612+261413  & 206.5509 & 26.2371 & MCXC J1348.8+2635 & A1795           & 0.0622 \\
NVSS J134756+255812  & 206.9839 & 25.9702 & MCXC J1348.8+2635 & A1795           & 0.0622 \\
NVSS J135011+271338  & 207.5472 & 27.2275 & MCXC J1348.8+2635 & A1795           & 0.0622 \\
NVSS J172026+265456  & 260.1105 & 26.9157 & MCXC J1720.1+2637 & RX J1720.1+2638 & 0.1644 \\
NVSS J005654+270909  & 14.2266 & 27.1527 & MCXC J0058.9+2657 & RX J0058.9+2657 & 0.0451 \\
NVSS J111018+292016  & 167.5758 & 29.3379 & MCXC J1110.7+2842 & A1185           & 0.0314 \\
NVSS J075248+294630  & 118.2031 & 29.7750 & MCXC J0753.4+2921 & A602            & 0.0621 \\
NVSS J004204+294712  & 10.5179 & 29.7869 & MCXC J0040.4+2933 & A0077           & 0.0712 \\
NVSS J154034+310239  & 235.1442 & 31.0442 & MCXC J1539.8+3043 & A2110           & 0.0980 \\
NVSS J015715+315417  & 29.3137 & 31.9049 & MCXC J0157.4+3213 & A278            & 0.0894 \\
NVSS J170011+323533  & 255.0491 & 32.5926 & MCXC J1659.0+3229 &                 & 0.0621 \\
NVSS J170013+322549  & 255.0566 & 32.4305 & MCXC J1659.0+3229 &                 & 0.0621 \\
NVSS J132218+325254  & 200.5753 & 32.8818 & MCXC J1320.2+3308 & NGC 5098        & 0.0362 \\
NVSS J160047+325445  & 240.1991 & 32.9126 & MCXC J1600.3+3313 & A2145           & 0.0880 \\
NVSS J101735+334031  & 154.3995 & 33.6754 & MCXC J1016.3+3338 & A961            & 0.1240 \\
NVSS J174018+350549  & 265.0763 & 35.0969 & MCXC J1740.5+3538 & RX J1740.5+3539 & 0.0428 \\
NVSS J174132+360724  & 265.3845 & 36.1235 & MCXC J1740.5+3538 & RX J1740.5+3539 & 0.0428 \\
NVSS J105544+374601  & 163.9356 & 37.7671 & MCXC J1057.7+3738 &                 & 0.0353 \\
NVSS J185527+374256  & 283.8652 & 37.7156 & MCXC J1857.6+3800 &                 & 0.0567 \\
NVSS J015823+413739  & 29.5992 & 41.6276 & MCXC J0157.1+4120 & A276            & 0.0810 \\
NVSS J084513+423918  & 131.3047 & 42.6553 & MCXC J0844.9+4258 &                 & 0.0541 \\
NVSS J171123+434842  & 257.8475 & 43.8118 & MCXC J1714.3+4341 & NGC 6329        & 0.0276 \\
NVSS J171202+431555  & 258.0089 & 43.2653 & MCXC J1714.3+4341 & NGC 6329        & 0.0276 \\
NVSS J171205+431701  & 258.0234 & 43.2839 & MCXC J1714.3+4341 & NGC 6329        & 0.0276 \\
NVSS J063932+474721  & 99.8868 & 47.7892 & MCXC J0638.1+4747 & ZWCL 0634.1+4750& 0.1740 \\
NVSS J102310+475146  & 155.7940 & 47.8628 & MCXC J1025.0+4750 & A1003           & 0.0520 \\
NVSS J090650+500334  & 136.7102 & 50.0597 & MCXC J0907.8+4936 & VV 196          & 0.0352 \\
NVSS J180947+492544  & 272.4500 & 49.4291 & MCXC J1811.0+4954 & ZWCL8338        & 0.0501 \\
NVSS J184203+502404  & 280.5142 & 50.4012 & MCXC J1843.6+5021 &                 & 0.1158 \\
NVSS J164934+535817  & 252.3950 & 53.9716 & MCXC J1649.2+5325 & ARP 330, SHK 016& 0.0290 \\
NVSS J160016+533945  & 240.0697 & 53.6625 & MCXC J1601.3+5354 & A2149           & 0.1068 \\
NVSS J124538+551134  & 191.4106 & 55.1929 & MCXC J1247.3+5500 & A1616           & 0.0830 \\
NVSS J134442+555313  & 206.1759 & 55.8870 & MCXC J1343.7+5538 & A1783           & 0.0766 \\
NVSS J114558+552222  & 176.4947 & 55.3730 & MCXC J1147.3+5544 & A1377           & 0.0510 \\
NVSS J142824+563611  & 217.1032 & 56.6032 & MCXC J1428.4+5652 & A1925           & 0.1051 \\
NVSS J172338+562857  & 260.9112 & 56.4827 & MCXC J1723.3+5658 & NGC 6370        & 0.0272 \\
NVSS J165438+583617  & 253.6584 & 58.6047 & MCXC J1654.7+5854 & A2239           & 0.0869 \\
NVSS J175341+654238  & 268.4246 & 65.7108 & MCXC J1751.1+6531 & NGC6505         & 0.0428 \\
NVSS J163540+655813  & 248.9171 & 65.9705 & MCXC J1635.8+6612 & A2218           & 0.1709 \\
NVSS J113025+662605  & 172.6060 & 66.4349 & MCXC J1133.2+6622 & A1302           & 0.1160 \\
NVSS J072754+665812  & 111.9751 & 66.9701 & MCXC J0724.9+6658 & A0578           & 0.0863 \\
NVSS J062025+673640  & 95.1045 & 67.6111 & MCXC J0618.6+6724 & A0554           & 0.1127 \\
NVSS J175049+680826  & 267.7083 & 68.1406 & MCXC J1754.6+6803 & ZW 1754.5+6807  & 0.0770 \\
NVSS J173339+683557  & 263.4159 & 68.5994 & MCXC J1736.3+6803 & ZW 1745.6+6703  & 0.0248 \\
NVSS J185609+680634  & 284.0397 & 68.1095 & MCXC J1853.9+6822 & A2312           & 0.0928 \\
NVSS J181702+694125  & 274.2606 & 69.6904 & MCXC J1814.2+6939 & A2301           & 0.0863 \\
NVSS J185101+732316  & 282.7568 & 73.3879 & MCXC J1847.2+7320 & A2310           & 0.1125 \\
NVSS J115626+730650  & 179.1125 & 73.1141 & MCXC J1156.0+7325 & A1412           & 0.0830 \\
NVSS J171206+774613  & 258.0265 & 77.7704 & MCXC J1718.1+7801 & A2271           & 0.0584 \\
NVSS J165135+782853  & 252.8989 & 78.4815 & MCXC J1703.8+7838 & A2256           & 0.0581 \\
\end{longtable*}
    \begin{tablenotes}
    \item
    Table columns: (1) Source name; (2) Right ascension (deg); (3) Declination (deg); (4) Name of tentatively-associated cluster; (5) Alternative cluster name; (6) Cluster redshift.
    Note that the source NVSS J160616+181459 appears twice, as it falls in the VS radius of two clusters.
    \end{tablenotes}
\end{center}

\putbib
\end{bibunit}

\end{document}

%% file: Definitions.tex
\newcommand{\ie}{\emph{i.e.} }
\newcommand{\eg}{\emph{e.g.,} }

\newcommand{\be}{\begin{equation}}
\newcommand{\ee}{\end{equation}}
\newcommand{\bea}{\begin{equation*}}
\newcommand{\eea}{\end{equation*}}
\newcommand{\beq}{\begin{equation} }
\newcommand{\eeq}{\end{equation}}
\newcommand{\beqr}{\begin{eqnarray} \nonumber}
\newcommand{\eeqr}{\end{eqnarray}}
\newcommand{\beqrb}{\begin{eqnarray}}
\newcommand{\eeqrb}{\nonumber \end{eqnarray}}
\newcommand{\fin}{\mbox{ .}}
\newcommand{\coma}{\mbox{ ,}}

\newcommand{\cm}{\mbox{ cm}}
\newcommand{\sr}{\mbox{ sr}}
\newcommand{\se}{\mbox{ s}}

\newcommand{\Gyr}{\mbox{ Gyr}}
\newcommand{\erg}{\mbox{ erg}}

\newcommand{\MHz}{\mbox{ MHz}}
\newcommand{\GHz}{\mbox{ GHz}}

\newcommand{\km}{\mbox{ km}}
\newcommand{\pc}{\mbox{ pc}}
\newcommand{\kpc}{\mbox{ kpc}}
\newcommand{\Mpc}{\mbox{ Mpc}}
\newcommand{\eV}{\mbox{ eV}}
\newcommand{\keV}{\mbox{ keV}}

\newcommand{\muG}{\mbox{ $\mu$G}}

\newcommand{\mJy}{\mbox{ mJy}}

\newcommand{\gama}{$\gamma$}

\newcommand{\te}{\theta}

\newcommand{\dgr}{^{\circ}}
